\documentclass[prd,preprint,showpacs,amssymb,superscriptaddress]{revtex4}
\usepackage[dvips]{graphics,color,epsfig,rotating}
\usepackage{multirow}
\newcommand{\BR}{\mbox{${\cal B}$}} 
\providecommand{\KPPNN}{\mbox{$K^+\to\pi^+\nu\bar{\nu}$}} 
\providecommand{\PNNSM}{\mbox {$(0.74\pm0.20)\times 10^{-10}$}} 
\providecommand{\KETHREE}{\mbox{$K^+\to\pi^0e^+\nu$}} 
\providecommand{\PIMUE}{\mbox{$\pi^+\to\mu^+\to e^+$}} 
\providecommand{\KPITWO}{\mbox{$K_{\pi2}$}}  
\providecommand{\KMUTWO}{\mbox{\mbox{$K_{\mu2}$}}}

\providecommand{\PIMU}{\mbox{$\pi^+\to\mu^+\nu_{\mu}$}} 
\providecommand{\MUE}{\mbox{$\mu^+\to e^+\nu_{e}\bar{\nu}_{\mu}$}} 
 
\providecommand{\PIMUE}{\mbox{$\pi^+\to\mu^+\to e^+$}} 
\providecommand{\KPIX}{\mbox{$K^+\to\pi^+X^0$}}  
 
\preprint{BNL/79257-2007-JA} 
\preprint{KEK-2007-34} 
\preprint{TRI-PP-07-18}
\preprint{TUHEP-EX-07-002} 
 
\begin{document}

 
\title{Measurement of the \KPPNN~Branching Ratio} 
 
\author{S.~Adler}\affiliation{Brookhaven National Laboratory, Upton, NY 11973}
\author{V.V.~Anisimovsky}\affiliation{Institute for Nuclear Research RAS, 60 October Revolution Pr. 7a, 117312 Moscow, Russia}
\author{M. Aoki}\altaffiliation{Present address: Department of Physics, Osaka University, Osaka 560-0043, Japan.}\affiliation{TRIUMF, 4004 Wesbrook Mall, Vancouver, British Columbia, Canada V6T 2A3}  
\author{M.~Ardebili}\affiliation{Joseph Henry Laboratories, Princeton University, Princeton, New Jersey 08544} 
\author{A.V.~Artamonov}\affiliation{Institute for High Energy Physics, Protvino, Moscow Region, 142 280, Russia} 
\author{M.~Atiya}\affiliation{Brookhaven National Laboratory, Upton, NY 11973}
\author{B.~Bassalleck}\affiliation{Department of Physics and Astronomy, University of New Mexico, Albuquerque, NM 87131} 
\author{A.O.~Bazarko}\affiliation{Joseph Henry Laboratories, Princeton University, Princeton, New Jersey 08544} 
\author{B.~Bhuyan}\altaffiliation{Also at Department of Physics, University of Delhi, Delhi 1100007, India} \affiliation{Brookhaven National Laboratory, Upton, NY 11973} 
\author{E.W.~Blackmore}\affiliation{TRIUMF, 4004 Wesbrook Mall, Vancouver, British Columbia, Canada V6T 2A3} 
\author{D.A.~Bryman}\affiliation{Department of Physics and Astronomy, University of British Columbia, Vancouver, British Columbia, Canada V6T 1Z1} 
\author{S.~Chen}\affiliation{Department of Engineering Physics, Tsinghua University, Beijing 100084, China}\affiliation{TRIUMF, 4004 Wesbrook Mall, Vancouver, British Columbia, Canada V6T 2A3} 
\author{I-H.~Chiang}\affiliation{Brookhaven National Laboratory, Upton, NY 11973} 
\author{I.-A.~Christidi}\altaffiliation{Present address: Department of Physics, Aristotle University of Thessaloniki, Thessaloniki 54124, Greece.}\affiliation{Department of Physics and Astronomy, Stony Brook University, Stony Brook, NY 11794} 
\author{M.R.~Convery}\affiliation{Joseph Henry Laboratories, Princeton University, Princeton, New Jersey 08544} 
\author{P.S.~Cooper}\affiliation{Fermi National Accelerator Laboratory, Batavia, IL 60510} 
\author{M.V.~Diwan}\affiliation{Brookhaven National Laboratory, Upton, NY 11973} 
\author{J.S.~Frank}\affiliation{Brookhaven National Laboratory, Upton, NY 11973} 
\author{T.~Fujiwara}\affiliation{Department of Physics, Kyoto University, Sakyo-ku, Kyoto 606-8502, Japan} 
\author{J.~Haggerty}\affiliation{Brookhaven National Laboratory, Upton, NY 11973} 
\author{J.~Hu}\affiliation{TRIUMF, 4004 Wesbrook Mall, Vancouver, British Columbia, Canada V6T 2A3}
\author{T. Inagaki}\affiliation{High Energy Accelerator Research Organization (KEK), Oho, Tsukuba, Ibaraki 305-0801, Japan}  
\author{M.M.~Ito}\affiliation{Joseph Henry Laboratories, Princeton University, Princeton, New Jersey 08544}  
\author{A.P.~Ivashkin}\affiliation{Institute for Nuclear Research RAS, 60 October Revolution Pr. 7a, 117312 Moscow, Russia} 
\author{D.E.~Jaffe}\affiliation{Brookhaven National Laboratory, Upton, NY 11973} 
\author{S.~Kabe}\affiliation{High Energy Accelerator Research Organization (KEK), Oho, Tsukuba, Ibaraki 305-0801, Japan} 
\author{M. Kazumori}\altaffiliation{Also at Graduate School of Science, The University of Tokyo, 
Tokyo 113-0033, Japan.}\affiliation{High Energy Accelerator Research Organization (KEK), Oho, Tsukuba, Ibaraki 305-0801, Japan} 
\author{Y. Kuno}\altaffiliation{Present address: Department of Physics, Osaka University, 
Osaka 560-0043, Japan.}\affiliation{High Energy Accelerator Research Organization (KEK), Oho, Tsukuba, Ibaraki 305-0801, Japan}
\author{M. Kuriki}\altaffiliation{Present address: Graduate School of 
Advanced Sciences of Matter,Hiroshima University, Hiroshima, 739-8530, Japan.}\affiliation{High Energy Accelerator Research Organization (KEK), Oho, Tsukuba, Ibaraki 305-0801, Japan}
\author{S.H.~Kettell}\affiliation{Brookhaven National Laboratory, Upton, NY 11973} 
\author{M.M.~Khabibullin}\affiliation{Institute for Nuclear Research RAS, 60 October Revolution Pr. 7a, 117312 Moscow, Russia} 
\author{A.N.~Khotjantsev}\affiliation{Institute for Nuclear Research RAS, 60 October Revolution Pr. 7a, 117312 Moscow, Russia} 
\author{P.~Kitching}\affiliation{Centre for Subatomic Research, University of Alberta, Edmonton, Canada T6G 2N5} 
\author{M.~Kobayashi}\affiliation{High Energy Accelerator Research Organization (KEK), Oho, Tsukuba, Ibaraki 305-0801, Japan} 
\author{T.K.~Komatsubara}\affiliation{High Energy Accelerator Research Organization (KEK), Oho, Tsukuba, Ibaraki 305-0801, Japan} 
\author{A.~Konaka}\affiliation{TRIUMF, 4004 Wesbrook Mall, Vancouver, British Columbia, Canada V6T 2A3} 
\author{A.P.~Kozhevnikov}\affiliation{Institute for High Energy Physics, Protvino, Moscow Region, 142 280, Russia} 
\author{Yu.G.~Kudenko}\affiliation{Institute for Nuclear Research RAS, 60 October Revolution Pr. 7a, 117312 Moscow, Russia} 
\author{A.~Kushnirenko}\altaffiliation{Present address: Institute for High Energy Physics, Protvino, Moscow Region, 142 280, Russia.}  \affiliation{Fermi National Accelerator Laboratory, Batavia, IL 60510} 
\author{L.G.~Landsberg}\altaffiliation{Deceased.}\affiliation{Institute for High Energy Physics, Protvino, Moscow Region, 142 280, Russia} 
\author{B.~Lewis}\affiliation{Department of Physics and Astronomy, University of New Mexico, Albuquerque, NM 87131} 
\author{K.K.~Li}\affiliation{Brookhaven National Laboratory, Upton, NY 11973} 
\author{L.S.~Littenberg}\affiliation{Brookhaven National Laboratory, Upton, NY 11973} 
\author{J.A.~Macdonald}\altaffiliation{Deceased.} \affiliation{TRIUMF, 4004 Wesbrook Mall, Vancouver, British Columbia, Canada V6T 2A3} 
\author{D.R.~Marlow}\affiliation{Joseph Henry Laboratories, Princeton University, Princeton, New Jersey 08544} 
\author{R.A. McPherson}\affiliation{Joseph Henry Laboratories, Princeton University, Princeton, New Jersey 08544}
\author{P.D.~Meyers}\affiliation{Joseph Henry Laboratories, Princeton University, Princeton, New Jersey 08544}  
\author{J.~Mildenberger}\affiliation{TRIUMF, 4004 Wesbrook Mall, Vancouver, British Columbia, Canada V6T 2A3} 
\author{O.V.~Mineev}\affiliation{Institute for Nuclear Research RAS, 60 October Revolution Pr. 7a, 117312 Moscow, Russia} 
\author{M. Miyajima}\affiliation{Department of Applied Physics, Fukui University, 3-9-1 Bunkyo, Fukui, Fukui 910-8507, Japan} 
\author{K.~Mizouchi}\affiliation{Department of Physics, Kyoto University, Sakyo-ku, Kyoto 606-8502, Japan} 
\author{V.A.~Mukhin}\affiliation{Institute for High Energy Physics, Protvino, Moscow Region, 142 280, Russia} 
\author{N.~Muramatsu}\affiliation{Research Center for Nuclear Physics, Osaka University, 10-1 Mihogaoka, Ibaraki, Osaka 567-0047, Japan} 
\author{T.~Nakano}\affiliation{Research Center for Nuclear Physics, Osaka University, 10-1 Mihogaoka, Ibaraki, Osaka 567-0047, Japan} 
\author{M.~Nomachi}\affiliation{Laboratory of Nuclear Studies, Osaka University, 1-1 Machikaneyama, Toyonaka, Osaka 560-0043, Japan} 
\author{T.~Nomura}\affiliation{Department of Physics, Kyoto University, Sakyo-ku, Kyoto 606-8502, Japan} 
\author{T.~Numao}\affiliation{TRIUMF, 4004 Wesbrook Mall, Vancouver, British Columbia, Canada V6T 2A3} 
\author{V.F.~Obraztsov}\affiliation{Institute for High Energy Physics, Protvino, Moscow Region, 142 280, Russia} 
\author{K.~Omata}\affiliation{High Energy Accelerator Research Organization (KEK), Oho, Tsukuba, Ibaraki 305-0801, Japan} 
\author{D.I.~Patalakha}\affiliation{Institute for High Energy Physics, Protvino, Moscow Region, 142 280, Russia} 
\author{S.V.~Petrenko}\affiliation{Institute for High Energy Physics, Protvino, Moscow Region, 142 280, Russia} 
\author{R.~Poutissou}\affiliation{TRIUMF, 4004 Wesbrook Mall, Vancouver, British Columbia, Canada V6T 2A3} 
\author{E.J.~Ramberg}\affiliation{Fermi National Accelerator Laboratory, Batavia, IL 60510} 
\author{G.~Redlinger}\affiliation{Brookhaven National Laboratory, Upton, NY 11973} 
\author{T.~Sato}\affiliation{High Energy Accelerator Research Organization (KEK), Oho, Tsukuba, Ibaraki 305-0801, Japan} 
\author{T.~Sekiguchi}\affiliation{High Energy Accelerator Research Organization (KEK), Oho, Tsukuba, Ibaraki 305-0801, Japan} 
\author{T.~Shinkawa}\affiliation{Department of Applied Physics, National Defense Academy, Yokosuka, Kanagawa 239-8686, Japan} 
\author{F.C. Shoemaker}\affiliation{Joseph Henry Laboratories, Princeton University, Princeton, New Jersey 08544}  
\author{A.J.S. Smith}\affiliation{Joseph Henry Laboratories, Princeton University, Princeton, New Jersey 08544}  
\author{J.R. Stone}\affiliation{Joseph Henry Laboratories, Princeton University, Princeton, New Jersey 08544}  
\author{R.C.~Strand}\affiliation{Brookhaven National Laboratory, Upton, NY 11973} 
\author{S.~Sugimoto}\affiliation{High Energy Accelerator Research Organization (KEK), Oho, Tsukuba, Ibaraki 305-0801, Japan} 
\author{Y.~Tamagawa}\affiliation{Department of Applied Physics, Fukui University, 3-9-1 Bunkyo, Fukui, Fukui 910-8507, Japan} 
\author{R.~Tschirhart}\affiliation{Fermi National Accelerator Laboratory, Batavia, IL 60510} 
\author{T.~Tsunemi}\altaffiliation{Present address: Department of Physics, Kyoto University, Sakyo-ku, Kyoto 606-8502, Japan}\affiliation{High Energy Accelerator Research Organization (KEK), Oho, Tsukuba, Ibaraki 305-0801, Japan} 
\author{D.V.~Vavilov}\affiliation{Institute for High Energy Physics, Protvino, Moscow Region, 142 280, Russia} 
\author{B.~Viren}\affiliation{Brookhaven National Laboratory, Upton, NY 11973} 
\author{N.V.~Yershov}\affiliation{Institute for Nuclear Research RAS, 60 October Revolution Pr. 7a, 117312 Moscow, Russia} 
\author{Y.~Yoshimura}\affiliation{High Energy Accelerator Research Organization (KEK), Oho, Tsukuba, Ibaraki 305-0801, Japan} 
\author{T.~Yoshioka}\altaffiliation{Present address: International Center for 
Elementary Particle Physics, University of Tokyo, Tokyo 113-0033, Japan.} 
\affiliation{High Energy Accelerator Research Organization (KEK), Oho, Tsukuba, Ibaraki 305-0801, Japan} 
\noaffiliation 
\newpage 
\begin{abstract} 
 
 
Experiment E949  at Brookhaven National Laboratory studied the 
rare decay \KPPNN\ and other processes with an exposure of  
$1.77\times 10^{12}$ $K^+$'s.  
The data were analyzed using a blind analysis technique 
yielding one candidate event with an estimated background of 
$0.30\pm0.03$ events. 
Combining this result with  the observation of two candidate events by 
the predecessor experiment E787  gave  the branching ratio 
${\cal B}(K^+\to\pi^+\nu\bar{\nu})=(1.47^{+1.30}_{-0.89})\times 10^{-10}$,  
consistent with the Standard Model prediction of \PNNSM. 
This is a more detailed report of results previously published in
Physical Review Letters.
 
\end{abstract} 
\pacs{13.20.Eb, 12.15.Hh, 14.80.Mz} 
\maketitle 
\tableofcontents 

\section{Introduction} 
\label{chap:intro} 
 
Although the Standard Model (SM) has successfully accounted for all 
low energy CP-violating phenomena thus far observed, it is insufficient as the 
source of CP-violation needed to explain the cosmological baryon 
asymmetry in our universe~\cite{buch}. According to 
Sakharov~\cite{sakharov}, one of the conditions necessary to generate such 
an asymmetry is that the elementary interaction violates charge 
conjugation symmetry (C) and the combined CP symmetry (where P is the 
parity symmetry). However, the size of the asymmetry needed for this 
cannot be derived in model calculations based on the SM~\cite{ruba} and 
new sources of CP violation have been sought for many years in particle  
physics experiments.  Prominent among these are the rare decays 
 $K \to \pi \nu\bar\nu$  which are sensitive to new physics  
involving both CP-violating and CP-conserving 
interactions. 
In this paper, we present a detailed description of the previously
reported measurement of the reaction $K^+ \to \pi^+ \nu\bar\nu$
performed by
Brookhaven National Laboratory (BNL) experiment BNL-E949~\cite{e949_prl}.
This paper is arranged as follows. We first briefly review CP 
violation and rare kaon decays, with an emphasis on  
$K^+ \to \pi^+\nu\bar\nu$ decays.  We then describe previous  
results on this reaction and discuss the sources of potential background   
and  the  methods for suppressing backgrounds. We  also  discuss the  
design of the $K^+$ beamline, the detector and the 
selection criteria used in data analysis and describe the  methods used for 
estimating  background levels and for evaluating the 
acceptance. After examining  the signal region, we  present  
the  method  used for extracting the branching ratio, making full use  
of our knowledge of the background in the signal region. In the last  
section, we show how the measurement of ${\cal B}(K^+ \to \pi^+ \nu\bar\nu)$  
impacts the search for  new physics beyond the SM. 
 
\subsection{CP Violation and the Rare Decay \KPPNN} 
 
Standard Model CP violation arises from 
a complex phase in the three-generation quark mixing matrix~\cite{CKM}. 
In the Wolfenstein  
parameterization~\cite{wolfenstein} of the Cabibbo-Kobayashi-Maskawa 
(CKM) matrix, the parameters  can be written in powers  
of $\lambda=\sin\theta_c\approx 0.22$: 
\begin{equation} 
V_{\mbox{CKM}}= 
\left( 
\begin{array}{ccc} 
V_{ud}&V_{us}&V_{ub}\\ 
V_{cd}&V_{cs}&V_{cb}\\ 
V_{td}&V_{ts}&V_{tb} 
\end{array}\right) 
\simeq  
\left( 
\begin{array}{ccc} 
1-\lambda^2/2 & \lambda & A\lambda^3(\rho-i\eta)\\ 
-\lambda & 1-\lambda^2/2 & A\lambda^2 \\ 
A\lambda^3(1-\rho-i\eta) & -A\lambda^2 & 1 
\end{array}\right). 
\label{eq:wolfenstein} 
\end{equation} 
Where $A$, $\lambda$, $\rho$ and $\eta$ are real numbers.  CP 
invariance of the Lagrangian for weak interactions is violated when 
the CKM matrix is complex.  The parameter $\eta$ quantifies CP 
violation in the SM. 
 
\begin{figure} 
\centering 
\epsfxsize 0.55\linewidth 
\epsffile{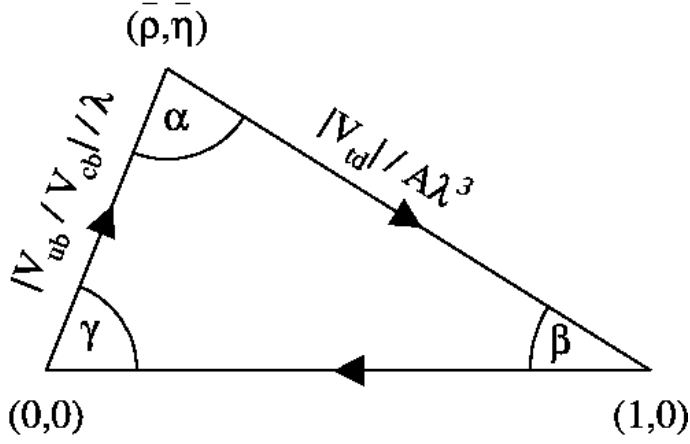} 
\caption{Unitarity triangles in the $\bar{\rho} - \bar{\eta}$ plane.  
Two sides of the triangle 
can be expressed by the CKM matrix elements $|V_{td}|/A\lambda^3$  
and $|V_{ub}/V_{cb}|/\lambda$, respectively, 
where $A$ and $\lambda$ are parameters in the  
Wolfenstein parameterization.} 
\label{fig:triangle} 
\end{figure}

The unitarity of the CKM matrix implies six unitarity 
conditions, which can be represented graphically in the form 
of triangles, all of which must have the same area. 
The area of these triangles is equal to one half of the Jarlskog 
invariant, $J_{\mbox{CP}}$~\cite{Jarlskog}. 
Applying the unitarity property $V^{\dagger}V=1$ to the 
CKM matrix in~(\ref{eq:wolfenstein}) yields 
\begin{equation} 
V^*_{ub}V_{ud}+V^*_{cb}V_{cd}+V^*_{tb}V_{td} 
\simeq V^*_{ub}-\lambda V^*_{cb}+V_{td}=0, 
\end{equation} 
where the approximations $V_{ud}\simeq V^*_{tb}\simeq 1$  
and $V_{cd}\simeq -\lambda$ have been made. 
This equation can be represented graphically,  
as shown in Fig.~\ref{fig:triangle},  
where we have divided all of the sides by $\lambda V^*_{cb}$.  
The apex of the triangle is given by two Wolfenstein 
parameters, $\bar{\rho}$ and $\bar{\eta}$,  
where $\bar{\rho}=\rho(1-\lambda^2/2)$ and  
$\bar{\eta}=\eta(1-\lambda^2/2)$~\cite{CKM_new}. 
 
\begin{figure} 
\centering 
\epsfxsize 0.6\linewidth 
\epsffile{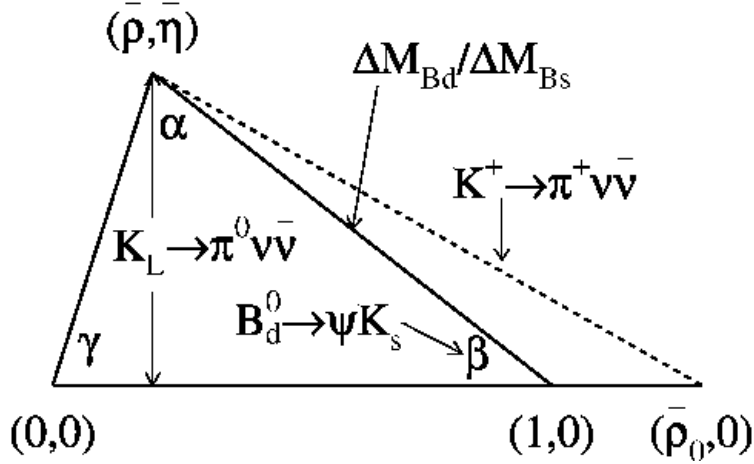} 
\caption{Unitarity triangle determined by $B$ and $K$ decays.  
The parameters $\bar{\rho}$ and $\bar{\eta}$ can be determined in two ways: 
the angle $\beta$ from the CP violating  
asymmetry in the decay $B^0_d\to J/\psi K^0_s$, 
and from the length of the side from $\Delta M_{B_s}/\Delta M_{B_d}$ 
in $B-\bar{B}$ mixing; 
the height of the triangle from \BR($K^0_L\to\pi^0\nu\bar{\nu}$) and  
the radius of a circle centered 
at $(\bar{\rho}_0,0)$ from \BR(\KPPNN).} 
\label{fig:triangle_B_K} 
\end{figure}

$B$'s and $K$'s are so far the only two mesons showing evidence of 
CP violation in their decay processes. Whether or not the observed 
CP violation can be completely explained by the CKM phase 
within the SM can be probed by the independent determination of  
$\rho$ and $\eta$, from $B$ and $K$ decays as shown in 
Fig.~\ref{fig:triangle_B_K}.  
Two sensitive methods for making the comparison are: 
\begin{itemize} 
\item A comparison of angle $\beta$ from the ratio 
\BR($K^0_L\to\pi^0\nu\bar{\nu}$)/\BR(\KPPNN) 
with that from the CP violating asymmetry (${\cal A}_{\mbox{CP}}$) in 
the decay $B^0_d\to J/\psi K^0_s$; and, 
\item A comparison of the magnitude $|V_{td}|$ from 
\KPPNN~with that from the mixing frequencies 
of $B_s$ and $B_d$ mesons, expressed in terms of 
the ratio of the mass differences, $\Delta M_{B_s}/\Delta M_{B_d}$. 
\end{itemize} 
 
\begin{figure} 
\centering 
\begin{minipage}{0.49\linewidth} 
\centering 
\epsfxsize 0.7\linewidth 
\epsffile{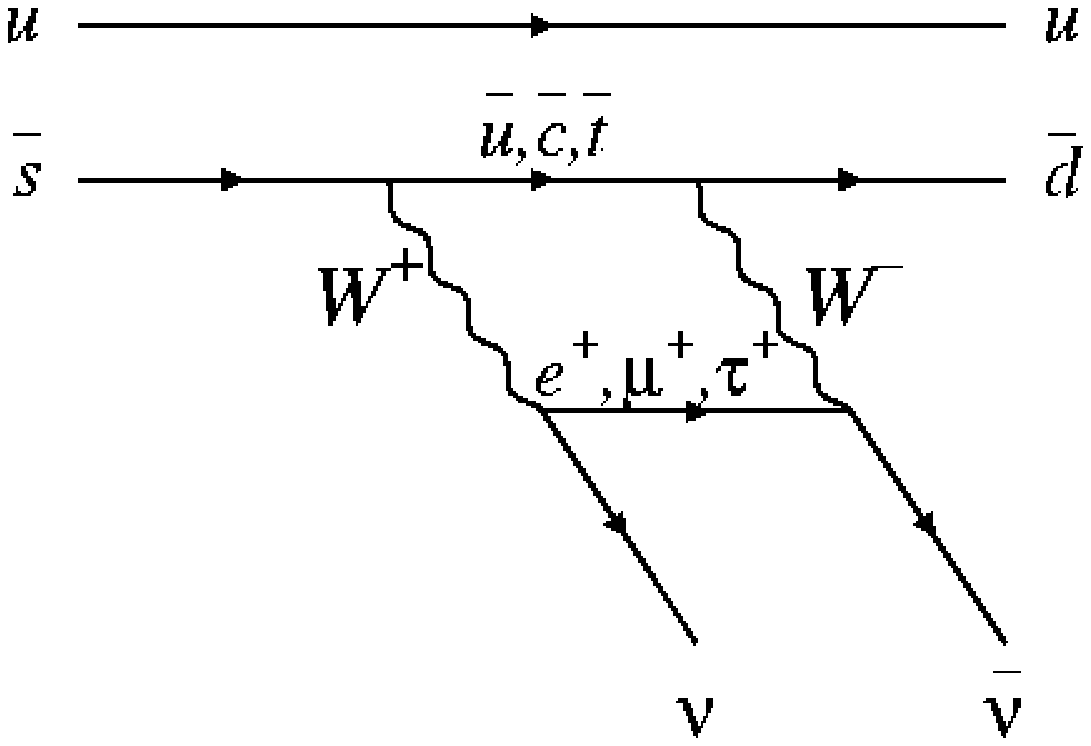} 
\end{minipage}\\[10mm] 
\begin{minipage}{0.49\linewidth} 
\centering 
\epsfxsize 0.7\linewidth 
\epsffile{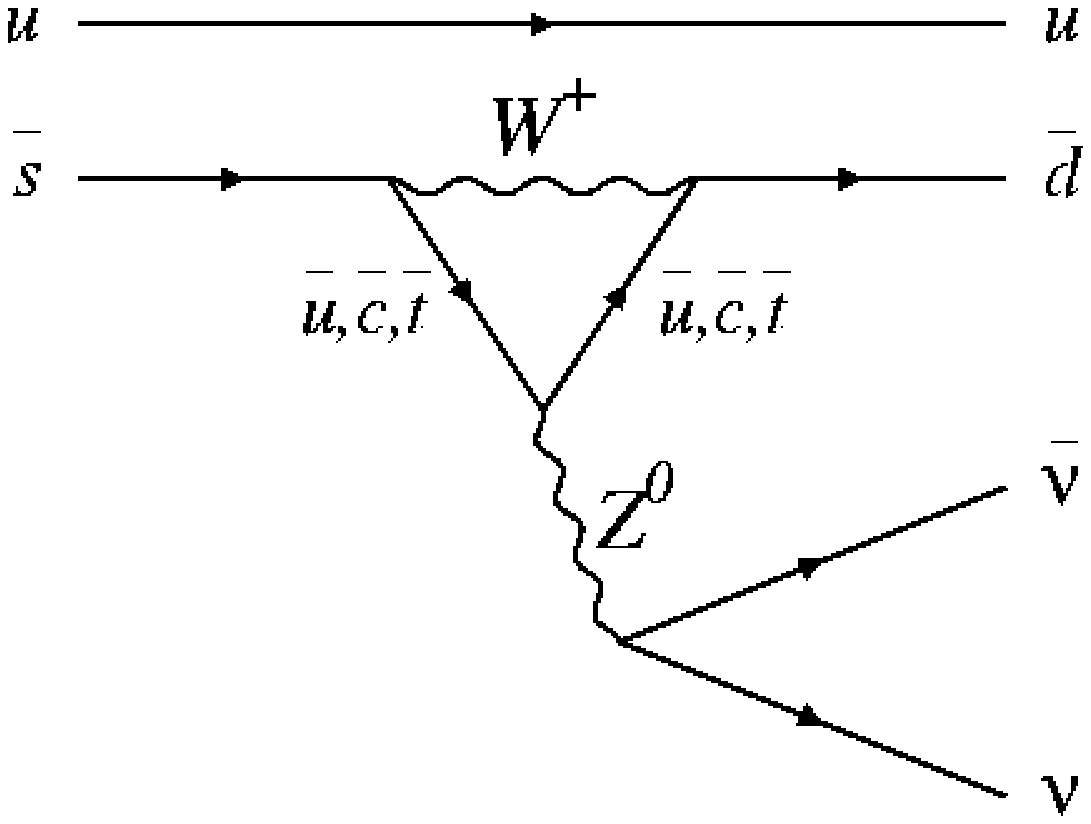} 
\end{minipage}\hfill 
\begin{minipage}{0.49\linewidth} 
\centering 
\epsfxsize 0.7\linewidth 
\epsffile{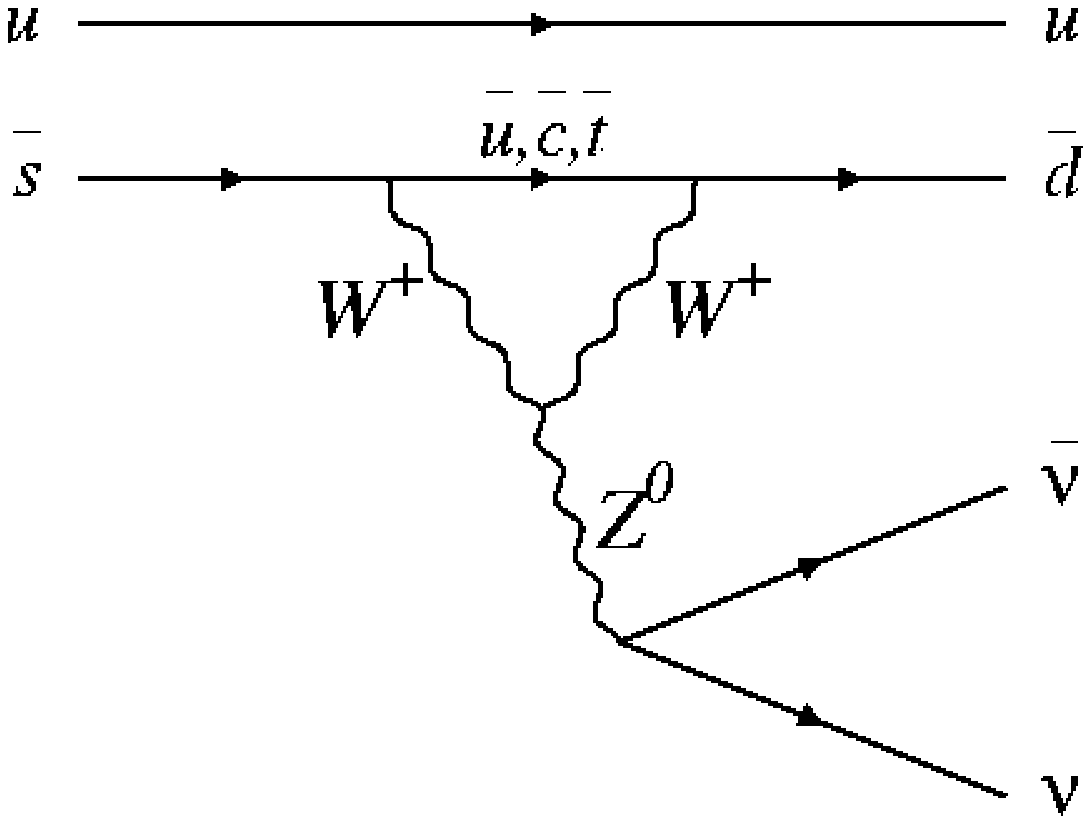} 
\end{minipage} 
\caption{Second-order weak processes that contribute 
to the \KPPNN~branching ratio: 
the ``Box" diagram (upper) and two ``$Z$-penguin" diagrams (bottom).} 
\label{fig:pnn-2nd} 
\end{figure}

Although the decay \KPPNN~is a flavor changing neutral current 
(FCNC) process prohibited at tree level in the SM, 
it is allowed at the one-loop level. In leading order, it 
is described by a ``Box'' diagram and two ``$Z$-penguin'' 
diagrams, as shown in Fig.~\ref{fig:pnn-2nd}. 
The weak amplitude for this process is represented as 
\begin{equation} 
{\cal M} \sim \sum_{i=u,c,t}V^{*}_{is}V_{id} 
\frac{\gamma^{\mu}q_{\mu}+m_i}{q^2-m^2_i}, 
\end{equation} 
where $V_{ij}$'s are the CKM matrix elements,  
$\gamma^{\mu}$'s are the Dirac matrices, 
$q_{\mu}$ is the momentum transfer, and $m_i$'s are quark masses. 
${\cal M}$ vanishes if all of the quark masses, $m_i$, are equal, 
because of the unitarity of the CKM matrix. However,  
the breaking of flavor symmetry, which results in the variation of  
quark masses, allows this decay to proceed at a very small rate.  
The top quark provides the dominant contribution to the 
\KPPNN~branching ratio due to its very large mass in spite of the small  
coupling of top to down quarks ($V_{td}$) in the CKM matrix. 
 
Following Ref.~\cite{buras_hep-ph0405132},  
the branching ratio for \KPPNN~is 
calculated as follows. 
The effective Hamiltonian can be written in the SM as 
\begin{equation} 
{\cal H}^{SM}_{eff} = \frac{G_F}{\sqrt{2}}\frac{\alpha} 
{2\pi\sin^2\Theta_W}\sum_{l=e,\mu,\tau} 
(V^*_{cs}V_{cd}X^l_{\rm NL}+V^*_{ts}V_{td}X(x_t))(\bar{s}d)_{V-A} 
(\bar{\nu_l}\nu_l)_{V-A}, 
\end{equation} 
in next-to-leading order (NLO), where 
\begin{equation} 
X(x_t) \equiv X_0(x_t) + \frac{\alpha_s(m_t)}{4\pi}X_1(x_t) \approx  
\eta_X\cdot X_0(x_t) 
\end{equation} 
and 
\begin{equation} 
X_0(x_t) \equiv C_0(x_t)-4B_0(x_t), \quad \eta_X=0.995. 
\label{eq:inami} 
\end{equation} 
$B_0(x_j)$ and $C_0(x_j)$ in (\ref{eq:inami}) are 
functions of $x_j \equiv m^2_j/M^2_W$, and were derived for 
the first time by Inami and Lim 
in 1981~\cite{InamiLim}. 
The coefficient $X^l_{\rm NL}$ and the function $X(x_t)$ 
are the charm and top quark contributions, 
including QCD corrections at NNLO~\cite{QCD1,QCD2,QCD3,QCD4,QCD5}. 
 
With the top quark mass in the minimal subtraction scheme  
$m_t(m_t)=(162.3\pm2.2)$~GeV~\cite{buras_hep-ph0405132}, 
\begin{equation} 
X(x_t)=1.464\pm0.025  
\end{equation} 
is obtained.  
 
The perturbative charm contribution gives the  
largest theoretical uncertainty 
and can be described in terms of the parameter 
\begin{equation} 
P_c(X) \equiv \frac{1}{\lambda^4}[\frac{2}{3}X^e_{\rm NL} 
+\frac{1}{3}X^{\tau}_{\rm NL}]=0.34\pm0.04, 
\label{eq:charm_cont} 
\end{equation} 
where the error is obtained by varying the charm mass, $m_c$, the scale 
factor, $\mu_c={\cal O}(m_c)$ and the coupling constant, $\alpha_s(M^2_Z)$, 
by reasonable amounts. One obtains 
\begin{equation} 
{\cal B}(K^+\to\pi^+\nu\bar{\nu}) = \kappa_+ 
\cdot\left[\left(\frac{{\rm Im}\lambda_t}{\lambda^5} X(x_t)\right)^2 
+ \left(\frac{{\rm Re}\lambda_c}{\lambda}( 
P_c(X)+ \delta P_{c,u})+\frac{{\rm Re}\lambda_t}{\lambda^5} X(x_t)\right)^2\right], 
\label{eq:pnn_br} 
\end{equation} 
where 
\begin{equation} 
\kappa_+ \equiv r_+\frac{3\alpha^2{\cal B} 
(K^+\to\pi^0e^+\nu)}{2\pi^2\sin^4\Theta_W}\lambda^8 
= (5.26\pm0.08)\times 10^{-11}\left[\frac{\lambda}{0.2257}\right]^8, 
\label{eq:kappa} 
\end{equation} 
$\delta P_{c,u} = 0.04 \pm 0.02$ comprises the long-distance contribution 
calculated in Ref.~\cite{Isidori:2005xm}, 
and the $\lambda_j$'s ($\equiv V^*_{js}V_{jd}$) are 
from the CKM matrix elements. 
The $r_+~(=0.901)$ represents isospin 
breaking corrections in relating \KPPNN~to 
the well-measured decay \KETHREE~\cite{ke3cancel}. 
In obtaining the numerical value in (\ref{eq:kappa}),  
we used~\cite{Yao:2006px} 
\begin{equation} 
\sin^2\Theta_W = 0.231,\qquad \alpha = \frac{1}{127.9},  
\qquad {\cal B}(K^+\to\pi^0e^+\nu_e)=(4.98\pm0.07)\times10^{-2}. 
\end{equation} 
Expression~(\ref{eq:pnn_br}) 
describes in the $\bar{\rho}-\bar{\eta}$ plane an  
ellipse with a small eccentricity, namely 
\begin{equation} 
(\sigma\bar{\eta})^2 + (\bar{\rho}-\bar{\rho}_0)^2 =  
\frac{\sigma{\cal B}(K^+\to\pi^+\nu\bar{\nu})} 
{\bar{\kappa}_+|V_{cb}|^4X^2(x_t)}, 
\label{eq:ellipse} 
\end{equation} 
where 
\begin{equation} 
\bar{\rho}_0 \equiv 1+\frac{\lambda^4 (P_c(X)+\delta  
P_{c,u})}{|V_{cb}|^2 X(x_t)},\qquad 
\sigma \equiv \left(1-\frac{\lambda^2}{2}\right)^{-2},\qquad 
\overline{\kappa}_+ \equiv \frac{\kappa_+}{\lambda^8}. 
\label{eq:rho_zero} 
\end{equation} 
 
Using (\ref{eq:pnn_br}) and varying $m_t$, $|V_{cb}|$,  
$P_c(X)$ and $|V_{td}|$, which is constrained 
by $|V_{ub}/V_{cb}|$ and $B-\bar{B}$ mixing  
in the $\bar{\rho}-\bar{\eta}$ plane, 
the branching ratio of \KPPNN~is predicted to be 
\begin{equation} 
{\cal B}(K^+\to\pi^+\nu\bar{\nu}) = (0.74\pm0.20)\times 10^{-10} 
\label{eq:BR_SM} 
\end{equation} 
within the SM.  
It should be noted that, of the uncertainty of 27\% in (\ref{eq:BR_SM}), 
the theoretical uncertainty is $\sim$6\% at present, 
mainly due to the uncertainty in the charm quark mass. 
 
Theoretically a precise measurement of \BR(\KPPNN) 
is one of the cleanest ways to extract $|V_{td}|$.  
This is due to the following factors: 
\begin{itemize} 
\item  the long-distance contributions to the  
branching ratio are small~\cite{LD} and under control, 
the most recent calculation gives a contribution of (+6$\pm$3)\% to the 
branching ratio~\cite{Isidori:2005xm}; 
\item the uncertainty from the hadronic matrix  
element has been reduced to $<$1\% by recent theoretical and experimental 
developments~\cite{Mescia:2007kn}, and 
\item the recent NNLO calculation~\cite{QCD4,QCD5} 
has reduced the total theoretical uncertainties to $\sim$6\%, 
 i.e. relatively small and reliably calculated as compared 
with the uncertainties present in other $K$ and $B$ decays. 
\end{itemize} 
If a precise measurement of the neutral analog $K^0_L \to 
\pi^0\nu\bar\nu$ could also be made, the intrinsic theoretical error 
on $|V_{td}|$ could be reduced to $\sim$1\%~\cite{QCD5}. 
 
As determinations of $B$-system parameters become increasingly precise, 
the uncertainty on the SM prediction for \KPPNN~ will approach the 
current theoretical accuracy of $\sim$6\%.  A correspondingly precise 
measurement of the \KPPNN~branching ratio therefore provides a stringent 
test of the SM and probes for new physics. 
There have been numerous predictions for \KPPNN~ in the models beyond 
the SM and applications of the measured branching ratio to constrain new 
models. These include the Minimal Supersymmetric Standard Model 
with~\cite{Buras:2004qb,Isidori:2006qy} and  
without~\cite{Isidori:2006qy,Buras:2000dm}  
new sources of flavor- or CP-violation, 
generic SUSY with minimal particle content~\cite{Buras:1999da}, 
SUSY with non-universal A terms~\cite{Chen:2002eh}, SUSY with 
broken R-parity~\cite{Bhattacharyya:1998be,Deandrea:2004ae}, 
topcolor~\cite{Buchalla:1995dp}, topcolor-assisted technicolor  
models~\cite{Xiao:1999cn,Xiao:1999pt}, multiscale walking  
technicolor~\cite{Xiao:1999ps}, four generation models~\cite{Hattori:1999ap}, 
leptoquarks~\cite{Agashe:1995qm}, Left-Right model with right-handed  
$Z^\prime$~\cite{He:2004it}, extension of the SM to a gauge theory with 
$J=0$ mesons~\cite{Machet:1999dj}, a multi-Higgs multiplet  
model~\cite{Grossman:1994jb}, light sgoldstinos~\cite{Gorbunov:2000cz}, 
universal extra dimensions~\cite{Buras:2002ej}, 5-dimensional split 
fermions~\cite{Chang:2002ww}, a Randell-Sundrum  
scenario~\cite{Burdman:2002gr}, a littlest Higgs  
model~\cite{Blanke:2006eb,Blanke:2007wr},  
non-standard neutrino interactions~\cite{Chen:2007cn},  
and a minimal 3-3-1 model~\cite{Promberger:2007py}.

\subsection{History of $K^+\rightarrow\pi^+\nu\bar\nu$ Experiments}

Searches for this process which began over 35 years ago have 
used stopped-$K^+$ beams.  It was believed at the time of the first of these 
that the branching ratio might be as high as a few  
$\times 10^{-5}$~\cite{Oakes:1969st}.   
It was recognized that even at this level, a poor-signature  
process such as $K^+ \to \pi^+ \nu\bar\nu$ would need effective 
particle identification, precise kinematic measurement and the ability to  
veto extra charged and neutral tracks to discriminate it from common 
decay modes such as $K^+ \to \mu^+ \nu_\mu$ and $K^+ \to \pi^+ \pi^0$  
(referred to as $K_{\mu2}$ and $K_{\pi2}$, respectively). 
The earliest published result was from a  
heavy liquid bubble chamber experiment~\cite{exp1} at the Argonne Zero 
Gradient Synchrotron, in which a 90\% CL upper limit  ${\cal B}(K^+ \to \pi^+  
\nu\bar\nu)<10^{-4}$ was obtained.  In that paper 
it was recognized that $K_{\pi2}$ decay in  
flight and hadronic $\pi^+$ interaction in 
the detector were dangerous sources of potential background.   
 
The final analysis of the Argonne experiment improved the limit to 
$5.7 \times 10^{-5}$~\cite{exp2}, but before it appeared in print, a 
subsequent counter/spark-chamber experiment at the Berkeley Bevatron 
improved the limit to $1.4 \times 10^{-6}$~\cite{exp3}.  However this 
experiment was sensitive to only the most energetic of $\pi^+$, 
whereas the bubble chamber experiment covered a wide kinematic range. 
In addition to the background from common $K^+$ decay modes, this 
experiment considered possible background from $K^+$ charge exchange 
in the stopping target followed by $K^0_L \to \pi^+ e^- \bar{\nu}_e$, and from 
beam $\pi^+$ which scattered into the detector.  The Chicago-Berkeley 
group continued their program with a setup sensitive to $\pi^+$ in the 
kinetic energy range 60--105~MeV, i.e. below that of the 
potential background process $K_{\pi2}$ rather than above it.   
This required reconfiguring their photon veto system so that it became 
nearly hermetic.  Combining results from the two configurations, the 
branching ratio upper limit was improved slightly to $5.6 \times 
10^{-7}$~\cite{exp4}. 
 
About a decade later, an experiment at the KEK Proton Synchrotron 
improved the limit to $1.4 \times 10^{-7}$~\cite{exp5}.   
The technique of waveform digitization to record the  
$\pi^+ \to \mu^+ \to e^+$ decay chain was introduced  
for the first time. This 
experiment was sensitive only to the $\pi^+$ with momenta greater 
than that from $K_{\pi2}$ (referred to as the ``$\pi\nu\bar{\nu}(1)$'' region)  
and its  
setup resembled that of Ref.~\cite{exp3}.  
 
\begin{figure} 
\centering 
\epsfxsize 0.9\linewidth 
\epsffile{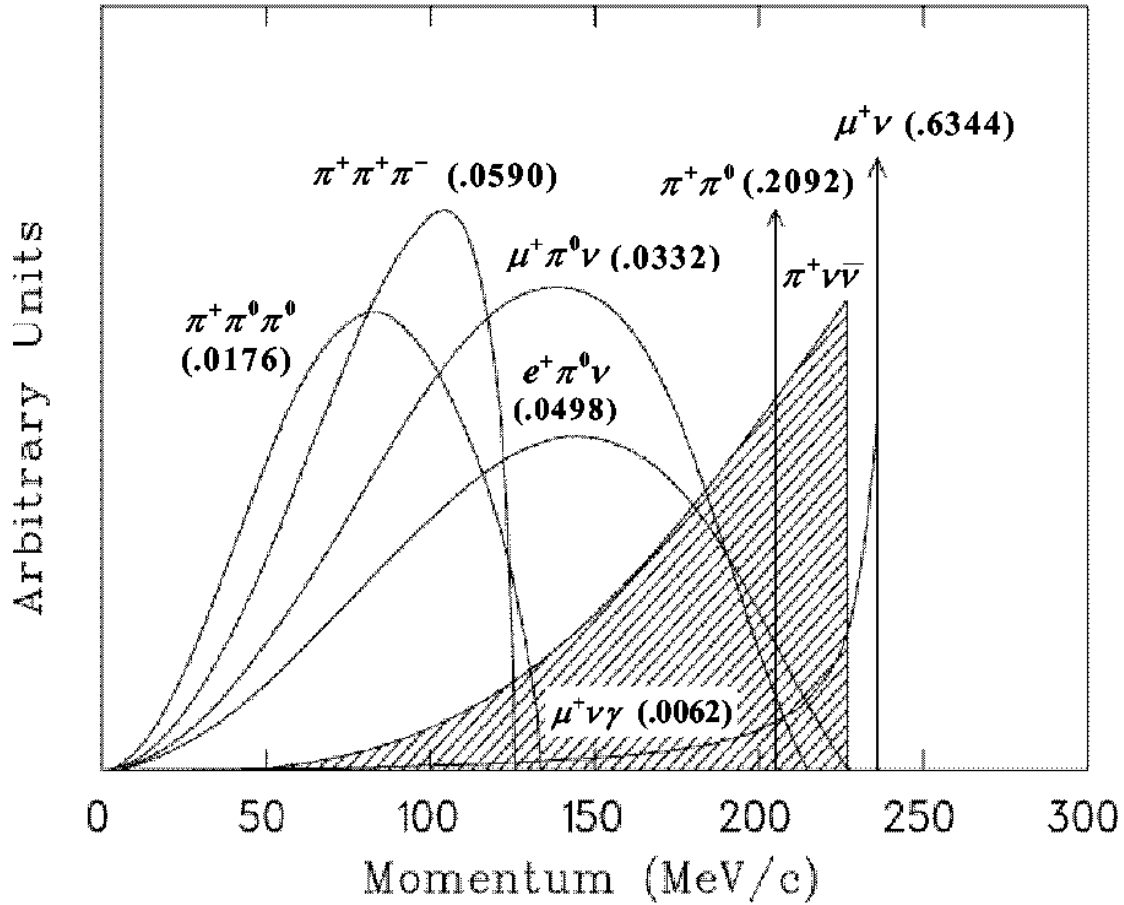} 
\caption{Momentum spectra (in~MeV/$c$) of charged  
particles from $K^+$ decays in the 
rest frame. The values in the parentheses represent the branching ratios of  
the decay modes~\cite{Yao:2006px}.  
The hatched spectrum shows the $\pi^+$ momentum from 
 \KPPNN~decay assuming the $V-A$ interaction.} 
\label{fig:mom_spec} 
\end{figure}

The BNL series of experiments was initiated in the early 1980's.  They
were based on a large-acceptance solenoidal spectrometer with a 
hermetic photon veto situated at the end of a highly pure, very 
intense stopped-$K^+$ beam~\cite{E787phase1NIM} from the 
BNL Alternating Gradient Synchrotron (AGS).   
The experimental signature of the \KPPNN~decay was a single $\pi^+$ 
track with $\pi^+$ momentum less than 227~MeV/$c$ plus  
no other particle from a $K^+$ decay. Fig.~\ref{fig:mom_spec} 
shows momentum spectra of major decay modes of $K^+$. 
  
The first phase of 
E787 in 1988-91 achieved a 90\% CL upper limit on the 
branching ratio of $2.4 \times 10^{-9}$~\cite{E787phase1}, using data 
from the $\pi\nu\bar{\nu}(1)$ region.  A separate limit of $1.7 \times 
10^{-8}$ at 90\% CL~\cite{exp8} was extracted  
from the kinematic region in which 
the $\pi^+$ is softer than that of the $\pi^+$ from $K_{\pi2}$  
(referred to as the ``$\pi\nu\bar{\nu}(2)$'' region).   
This program completed the 
identification of backgrounds needed to reach the $10^{-10}$ level of 
sensitivity and developed methods to reliably measure them. 
 
A major upgrade of both the beam line and the detector was undertaken 
between 1992 and 1994.  The search for $K^+ \to \pi^+ \nu\bar\nu$  
resumed in 1995 and continued through 1998.  The limit on the branching 
ratio from the $\pi\nu\bar{\nu}(2)$ region was  
improved by an order of magnitude to $2.2  
\times 10^{-9}$ at 90\% CL~\cite{e787_pnn2}, but  
the major output of this series of runs was 
the observation of two clean $K^+ \to \pi^+ \nu\bar\nu$ 
events~\cite{E787-1998} in the $\pi\nu\bar{\nu}(1)$ region and a 
measurement of the branching ratio ${\cal B}(K^+ \to \pi^+ \nu\bar\nu) 
=(1.57^{+1.75}_{-0.82}) \times10^{-10}$. The BNL-E787 
detector was upgraded again over the period from 1999--2001. 
The E949 experiment was proposed to use this detector to run for 60 weeks. 
After the first 12 weeks of running in 2002 no further funds
were provided to complete the experiment. 
Based on the collected BNL-E949 data, the first result was 
already published in 2004~\cite{e949_prl}. 
This paper provides an extended and detailed description of the
detector and data analysis
techniques used to produce the E949 result.
 
\section{Experimental Method} 
\label{chap:experiment} 
  
\subsection{Overview} 
\label{subsect:overview}
 
E949  (BNL-E949) which succeeded  BNL-E787 had a  
sensitivity goal of detecting ten SM signal 
events~\cite{e949}. E949 employed a low momentum beam of $K^+$'s  
which were degraded and stopped in the detector. 
Measurement of the \KPPNN~ decay involved  observation of the  
daughter $\pi^+$ in the absence of other coincident  
activity. The $\pi^+$ was identified by its kinematic features  
obtained from  energy, momentum and range measurements,  
and by the observation of a \PIMUE~ decay sequence. 
Since the signal  was expected at the  $10^{-10}$ level,  
the detector was designed to have powerful $\pi^+$ identification for  
rejection of  $K_{\mu2}$ and $K^+ \to \mu^+\bar{\nu}_\mu\gamma$ decays 
($K_{\mu2\gamma}$), 4-$\pi$ solid angle photon detection  
coverage for vetoing \KPITWO~decays, and efficient 
$K^+$ identification system  
for eliminating beam-related backgrounds. 
 
\begin{figure} 
\centering 
\begin{minipage}[b]{0.5\linewidth} 
\epsfxsize 1.\linewidth 
\epsffile{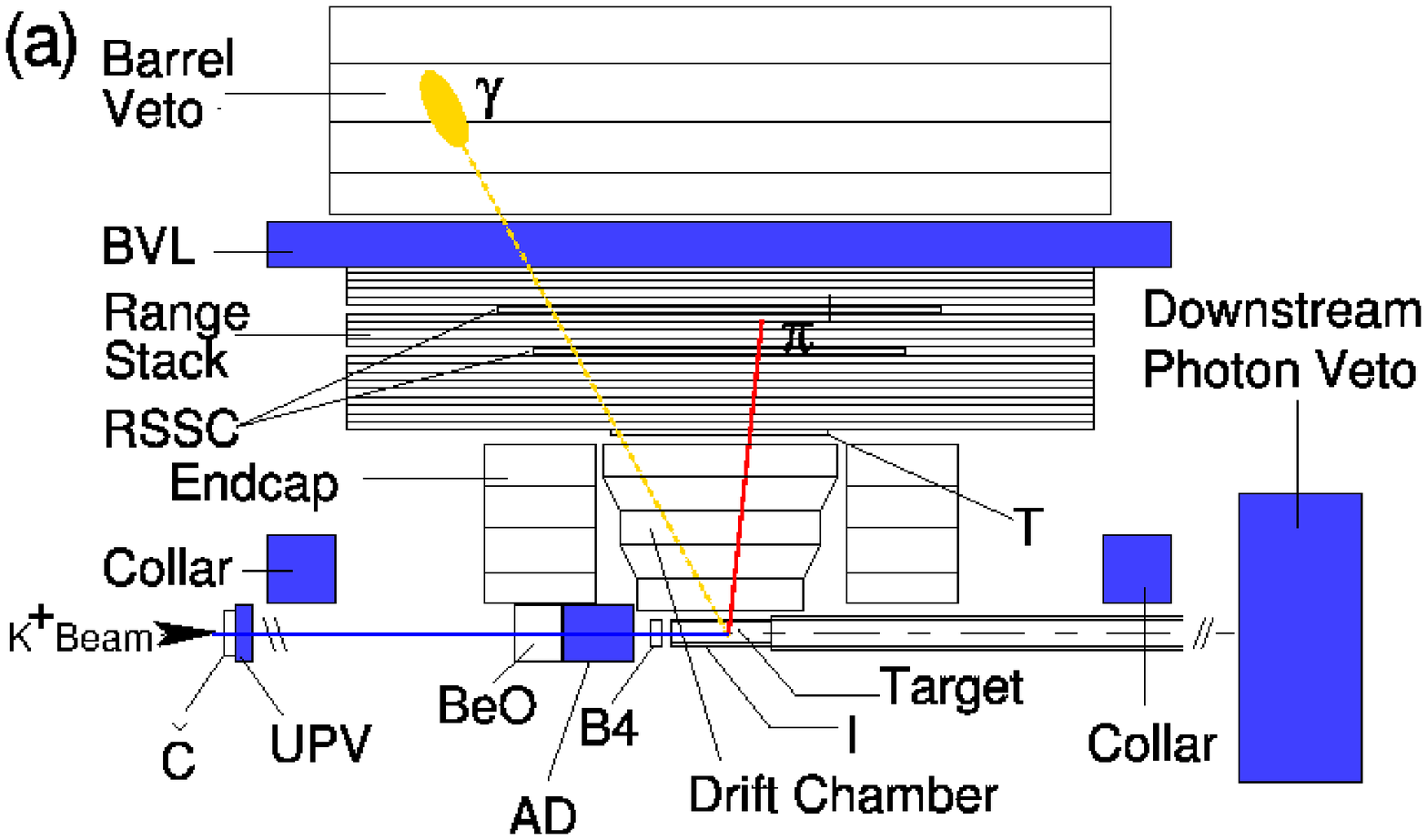} 
\end{minipage} 
\hfill 
\begin{minipage}[b]{0.45\linewidth} 
\epsfxsize 0.9\linewidth 
\epsffile{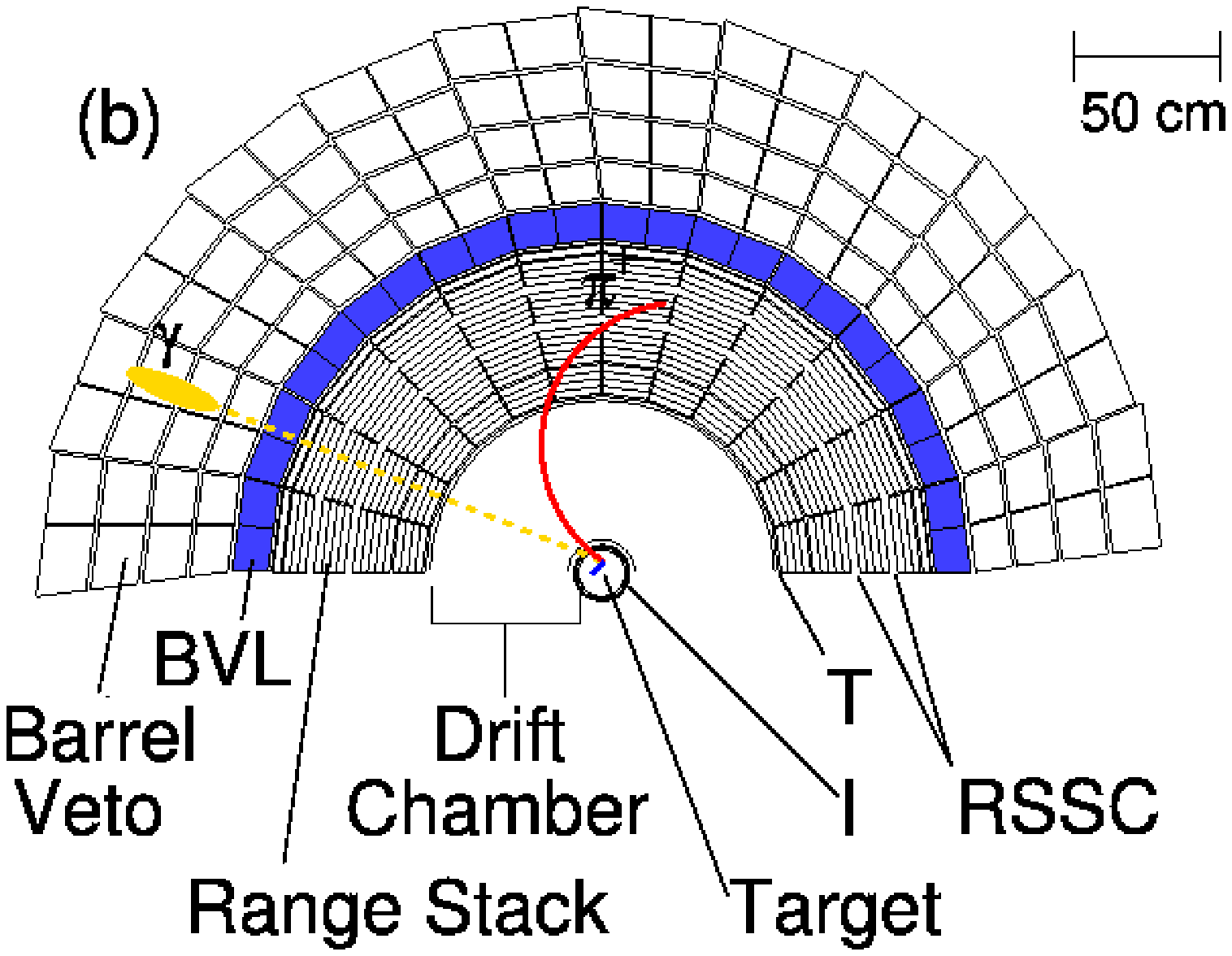} 
\end{minipage} 
\caption{Schematic side (a) and end (b) views of the upper half 
of the E949 detector. Illustrated in this figure, 
an incoming $K^+$ that traverses all the beam 
instruments, stops in the target and undergoes the decay 
$K^+\to\pi^+\pi^0$. The outgoing $\pi^+$ and one photon from 
$\pi^0\to\gamma\gamma$ are also shown.  
The detector elements and acronyms are  
described in detail in the text. 
} 
\label{fig:gamma_detectors} 
\end{figure}

The entire E949 spectrometer was surrounded by a 1 Tesla solenoidal  
magnetic field along the beam direction. The coordinate of detector used
a Cartesian coordinate system in which the origin was at the center of the 
target; the $+z$ axis was along the  
incident beam direction and the $+y$ axis in the vertical up direction 
as shown in Fig.~\ref{fig:gamma_detectors}. 
Under this coordinate system,  
the azimuthal angle of a track 
was defined as the arctangent of $y/x$ and, the  
polar angle $\theta$ was defined as the angle with the $+z$ axis. 
Many detector components have been discussed  
elsewhere~\cite{ccd,TD,utc,rssc,endcap,finemesh}. 
Fig.~\ref{fig:gamma_detectors} shows the E949 detector after upgrades 
(1999--2001~\cite{e949,L0-IEEE}) with the following improved components: 
photon veto detection efficiency, tracking and trigger efficiency, and 
data acquisition (DAQ) live time.
E949 was designed to run at the same instantaneous rate as  
E787, and to achieve a factor of five improvement in sensitivity,  
through the use of a higher duty factor and reduced $K^+$ 
momentum for a higher stopping fraction.  
The higher duty factor was not achieved in the
engineering run in 2001 or the first physics run in 2002 due to a
broken motor generator set that supplied power to the AGS.
The regular supply was removed from operation on August 3, 2001 and
the backup was used during the rest of 2001 and 2002.
E949 ran at about twice the beam rate 
of E787.  
\subsection{Accelerator and $K^+$ Beam Line} 
 
The $K^+$ beam was produced by a high-intensity proton beam from the 
AGS at BNL: the entire AGS beam of 65$\times10^{12}$ protons (Tp/spill)  
at a momentum of 21.5~GeV/$c$ was 
delivered to the E949 $K^+$ production target. 
Prior to 2001 the AGS 
typically ran at 24~GeV/$c$, but at this momentum the longest 
spill achievable was 0.5 sec.  Combined 
with the longer cycle time (3.2 s between 
spills, as compared to 2.3~s), the duty factor at 
24~GeV/$c$ was unacceptably low. By lowering the proton momentum to 
21.5~GeV/$c$, the spill length was increased to a maximum of 2.2~s, 
resulting in a duty factor of 2.2~s/5.4~s. At this lower proton 
momentum the production of 710~MeV/$c$ $K^+$'s was reduced by 10\%. The 
$K^+$ production target was made of 2/3 of an interaction length of 
platinum (6~cm along the beam direction), and was located on a 
water-cooled copper base. At the typical AGS running condition 65~Tp 
on the production target per 2.2~s spill, the 
maximum target temperature was measured to be $\sim$700$^\circ$C. 
 
The Low Energy Separated Beam~\cite{LESBIII} (LESB III) 
collected and transported $K^+$'s emitted at 0$^\circ$ 
(along with 500 $\pi^+$'s and 500 protons per $K^+$), 
and momentum-selected by the first dipole magnet.  
Two electro-magneto-static separators 
swept $\pi^+$'s and protons 
out of the $K^+$ beam axis.   
The resulting beam was further selected by a second 
dipole magnet. LESB III  
had a total length of 19.6~m 
from the production target to the E949 target with an angular 
acceptance of 12~msr and a momentum acceptance of 4.5\% FWHM 
at a mean momentum of 710~MeV/$c$.  
During most of the 
2002 running period the first separator voltage was lowered from the 
standard voltage of 600~kV to $\sim$250~kV due to high voltage 
discharges. Under 
these conditions a $K^+:\pi^+$ ratio in the beam of 3:1 was achieved 
with a 40\% loss in $K^+$ flux (typically E787 ran with 4:1).  Proton 
contamination was suppressed to a negligible level by 
the separators.  At the same typical AGS running condition described previously, 
$1.3\times 10^7$ $K^+$'s were transported 
through the beam line. 

The typical conditions during the 2002 run had 3.5$\times 10^6$ $K^+$'s 
entering the E949 target every spill. This corresponded to a rate of 
1.6$\times 10^6$ $K^+$'s/sec during the 2.2~sec spill.  The effective 
spill length was actually 8\% shorter (2.0~sec) due to some residual 
modulation of the beam intensity out of the AGS. The typical 
instantaneous rate of beam particles at the $\check{\rm C}$erenkov  
counter was 
6.3~MHz of $K^+$'s and 1.5--2.5~MHz of $\pi^+$'s. 
 
\subsection{Beam Counters} 
 
The incoming $K^+$ beam traversed a scintillation counter (B0), 
a $\check{\rm C}$erenkov counter, two beam wire  
proportional chambers (BWPC's), a passive  
BeO degrader, an active degrader (AD) and  
a beam hodoscope (B4) as depicted in Fig.~\ref{fig:gamma_detectors}. 
The BWPC's and B0 counter are not shown in  
Fig.~\ref{fig:gamma_detectors}. 
  
The B0 counter, which was a 30.5-cm long, 0.6-cm thick and 7.6-cm wide 
Bicron BC408 plastic scintillator, 
was located just downstream of the last quadrupole 
magnet and counted all charged particles in the beam.  It was 
read out by an analog-to-digital converter (ADC), 
a time-to-digital converter (TDC) and a 
500~MHz transient digitizers based on gallium-arsenide 
charge-coupled device (CCD)~\cite{ccd}.
The $\check{\rm C}$erenkov counter~\cite{E787phase1NIM} 
located just downstream of the B0 counter identified particles  
as $K^+$'s or $\pi^+$'s. 
The $\check{\rm C}$erenkov light from the $K^+$ ($\pi^+$) was transmitted 
(internally reflected) at the downstream surface of the $\check{\rm 
C}$erenkov radiator and read out with 14 ``$K$ $\check{\rm C}$erenkov''
($C_K$) and 14 ``$\pi$ $\check{\rm C}$erenkov'' ($C_\pi$) EMI9954KB 
PMT's.
The PMT signals were split, with 90\% sent  
to TDC's via fast LRS3412 
discriminators and 10\% to a $\times$10 amplifier.  
The amplifier output was sent to  
CCD's. The pulse-height 
information in every 2~ns interval was recorded to reproduce 
the time development of the pulses and to detect two particles close 
in time to each other. The multiplicity output of $C_K(C_\pi)$ PMT's  
was discriminated (typical threshold: $n>5$)
to identify $K^+$'s and $\pi^+$'s in the trigger ($KB$ and $\pi B$ 
defined in Section~\ref{sec:trigger}). 
 
The two BWPC's were located downstream of the 
$\check{\rm C}$erenkov counter to monitor 
the beam profile and identify 
multiple incoming particles.  
The first chamber (BWPC1) was located 168.5 cm upstream of 
the target entrance and contained three planes of sense wires:  
vertical ($x$-plane) and 
$\pm45^\circ$~to the vertical ($u$- and $v$-planes). The sense wires were 
12-$\mu$m-diameter gold-plated tungsten.  The $x$-, $u$- and $v$-planes had 
144, 60 and 60 readout channels, respectively, with a 1.27~mm wire 
spacing. In the $u$- and $v$-planes, pair of wires were multiplexed in one 
readout channel.  
The active area was 
17.8~cm (horizontal) by 5.08~cm (vertical).  The cathode foils were 
25-$\mu$m thick aluminized mylar coated with carbon.  The anode-cathode 
distance was 3.18~mm, and the total thickness of BWPC1 was 
approximately 56~mm. The second chamber (BWPC2) was located 1.0~m 
downstream of BWPC1 and also contained three planes ($x$, $u$ and $v$). The 
direction of the sense wires was vertical ($x$-plane) and 
$\pm60^\circ$~to the vertical ($u$- and $v$-planes). Each plane had 
120 active sense wires with a 0.8-mm wire spacing. 
Among the 120 wires, the central 72 ones were multiplexed 
by 3 and the remaining were multiplexed by 6 in the readout channels,
yielding a total of 32 readout channels for each plane.  
The cathode foils were 8-$\mu$m 
single-sided aluminized mylar coated with carbon.   
The anode-cathode distance was 1.6~mm.  
Both chambers were filled with a recirculated mixture of 
CF4 (80\%) and Isobutane (20\%). 

Downstream of the BWPC's a degrader slowed the $K^+$'s so that they  stopped 
in the center of the scintillator fiber target.  
The upstream section of the degrader was 
inactive, consisting of 11.11~cm of beryllium oxide (BeO) and 4.76~mm of 
Lucite. The high density (3.0 g/cm$^3$) and 
low atomic number of BeO was used to minimize multiple scattering.  
The AD consisting of 
40 layers of 2~mm thick disks of Bicron BC404 scintillator 
(13.9~cm diameter) alternating with 2.2-mm thick copper disks 
(13.6~cm diameter) was divided into 12 azimuthal segments with 
readout to a single Hamamatsu 
R1924 PMT through fourteen 1-mm-diameter Bicron BCF99-29-AA-MC  
wave length shifting (WLS) fibers. The PMT outputs were  
provided to TDC's, CCD's and a 4-fold analog sum that 
was provided to an ADC.  
These measurements enabled the AD to identify the  
beam particles and to detect 
activity coincident with $K^+$ decays.  
  
Downstream of the degrader the B4 hodoscope 
detected the entrance position of the incoming 
particle in the target   
and identified the particle type by measuring its energy 
loss.  
The B4 hodoscope consisted of two planes, $u$ and $v$, with about a 11.8-cm 
diameter oriented at a $\pm33.50^\circ$ angle  
with respect to the horizontal axis.   
Each plane had 16 Bicron BC404 scintillator fingers with a 7.2-mm pitch.   
The cross section of 
each finger had a `Z-shape'  
with a 6.4-mm thick middle part and 3.2-mm thick edges.  This shape 
reduced inactive regions and improved the spatial resolution. 
Three Bicron BCF99-29-AA-MC WLS fibers were embedded in each finger 
and connected to a single Hamamatsu H3165-10 PMT that was read out by TDC's, 
ADC's and CCD's.   
At the same position as the B4 
hodoscope, but at larger radius was an annular scintillator counter, the 
ring veto (RV). The RV was designed to veto particles that passed 
through perimeter of the B4 hodoscope. The RV was composed of two 
180$^\circ$ arcs of 3.3~mm thick Bicron BC404 scintillator  
with an inner diameter 
varying from 11.9~cm to 12.0~cm and an outer diameter of 14.6~cm. The 
two RV elements were readout by Hamamatsu H3165-10 PMT's and the  
signals were split 
three ways to ADC's, TDC's and CCD's. 
 
\subsection{Target} 
 
The target consisted of 413 Bicron BCF10 scintillating fibers of 5-mm  
square cross section and 3.1-m length 
that were bundled to form a 12-cm-diameter cylinder.  
A number of 1-mm, 2-mm and 3.5-mm square  
scintillating fibers (called ``edge fibers")  
filled the gaps near the outer edge of the target.   
Each of the 5-mm fibers was connected to a Hamamatsu R1635-02 PMT,  
whereas the adjacent edge fibers were grouped onto 
16 PMT's, providing signal read out by ADC's, TDC's and CCD's.  
 
The fiducial region of the target  
was defined by two layers of six plastic-scintillating counters 
surrounding the target. 
The inner scintillators, called I-Counters (IC's),  
helped to define the fiducial volume 
and the logical OR of the six IC's ($IC$ for trigger condition) used 
by the trigger for this purpose. 
The IC's were 6.4-mm thick (with an inner radius of 6.0~cm) 
and extended 24~cm from the upstream face of the target.  
The outer scintillators, called V-Counters (VC's), overlapped  
the downstream edge of the IC's by 6~mm, and served to detect  
particles that decayed downstream of the fiducial region of the target.  
The VC's consisted of six 5-mm thick and 1.96-m long  
scintillators, and were staggered  
azimuthally with respect to the IC's. 
Each IC and VC element was instrumented with  
an EMI 9954KB PMT 
which was read out by an ADC, TDC and a 500~MHz transient  
digitizer (TD) based on a flash ADC~\cite{TD}.   
 
Approximately 27\% of the incident $K^+$'s (typically $3.5\times10^6$ 
$K^+$'s/spill) penetrated far enough into the target to satisfy the 
online target criteria for $KB$ defined in Section~\ref{sec:pnn_trigger}. 
The remaining $K^+$'s either 
underwent decay-in-flight, nuclear interaction in the degrader or 
scattered in the material of the beam instrumentation and did not 
reach the target. 
It should be noted that some of the $K^+$'s ($<$ 25\%) that
satisfied the online $KB$ requirement didn't stop in the target,
and a factor for the stopping fraction of $K^+$'s was introduced
as described in Section~\ref{sec:fs}.
The $K^+$'s deposited an average energy of 100~MeV 
in the scintillating fiber target when coming to rest in the 
center of the target fiducial volume.  The low velocity $K^+$'s 
typically lost 5--40~MeV in each fiber, while the nearly minimum 
ionizing (MIP) $\pi^+$'s from $K^+$ decays deposited about 1~MeV per 
fiber, as they passed transversely through the fibers. 
 
\subsection{Drift Chamber} 
\label{sec:track_chamber} 
 
The drift chamber,   
called the ``Ultra Thin Chamber" (UTC)~\cite{utc},  
was located just outside of the IC. The primary functions of the  
UTC were to measure the momenta of charged 
particles and to provide tracking between the target and the Range Stack (RS). 
The UTC had inner and outer  
radii of 7.85~cm and 43.31~cm, respectively.  
Twelve layers of 5--8 mm drift cells were grouped into three super-layers 
(each consisting of four layers) with  
active lengths of 38.8~cm (inner), 44.8~cm  
(middle) and 50.8~cm (outer).  
The super-layers were filled with a 49.8\%:49.8\%:0.4\% mixture of 
argon, ethane and ethanol.  
Each anode wire was instrumented with an ADC and a TDC.  
The drift time to the anode wires was used to determine the  
($x,y$) positions for the charged track.  
At the inner and outer radii of each super-layer were cathode foils,  
with 7~mm helical cathode strips at a $\sim45^\circ$ pitch angle.  
Each cathode strip was instrumented with an ADC and a TDC. The  
combination of the charge centroid on the cathode strips 
and anode hits provided the $z$ hit position.
There were two inactive regions filled with nitrogen gas between  
the three super-layers. Differential pressures of $\sim$2 mbar 
in the five gas volumes  
supported the cathode foils. The total mass of   
the UTC (excluding the inner and outer support tubes with the 
attached foils) amounted to 2$\times 10^{-3}$ radiation lengths. 
The UTC position resolutions were approximately 175 $\mu$m for $x$  
and $y$ and 1~mm for $z$. 
 
\subsection{Range Stack}  
\label{sec:rs}

Located just outside the UTC at an inner radius of 45~cm 
and an outer radius of 84~cm, the RS consisted of  
both scintillation counters and  
embedded straw chambers, providing energy and range measurement 
of the charged particles, information on the \PIMUE~decay sequence and  
measurement of photon activity. 
 
\subsubsection{Scintillation Counters}

The RS consisted of 19 layers 
of Bicron BC408 plastic scintillator, azimuthally segmented into 24 sectors  
as shown in Fig.~\ref{fig:gamma_detectors}.  
Layers 2--18 were 1.9~cm thick and  
1.82~m long. Layer 19 was 1.0~cm thick and 
was mainly used to veto  
longer range muons. 
Each scintillator in layers 2--19 
was coupled through Lucite light guides to EMI 9954KB PMT's 
at both upstream and downstream ends.  
The innermost counters were 6.4~mm thick and  
52~cm long trigger-counters (T-counters), defining the fiducial 
volume for charged $K^+$ decay products.  
The T-counters were thinner than layers 2--19 to  
suppress rate from photon conversions. 
Seventeen 1-mm-diameter WLS fibers (Bicron multi-clad BCF-92) with a pitch  
of 6.9~mm were embedded in each scintillator and  
coupled to a single Hamamatsu R1398 PMT 
at each end. 
 
The signal from each PMT of the RS scintillators was passively split 
1:2:2 for ADC's, discriminators and fan-in modules, respectively.  
The discriminator output was sent to  a 
TDC and the trigger. Each PMT was read by an ADC and a TDC. 
The TDC's (LeCroy 3377) recorded up  
to 16 hits in a 10.5 $\mu$s time window, 
and thus allowing for efficient detection of the $\mu^+\to e^+$ 
decay. The analog fan-in summed the signals from 4 PMT's  
on each end in the same hextant (four adjacent sectors) and layer. 
This analog sum was read out by a single TD~\cite{TD} and  
was provided to  
mean-timers~\cite{L0-IEEE} 
for good timing on the photon veto in the trigger and a $z$-measurement 
for each layer of the track. 
The mean-timer output 
from each layer in a hextant was ORed to provide input to the hextant 
photon veto algorithm in the trigger (vetoing more than one  
non-adjacent hextant). 
The TD's recorded the charge in 2~ns intervals (500~MHz sampling)  
in a 2.5 $\mu$s time window with  
a resolution of 8 bits. 
The 500~MHz sampling provided sufficient pulse shape information  
to separate pulses from different events as 
close as 5~ns apart, and enabled the detection of the $\pi^+\to\mu^+$ decay 
as described in Section~\ref{sec:pimue_id}. 
The time window of the TD's was narrower than that of the  
TDC's in order to reduce the data size.  

\subsubsection{Range Stack Straw Chambers} 
 
\begin{figure} 
\centering 
\epsfxsize 0.9\linewidth 
\epsffile{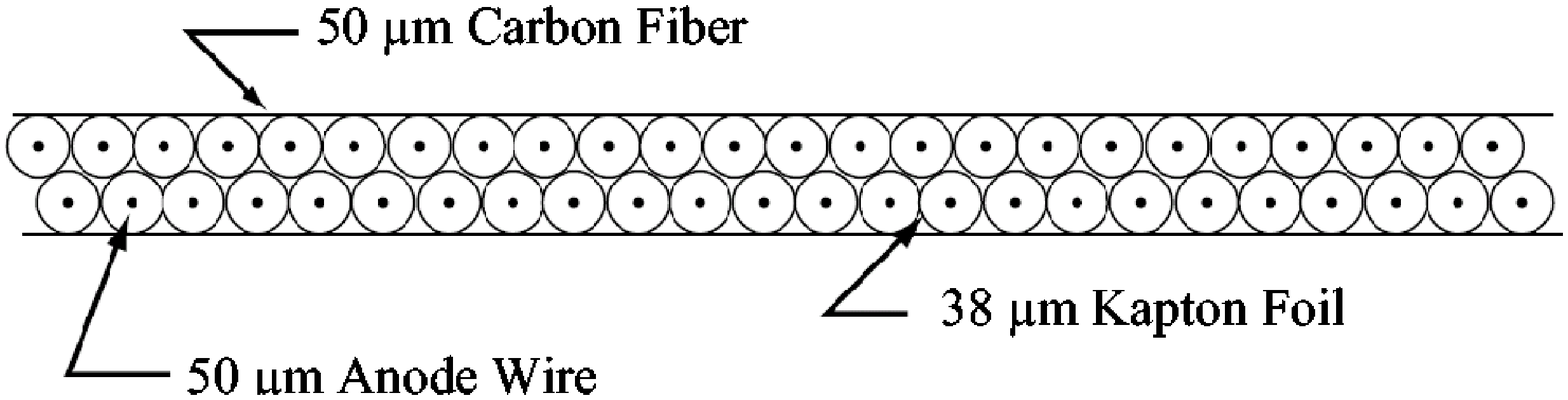} 
\caption{Schematic end view of the inner RSSC. The tubes run 
through the beam direction.} 
\label{fig:rssc} 
\end{figure}

Two range stack straw chambers (RSSC's) were 
located outside RS layer 10 and 14~\cite{rssc}.  
The inner (outer) RSSC  
consisted of two staggered layers of 24 (28) straws per sector  
with a length 97.8 (113.0)~cm.  
The average density of an RSSC was 
0.054~g/cm$^2$. Each straw tube was 3.4~mm in 
radius with a 50-$\mu$m-diameter  
gold-coated tungsten anode wire at the center. A schematic drawing is given  
in Fig.~\ref{fig:rssc}. 
The straw chambers were operated  
with 67\% argon and 33\% isobutane mixture with a trace of water 
in a self-quenching streamer mode at 3,450 V. 
The local $x$-axis of a chamber was defined  
to be along the width of the chamber. There were a total  
of 48 chambers, with 2,496 straws installed in the E949 experiment. 
Because of access restriction, a pair of straws from  
the right and left halves in the same layer  
were connected at the upstream end to allow downstream-only readout.  
 
The position of the hit straws   
provided $x-y$ position information, while the end-to-end time  
differences provided the $z$ measurement. 
The $z$ resolution of the E787 RSSC's was degraded due to a   
pulse-height dependent time-walk effect. For E949  
an amplifier and two discriminators  were installed on 
each channel to allow for low threshold timing discriminator  
and a high threshold logic discriminator (above the noise level).  
The $z$ resolution was improved  
from 3.0~cm (RMS) observed in E787 to 1.5~cm (RMS). 
 
\subsection{Photon Veto Counters} 
 
The detection of activity coincident with the charged track was 
crucial for suppressing background processes that can mimic \KPPNN.  
Photons from \KPITWO~and other radiative decays were detected by the  
hermetic photon system as shown in Fig.~\ref{fig:gamma_detectors}. 
The photon detectors surrounding the $K^+$ decay vertex with 
a 4$\pi$ solid angle coverage were located  
in the barrel, upstream and downstream  
end caps, and near the beam line. 
The photon system consisted of essentially every scintillator  
detector in experiment: the 
Barrel Veto (BV), the Barrel Veto Liner (BVL), the RS,  
the upstream and downstream End Caps (EC's),  
the Upstream Photon Veto (UPV),  
the upstream and downstream  
Collar detectors (CO),  
the downstream Microcollar detector ($\mu$CO),  
the Downstream Photon Veto (DPV), the RV, the IC, the VC,  
the AD and the target. The regions of the target, IC and RS  
traversed by the charged 
track were excluded from the photon veto. The AD and DPV  
were part of the E949 detector upgrade but  
were only used in the $\pi\nu\bar{\nu}(2)$ 
analysis, where photon veto near the beam axis was more important.  
 
The 1.9-meter-long, 14.3 radiation length thick  
BV covering 2/3 of the 4$\pi$ sr solid 
angle was located in the outermost barrel region  
with an inner radius of 94.5~cm and an outer radius of  
145~cm~\cite{E787phase1NIM}. 
The BV was divided into  
48 azimuthal sectors.  
Each sector consisted of  
four radial layers, in which there were 16 (innermost), 18, 20,  
21 (outermost)  
layers of 1-mm thick lead and 5-mm thick Bicron BC408 plastic scintillator.  
The light collected in the scintillators accounted for 30\% 
of the total energy deposit in the BV. 
The azimuthal boundaries of each sector were tilted so that there were no  
projective cracks for photons from the decay vertex. 
Both ends of every module were read out by an EMI 9821KB PMT 
and the signals were recorded by an ADC and a TDC.  
The time resolution of individual BV counter was 
measured to be 1.2~ns. 
 
\begin{figure} 
\centering 
\epsfxsize 0.8\linewidth 
\epsffile{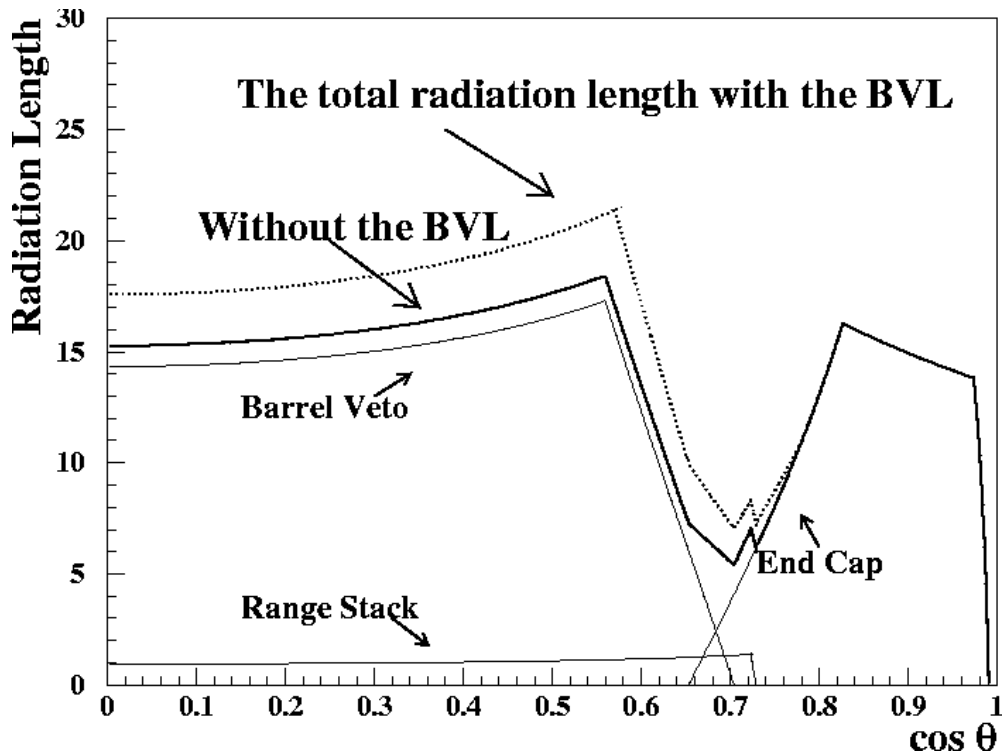} 
\caption{Radiation length with (solid curve) and  
without (dotted curve) the BVL as a function of the cosine of the 
polar angle. These curves  account for the contribution from  
RS, BV, EC and BVL. Other photon veto counters are not accounted for in  
this plot.} 
\label{fig:bvl} 
\end{figure}

In order to improve the photon veto capability, the BVL, 
located between the RS and the BV,  
replaced  the outermost layers  
20 and 21 of the RS in E787. 
Each BVL counter was 10~cm wide and 2.2~m long.  
There were 48 azimuthal sectors, each  
with 12 layers of 1-mm thick lead and 5-mm  
thick Bicron BC408 plastic scintillator, for a total  
thickness of 2.29 radiation lengths. Both ends of the  
BVL modules were read out by EMI 9821KB PMT's  
and the signals were recorded by ADC's and TDC's.  
The eight adjacent sectors (hextant) in each end  
were grouped and read out by TD's. 
The timing resolution of an individual BVL counter was 0.7~ns. 
A comparison of radiation length coverage 
with and without the BVL is shown in Fig.\ref{fig:bvl}. A factor 
of two improvement in the photon veto rejection of K$_{\pi2}$  
decays was expected from the BVL. 
 
The EC's had 
roughly one-third of the 4$\pi$ sr  
photon coverage~\cite{endcap}.  
The upstream EC detector consisted of seventy-five 25~cm long  
(13.5 radiation lengths)  
undoped Cesium Iodide (CsI)  
crystals segmented into  four rings,  
and the downstream EC detector consisted of 68 crystals in four rings.  
To maximize light collection the PMT's were directly coupled to the crystals 
through a Sylgard cookie formed over the PMT and a ultraviolet  
transmitting optical filter 
that selectively passed the fast component of the CsI scintillation  
light (with a decay time of  
a few tens of nanoseconds at a wavelength of 305~nm) and blocked the  
slow component.  
Since the PMT's were situated in the magnetic field,  
high-field PMT's 
were used~\cite{finemesh} (Hamamatsu R5543 3'' PMT's for  
the outer three rings and  
Hamamatsu R5545 2'' PMT's 
for the smaller inner ring crystals); the signals were  
split to ADC's, constant fraction discriminators (CFD), and  
CCD's.  
The CFD's outputs were ORed to provide an online photon veto  
signal and were also 
sent to TDC's. The pulses recorded in the CCD's were  
analyzed offline by a pulse-finding algorithm to provide the  
timing of the EC signals and to  
separate two pulses close in time (to reduce the possibility of  
accidental hits reducing the efficiency of photon detection). 
Since the EC was exposed to a high counting-rate 
environment near the beam line, good timing was required to  
reduce acceptance losses and the masking of \KPITWO photons from 
early accidental hits. 
 
The UPV mounted to the downstream face of the $\check{\rm C}$erenkov  
counter was 3.1 radiation 
lengths thick, with an outer dimension of 28.4~cm $\times$ 28.4~cm and an 
inner hole for the beam of 17.5~cm (horizontal) $\times$ 4~cm (vertical). 
It consisted of twelve layers of 2~mm thick Bicron BC404 plastic  
scintillator and 1--2~mm thick lead plate. The scintillation light was 
read out by $12\times21$ WLS fibers coupled to two Hamamatsu  
R1924 PMT's. The UPV signals were sent to ADC's, TDC's and CCD's. 
 
The upstream (downstream) CO detector was located just upstream (downstream) of the EC's~\cite{e949_co}.   
Both of the CO's consisted  of twenty-four 2-mm thick lead sheets  
alternating with 25 layers of 5-mm thick Bicron 
BC404 scintillator sheets, 
providing about 9 radiation lengths along the beam direction. 
Each scintillator layer was segmented into 12 wedges forming 12  
identical azimuthal sectors. Light from the wedges was readout by  
WLS multi-clad fibers (Bicron BCF99-29AA). Each wedge had 16 fibers 
glued in grooves in the BC404 scintillator layer. One end of the fiber
was polished and aluminized to provide reflective mirror surface. 
All fibers for each sector (16$\times$25=400) were terminated in a 
connector which was coupled to a 1.17-m long Lucite rod via a silicone 
cookie. The rods transmitted the light to EMI 9954KB PMT's in a low field 
outside the magnet. Signals were  
sent to ADC's and TDC's. The CO's detected photons emitted with a  
small polar angle ($0.970<|$cos$\theta|<0.995$)  
in both the upstream and downstream regions. 

The $\mu$CO was installed between the inner face of the magnet 
end-plate and the target downstream of the downstream CO to give more photon 
detection coverage. The $\mu$CO had an inner (outer) diameter of 
15.6~cm (20.0~cm) and a length of 53~cm. 
The eight azimuthal sectors contained eight layers of 2~mm scintillating 
fibers (from the original E787 target~\cite{E787phase1NIM}) separated by 
seven layers of 60~$\mu$m Pb. The 536 fibers from two adjacent sectors 
were read out by one EMI9954 PMT into ADC's and TDC's.

\subsection{Trigger} 
\label{sec:trigger} 
 
The E949 trigger selected \KPPNN\ events from the large number 
of $K^+$ decays and scattered beam particles with requirements on the 
range of the $\pi^+$ track, the presence of a \PIMU\ decay in the RS,  
the absence of other activity at the time of the $\pi^+$ and the presence of a 
$K^+$ at an appropriately earlier time. 
 
The trigger was composed of two stages: a fast Level-0 trigger with a 
decision time of $\sim$100~ns and a slower Level-1
trigger with a decision time of 10--100~$\mu$s.  The Level-0 trigger 
was based on signals from the beam, target, RS 
and photon veto systems, processed with a combination of commercial and  
custom-built boards. The Level-1.n trigger was composed of two 
parts running in parallel, the Level-1.1 and Level-1.2 triggers, 
that involved the partial processing of TD and ADC data, and operated  
only on events that passed the Level-0 trigger.   
 
The original E787 trigger has been described 
previously~\cite{E787phase1NIM}.
The trigger was upgraded for E949 with the 
addition of a new programmable Level-0 trigger board and mean-timers 
for the photon veto signals~\cite{L0-IEEE}. 
 
E949 collected data with triggers for the \KPPNN~signal and for other
physics studies and calibrations. In this paper, 
the triggers for calibration and other physics are referred to as the 
``monitor'' samples. 
 
\subsubsection{Trigger Architecture} 
\label{sec:triggers} 
 
For physics and most calibration triggers, Level-0 required 
a signal from the beam system and 
a charged track in the RS. A beam $K^+$ or $\pi^+$ was 
identified in NIM logic by a coincidence of hits from the 
$\check{\rm C}$erenkov counter, B4 ($B4$), target ($Target$) 
and AGS spill gate ($Spill$). This $KB$ signal served as the 
Beam Strobe ($BS$) for the trigger. A 
charged track $T\cdot2$ was identified by a 
coincidence of the $IC$ and T-counter and second 
layer (mean-time of the two ends) in a single RS sector.  Additional 
requirements were placed on the crude estimate of the track range, 
which was the track length derived from the deepest layer of the 
RS hit in coincidence with the $T\cdot2$ from that sector or the two 
proceeding sectors (for a positively charged track). Another 
requirements were placed on the 
photon veto (EC, RS, BV, BVL) and the delayed-coincidence ($DC$). The 
$DC$ was an OR over the six coincidences of the individual  
IC's and the delayed $KB$ signal such 
that a track in the IC must typically have occurred at least  
1.5~ns later than a prompt signal. The 
$T\cdot2$ signal, an OR of the 24 individual sector $T\cdot2$'s, 
typically served as the Detector Strobe ($DS$) that gated all of the 
ADC's not associated with the beam system or target and provided a 
common stop for many of the TDC's; the $T\cdot2$ signal introduced 40~ns of 
dead time.  When an event passed the Level-0 trigger, the dead time 
was extended to 100~ns to allow further processing. Additional 
triggers that did not contain both a beam signal and a $T\cdot2$ were 
produced on a separate trigger board that fed directly into the 
trigger bus (with external coordination of the dead time). 
 
The Level-1.1 trigger ($L1.1$) was based on information from the TD system for 
the hextant containing  
the RS counter in which the charged track was determined to have 
stopped.  Stopped $\pi^+$'s were preferentially selected (over 
$\mu^+$'s) by looking for the $\pi \rightarrow \mu$ decay by comparing 
the pulse height to the pulse area (for early decays) or by looking 
for a second detached pulse.  This was done by a custom-built ASIC 
which had access to the TD memories. The ASIC could also reduce the TD 
data size for readout by discarding the waveform samples outside the 
``prompt'' time window, keeping instead a calculation of the pulse 
area and leading-edge time. The ``prompt'' window typically extended from 
0.5 (2) $\mu$s before (after) the $\pi^+$ track. Level-1.1 typically 
provided a decision in about 10--20~$\mu$s. 
 
The Level-1.2 ($L1.2$) used data digitized by the ADC's and 
had two components. One of them, the ``Level-1.1 
afterburner'' rejected events with an accidental hit near the stopping 
counter. Such hits might defeat the Level-1.1 with an apparent double 
pulse.  The other component, the ``$HEX$ afterburner'', was used for 
the photon veto, rejecting events with hits in both of the two 
adjacent hextants when the $T\cdot2$ counter and the stopping counter 
were found in the same hextant. This usually indicated that a single 
hextant fully contained the trajectory, thus those observed in 
the other hextant came from another particle.  The Level-1.2 trigger
introduced a dead time of up to 100~$\mu$s per Level-1.1 trigger. 
  
\subsubsection{\KPPNN\ Triggers} 
\label{sec:pnn_trigger} 
 
The triggers as described below were valid for the bulk of the 2002 
running period.  The trigger conditions for \KPPNN\ were designed 
differently according to the $\pi^+$ momentum. For high $\pi^+$ 
momentum, the $\pi\nu\bar{\nu}(1)$ trigger condition was defined as: 
\begin{eqnarray} 
\pi\nu\bar{\nu}(1) 
&=&KB\cdot DC\cdot(T\cdot2)\cdot(6_{ct}+7_{ct})\cdot \\ \nonumber 
& &\overline{19_{ct}}\cdot \overline{zfrf}\cdot L0rr1\cdot 
\overline{(BV+BVL+EC)}\cdot\\ \nonumber 
& & HEX\cdot L1.1\cdot L1.2, 
\end{eqnarray} 
where 
\begin{equation} 
KB=C_K\cdot B4\cdot Target \cdot Spill. 
\end{equation}  
The $6_{ct}+7_{ct}$ required that the $\pi^+$ reached the $6^{\rm th}$ 
or $7^{\rm th}$ layer of the RS and suppressed the copious 3-body 
$K^+\rightarrow\pi^+\pi^-\pi^+$ and $K^+\rightarrow\pi^+\pi^0\pi^0$ 
backgrounds. The $\overline{19_{ct}}$ signal required that the charged 
track should not reach the $19^{\rm th}$ layer in order to suppress 
$K_{\mu2}$ background.  The ``$ct$'' 
designated the RS sectors that were associated with a $T\cdot2$ 
($T\cdot2$ sector plus the next two clockwise sectors: this was the 
direction that a positive particle moved in the magnetic field). The   
$\overline{zfrf}$ condition was a fiducial cut on the $z$-position of 
the charged track in each layer, vetoing tracks 
that exited the fiducial volume.  
The $L0rr1$ was a refined calculation of the charged track range, 
which included the number of target fibers hit and a measurement of 
the $z$-position of the track from flash TDC's on layers 3 and 11--13 as well 
as the deepest layer of penetration of the track; this rejected events 
with long range such as the $\mu^+$ from $K_{\mu2}$ decay. The photon 
veto $\overline{BV+BVL+EC}$ and $HEX$ were from the BV, BVL, EC  
and RS, respectively, which removed events with 
photons such as $K_{\pi2}$, $K_{\mu3}$ and $K_{\mu2\gamma}$. 
  
The number of $K^+$'s which met  
the $KB$ trigger requirement when 
the detector was live ($N_K$), was recorded at the end of each AGS spill. 
The total exposure from the $\pi\nu\bar{\nu}(1)$ trigger data stream was  
\begin{equation} 
N_K=1.77\times 10^{12}. 
\end{equation} 

E949 also provided a $\pi\nu\bar{\nu}(2)$ trigger for  
lower momentum $\pi^+$'s. Since the study involves  
different background mechanisms and a different kinematic region,  
the result will be presented in a separate paper. 

\subsubsection{Monitor Triggers} 
\label{sec:moni} 
 
In addition to the $K^+\rightarrow\pi^+\nu\bar\nu$ triggers, there 
were monitor triggers for calibration and normalization, as well as 
triggers for other physics modes. All triggers were prescaled to reduce the 
impact on the total dead time except 
for the \KPPNN\ triggers and the ``$\gamma$" trigger which required the  
presence of photons in the barrel region for studying  
$K^+\to\pi^+\gamma\gamma$ and $K^+\to\pi^+\gamma$ decays~\cite{two_gammas}. 
 
To monitor detector performance several processes were employed, 
including $K_{\pi2}$ and $K_{\mu2}$ decays, beam particles  
scattered into the detector fiducial volume, 
beam $K^+$, charge exchange events and cosmic rays. These 
monitor samples were used for calibration and acceptance studies. They 
were taken simultaneously with the $\pi\nu\bar\nu(1)$ and  
$\pi\nu\bar\nu(2)$ triggers to reflect any condition changed into 
the acceptance and background calculations. 

The $K_{\mu2}$ decay has the largest branching ratio.  Since the final 
state does not contain photons or additional tracks and the daughter 
$\mu^+$ does not interact strongly, it was a convenient sample for 
a variety of acceptance measurements as well as several calibrations. 
This sample was also used for the normalization of the 
experiment; the measurement of the $K_{\mu2}$ branching ratio 
effectively normalized the counting of $K^+$ stops to the well-known 
$K_{\mu2}$ branching ratio. The $K\mu2$ trigger condition for $K_{\mu2}$
was defined as follows
\begin{equation} 
K\mu2 
=KB\cdot(T\cdot2)\cdot(6_{ct}+7_{ct})\cdot(17_{ct}+18_{ct}+19_{ct}). 
\end{equation} 
 
The final state of the $K_{\pi2}$ decay mode contains one charged $\pi^+$ 
and two photons from $\pi^0$ decay. Since the $\pi^+$ momentum is 
monochromatic, the  $K_{\pi2}$ sample was used to check  
the measurement of the charged 
track momentum, range and energy.  Also the $\pi^+$'s were used to study 
particle identification, while the photons were used to study the 
photon veto.   
Two $K_{\pi2}$ triggers were defined, a loose one, $K\pi2(1)$: 
\begin{equation} 
K\pi2(1) 
=KB\cdot(T\cdot2)\cdot(6_{ct}+7_{ct})\cdot\overline{19_{ct}}. 
\end{equation} 
and a tighter one, $K\pi2(2)$: 
\begin{equation} 
K\pi2(2)=K\pi2(1)\cdot HEX\cdot L1.1\cdot L1.2. 
\end{equation} 
The $K\pi2(2)$ monitor trigger was used only for calibration in this analysis. 
 
Among the incoming beam particles there were many $\pi^+$'s, including 
some that scattered into the fiducial volume of the RS.  These 
$\pi_{scat}$ events were identified as $\pi^+$'s by the $C_\pi$  
and had an in-time track in the RS.  The kinematic features of 
this $\pi^+$ sample were almost the same as the \KPPNN\ signal except 
that the target pattern was different. This sample was suitable for 
calibrating the ionization energy loss  
($dE/dx$) of $\pi^+$'s and for studying the acceptance. 
The trigger condition for scattered beam particles was defined as: 
\begin{eqnarray} 
\pi_{scat} 
&=&\pi B\cdot\overline{DC} 
\cdot(T\cdot2)\cdot(6_{ct}+7_{ct})\cdot\overline{(BV+BVL+EC)}\cdot HEX. 
\end{eqnarray} 
where 
\begin{equation} 
\pi B=C_\pi\cdot B4\cdot Target \cdot Spill. 
\end{equation} 
 
The charge exchange process (CEX), $K^+ n\rightarrow p + K^0$ 
followed by $K^0_L\rightarrow\pi^+ l^-\bar{\nu}_l$ can mimic 
\KPPNN\ events when the charged lepton 
$l^-$ from the $K^0_L$ decay had a low momentum and was undetected. The 
largest uncertainty in this background was the determination of the 
reaction rate. Since $K^0$ has an equal fraction of $K^0_L$ and 
$K^0_S$, the $K^+ n\rightarrow p + K^0_S$ process with a $K^0_S$ decay 
to $\pi^+\pi^-$ can be used to measure the reaction rate for the 
determination of the background from $K^+ n\rightarrow p + 
K^0_L$. Since the $K^0_S\rightarrow\pi^+\pi^-$ could be cleanly 
identified, a $CEX$ trigger with two charged tracks was defined as: 
\begin{equation} 
CEX 
=KB\cdot\overline{DC}\cdot2(T\cdot2)\cdot(6_{ct}+7_{ct}) 
\cdot\overline{EC+\pi B}. 
\end{equation} 
 
In addition, we also defined triggers for beam $K^+$ and cosmic ray,  
which were used in trigger efficiency measurement and detector  
geometrical alignment. 
 
\subsection{Data Acquisition} 
 
Analog- and discriminated- signals from the detector were digitized by 
commercial ADC and TDC, and custom-built waveform digitizer 
(TD and CCD) systems. 
When an event was accepted 
by the trigger system, the digitized data for the event were 
transferred to a buffer module or a local crate controller. 
At the end of each spill, 
the data for the spill were transferred to a host computer.   
A summary of the digitizing electronics is shown 
in Table~\ref{tab:digitizers}. 
\begin{table} 
  \centering 
  \begin{tabular}{c c c c c}\hline\hline 
  Type & Model & Standard & Resolution & Subsystems \\ \hline 
  \multirow{2}{*}{ADC}  & LRS 4300B & CAMAC   & 10 bits & RS,BV,BVL,EC,Beam\\ 
                        & LRS 1881  & Fastbus & 13 bits & Target,UTC \\ 
  \hline 
  \multirow{3}{*}{TDC}  & LRS 3377 & CAMAC   & 0.5 ns & RS, BVL \\ 
                        & LRS 1879 & Fastbus & 2 ns & UTC,BV,Target\\ 
      & LRS 1876 & Fastbus & 1 ns & EC,RSSC,Beam\\ 
  \hline 
  \multirow{4}{*}{WFD} &\multirow{2}{*}{TD} & \multirow{2}{*}{Fastbus} & 500~MHz sampling  
   &\multirow{2}{*}{RS,BVL,IC} \\ 
                       &    &         & 8 bits, up to 10 $\mu$s depth  & \\ 
                       &\multirow{2}{*}{CCD}& \multirow{2}{*}{Fastbus} & 500~MHz sampling  
   &\multirow{2}{*}{Beam,Target,EC}\\ 
                       &     &         & 8 bits, 256 ns depth  & \\ 
  \hline\hline 
  \end{tabular} 
  \caption{Digitizing electronics for E949.} 
  \label{tab:digitizers} 
\end{table}

For the Fastbus systems, SLAC Scanner Processor (SSP) modules~\cite{SSP}  
served as crate controllers and also to read out, reformat and buffer 
the data from the front-end after each trigger accepted.  The CAMAC ADC's 
were read out through the FERA bus by a Struck 370 QDP 
DSP (Fastbus) module. The CAMAC TDC's 
were read out by custom-built DYC3 modules~\cite{DYC} which  
pushed the data into VME memory boards.  The readout time per  
event (as determined by the slowest crate) was typically 850 $\mu$s. 
 
At the end of each spill, the data from the Fastbus buffer memories were read 
out via the cable segment (12-15 MB/sec) by Struck 340 
SFI modules, each controlled by a MVME 2604 single-board computer (SBC) 
running VxWorks.  The 
VME memory boards were read out by a separate SBC.  Data were 
transferred from the SBC's to the host computer (SGI Origin 200) 
via Ethernet (9 MB/sec per link) through a simple network switch. 
Event fragments from the readout segments were combined by  
Event Builder processes running on the host computer.   
Complete events were distributed  to 
``consumer'' processes which included data logging and online 
monitoring. The \KPPNN\ triggers were written to two DLT-7000 drives 
at 5 MB/sec per drive; a third DLT drive was used to log monitor triggers. 
 
A slow control system, based on the MIDAS~\cite{MIDAS} framework, ran 
independently of the main DAQ system and was used to monitor a variety 
of experiment conditions, including crate voltages and  temperatures. 
 
Under typical running conditions, we wrote $\sim$300 events per spill 
with a typical event size of $\sim$80~kB.   
This was well within the maximum 
throughput of the system of about 50~MB/spill.  The DAQ dead time was 
due entirely to the speed of the event-by-event readout of the 
front-end electronics at the crate level. 
The total dead time introduced by the trigger and DAQ was typically 
26\%. 
The E949 experiment collected its physics data for 12 weeks 
from March through June of 2002 or about 20\% out of the total beam 
time approved by DOE's Office of High Energy Physics.

\section{Data Analysis} 
\label{chap:analysis} 
 
The branching ratio of the \KPPNN~decay in the SM is 
$\sim$$10^{-10}$ as discussed in Section~\ref{chap:intro}.  
Unlike the $K_{\mu2}$  
and $K_{\pi2}$ backgrounds, the \KPPNN~signal  
is continuous with no peak.
To establish that any possible observed candidate
event was really from \KPPNN, we required that the backgrounds were
suppressed to a level substantially below one event with small
uncertainty.    
All the detectors were calibrated before the background 
and acceptance studies using $K\mu2$, $K\pi2(1)$, $K\pi2(2)$ and cosmic-ray 
and beam trigger events. 
After a brief overview of the background sources, 
the data analysis technique will be  
described: 
including  selection criteria, background evaluation,  
acceptance measurement and the signal candidate search. 
 
\subsection{Overview of Background} 
\label{sec:bkg_overview} 
 
\begin{figure} 
\centering 
\epsfxsize 0.95\linewidth 
\epsffile{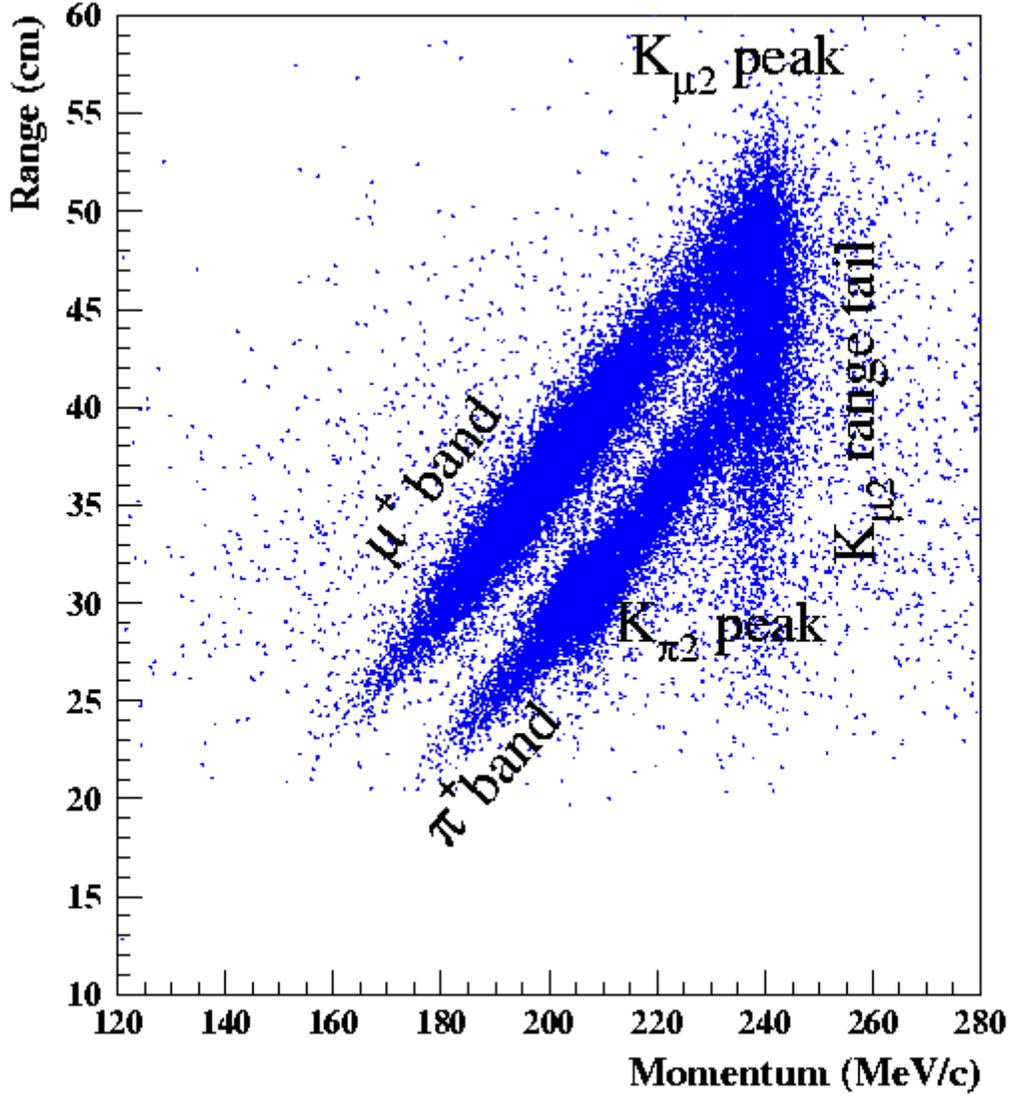} 
\caption{Range in  plastic scintillator (cm) 
 versus the momentum (MeV/$c$)  
of the charged particles for events  
that passed the $\pi\nu\bar{\nu}(1)$ trigger. These data represent 
$\sim$0.3\% of $N_K$.}  
\label{fig:r_vs_p_kp21} 
\end{figure}

Data selected by  the $\pi\nu\bar{\nu}(1)$ trigger  
were primarily from background  
events as shown in Fig.~\ref{fig:r_vs_p_kp21}.  
These events were classified into stopped-$K^+$-decay-related and  
beam-related backgrounds.   
 
\subsubsection{Origins of Stopped-$K^+$-Decay Background} 
 
The stopped-$K^+$-decay backgrounds were  
categorized into two types: $\pi^+$-related, $\mu^+$-related 
backgrounds. As can be seen in Fig.~\ref{fig:mom_spec}, multi-body $K^+$  
decays were suppressed by setting a signal momentum region  
higher than the \KPITWO~peak but lower than the \KMUTWO~peak.  
However, since \KPITWO~and \KMUTWO~ have such 
 large branching ratios (20.92\% and 63.44\%~\cite{Yao:2006px}),  
migration into the signal region  
through either resolution  or scattering effects was a significant background. 
Radiative \KMUTWO~decay and  $K^+\to\mu^+\pi^0\nu$ ($K_{\mu3}$) decays 
accounted for the majority of the observed $\mu^+$ band  
events in Fig.~\ref{fig:r_vs_p_kp21}. 
For the \KMUTWO~decay mode, the background originated from a \KMUTWO~peak 
event or a \KMUTWO~ low momentum (``tail'') event if  the  
particle identification 
was fooled.  This was also applicable to  
the \KPITWO~decay mode when the photons escaped detection. 
Because of  phase 
space limits, the $\pi\nu\bar{\nu}(1)$ analysis  
only needed to consider the $K^+$-decay-related  
backgrounds from the $K_{\mu2}$, $\mu^+$ band and $K_{\pi2}$ decay. 
 
\subsubsection{Origins of Beam Background} 
 
The beam-related backgrounds were 
categorized into three types: single-beam background,  
double-beam background and CEX background.  
The first two beam backgrounds accounted for most of  
the $\pi^+$ band shown in  
Fig.~\ref{fig:r_vs_p_kp21}.  
 
\begin{figure} 
\centering 
\epsfxsize 0.9\linewidth  
\epsffile{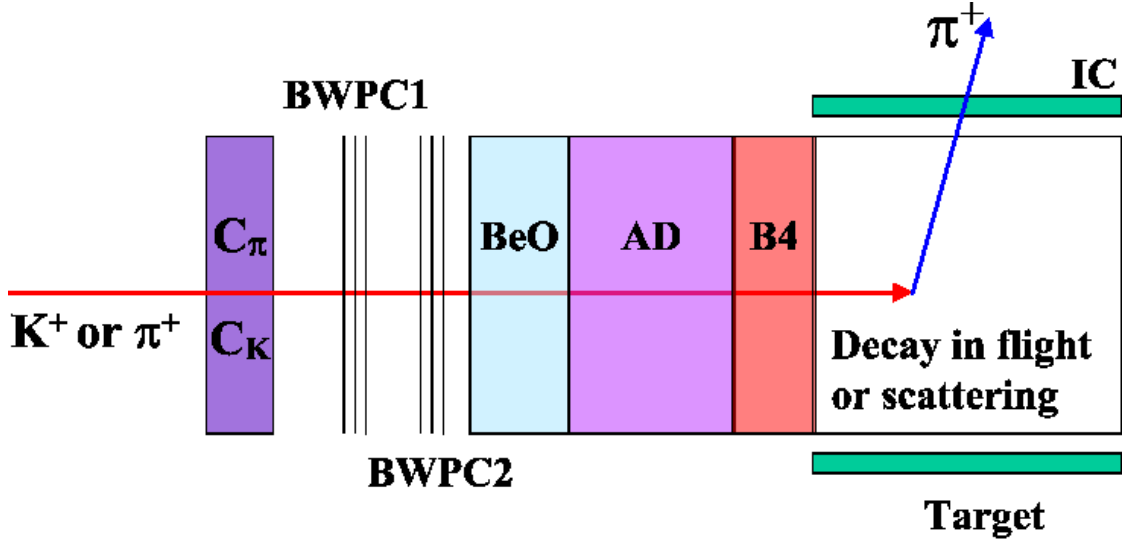} 
\caption{Schematic diagrams of the single beam background: 1) single beam  
$K^+$ background and 2) single beam $\pi^+$ background.  
The various detector elements and acronyms are  
described in the text. } 
\label{fig:single_beam} 
\end{figure}

The single-beam background consisted of the following components. 
(1) A $K^+$ entered the target and decayed in  
flight to a $\pi^+$ plus a $\pi^0$ as illustrated in   
Fig.~\ref{fig:single_beam}. The kinematic values  
of the $\pi^+$ were shifted upward to the 
signal region by the Lorentz boost, faking a \KPPNN~signal event. 
(2) A $\pi^+$ in the beam was misidentified as a $K^+$,   
scattered in the target and  entered the  
fiducial region of the detector as illustrated in  
Fig.~\ref{fig:single_beam}.  
This $\pi^+$ (referred to as scattered $\pi^+$)  
could mimic the target fiber pattern for signal 
and have kinematic values in the signal region.  
Rejection of these two background types required both good $\pi^+/K^+$ beam  
identification  and time resolution for delayed  
coincidence measurements.  
\begin{figure} 
\centering 
\epsfxsize 0.9\linewidth 
\epsffile{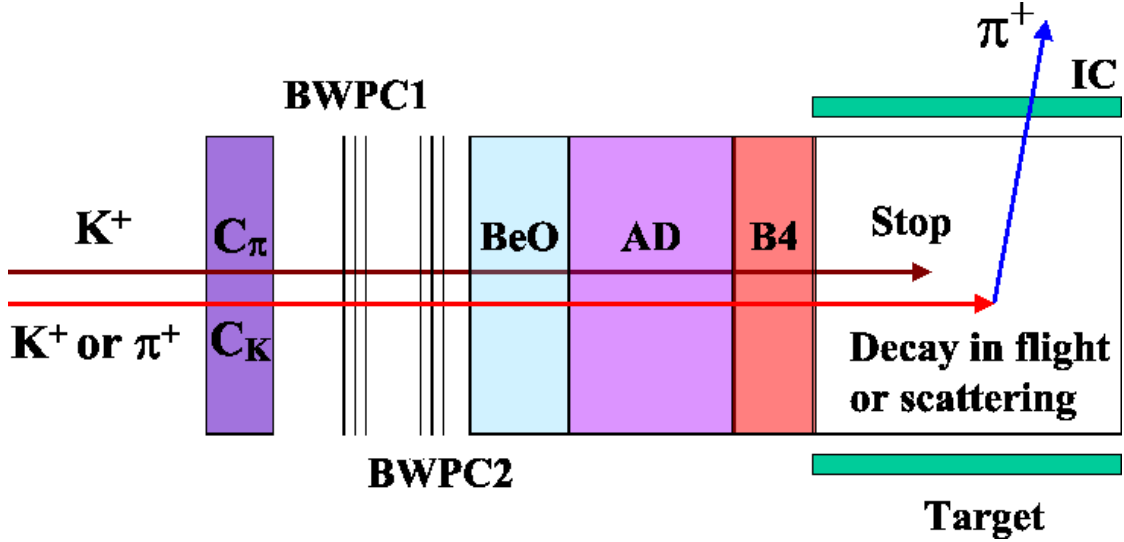} 
\caption{Schematic diagrams of the double beam background:  
1) double beam $K^+-K^+$ background and 2) double beam $K^+-\pi^+$ background.  
The various detector elements and acronyms are  
described in the text.} 
\label{fig:double_beam} 
\end{figure} 
\begin{figure} 
\centering 
\epsfxsize 0.9\linewidth 
\epsffile{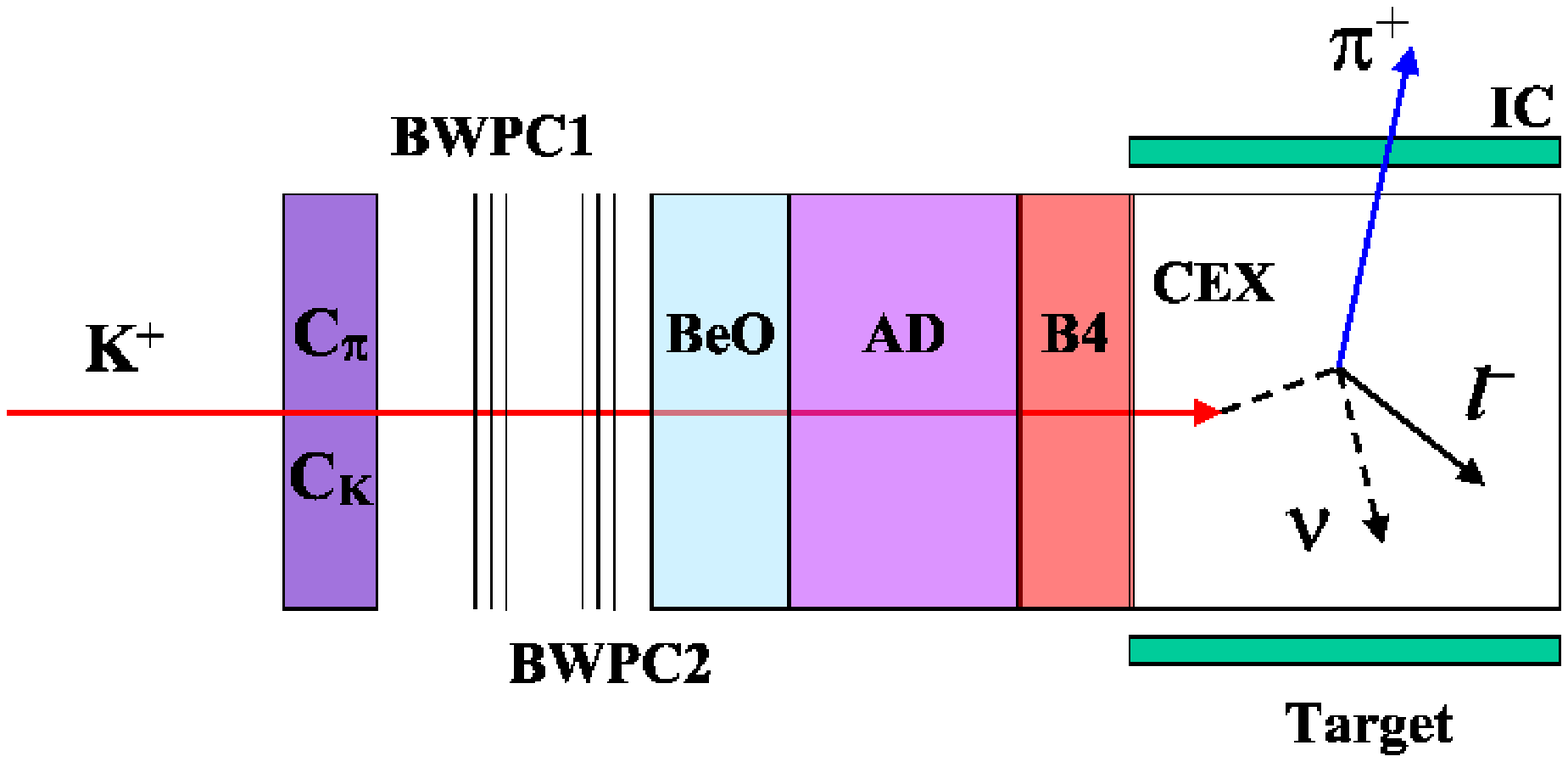} 
\caption{Schematic diagram of the charge exchange interaction background. 
The various detector elements and acronyms are  
described in detail in the text.} 
\label{fig:charge_exchange} 
\end{figure} 
 
The following cases were classified as the double-beam background.  
A $K^+$ came to rest in the target accompanied by  
another $K^+$ entering the target that decayed in flight  
to a $\pi^+$, which traversed  the fiducial  
region of the detector (Fig.~\ref{fig:double_beam}).  
The second case was similar,  
except that the beam $\pi^+$ scattered in the target and entered  
the fiducial region (Fig.~\ref{fig:double_beam}). 
Both cases could imitate a \KPPNN~signal if the decay products from  
the first $K^+$ were missed. Rejection of  these two backgrounds  
relied on the ability to observe extra activity 
the beam instrumentation, target and RS
coincident with the delayed decay.
 
The CEX background could occur if a $K^+$ produced a $K^0$
in the target and if the $K^0$ turned into
a $K^0_L$ that subsequently underwent semileptonic
decay. This process could produce background if the charged lepton from 
a $K^0_L$ decay was undetected  
and the $\pi^+$ satisfied the kinematics of the signal 
region (Fig.~\ref{fig:charge_exchange}).  
Rejection of the CEX background was achieved by using the fact that  
a $K^0_L$ usually did not deposit energy along the path in the  
target, leaving a gap observed between $K^+$ and $\pi^+$ fibers in the target. 
Also exploited was matching between the reconstructed $z$-position of the decay 
and the energy deposited by the incoming $K^+$ and the short flight time of $K^0_L$ in the  
target.

\subsection{Analysis Method and Strategy} 
\label{sec:analysis_method}

Disentangling the $K^+ \to \pi^+\nu\bar{\nu}$  decay  
from background in this experiment was challenging 
due to the poor kinematic signature and very small expected rate of  
the $K^+ \to \pi^+\nu\bar{\nu}$ signal. 
These necessitate enormous suppression of background events by rejecting
events with very low levels of extraneous activity. This high level of
veto makes it impractical to accurately simulate
the background rejection power of the detector at the required
$\sim\!10^{-10}$ level of sensitivity. Therefore, an accurate estimate of
the background must be obtained from the data.

\subsubsection{Blind Analysis Method}

A ``blind'' analysis method was developed to search for the $K^+ \to 
 \pi^+\nu\bar{\nu}$ signal in data samples.  
In this method, background sources were 
 identified {\it a priori}.  The signal region for the 
 $\pi\nu\bar{\nu}(1)$, determined so that the sensitivity was 
 optimized, was ``blinded'' or hidden until the background and 
 acceptance analysis was completed. When possible, selection criteria 
 were developed using the monitor samples to avoid examining the 
 signal region.  If monitor samples were inadequate and the 
 $\pi\nu\bar{\nu}(1)$ trigger sample were required, at least one 
 selection criterion distinguishing signal from backgrounds was 
 inverted (i.e. used to select a background region) to avoid examining 
 the signal region. In addition, the final background estimates 
 were obtained from 
 different samples than that used to determine the selection 
 criteria.  Each set of three consecutive $\pi\nu\bar{\nu}(1)$ events  
 were selected for ``1/3'' and ``2/3'' sample groups. 
 The 1/3 group was used to determine the selection criteria and an 
 unbiased background estimate was obtained from the 2/3 group.  The 
 signal region was examined only after the background analysis was 
 completed. 
 
\subsubsection{Bifurcation Method for Evaluating Background} 
\label{sec:bifurcated} 
 
\begin{figure} 
\centering 
\epsfxsize 16cm 
\epsffile{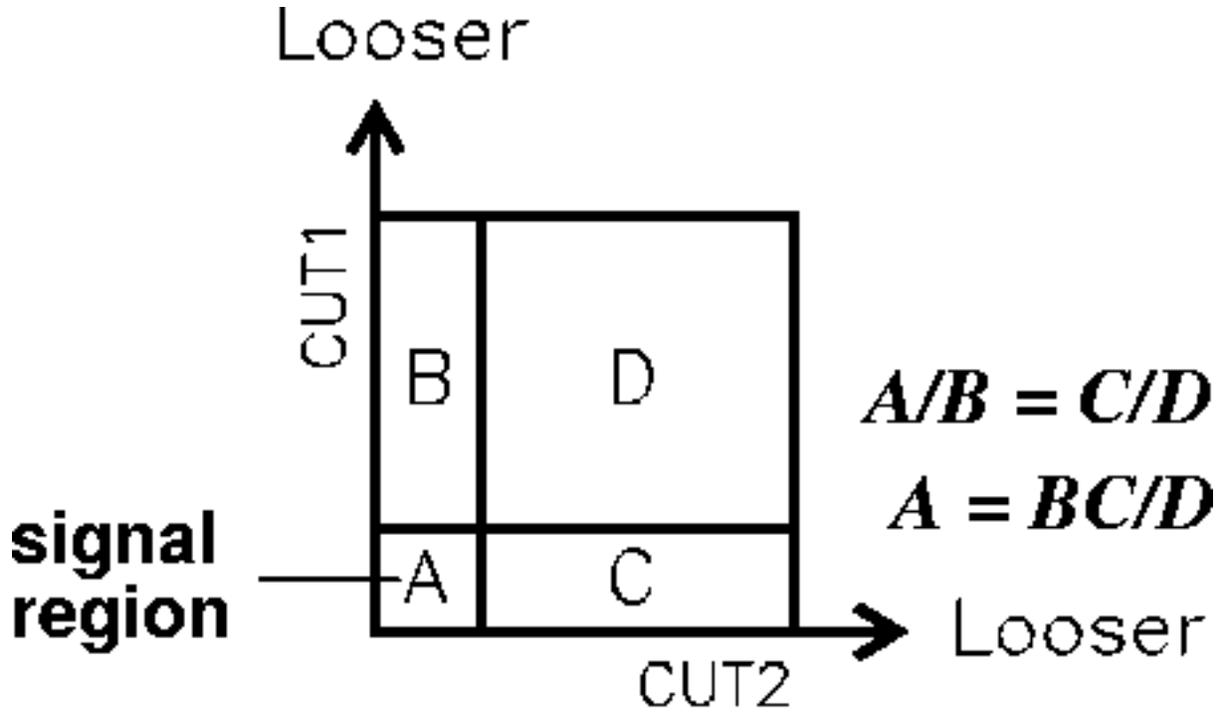} 
\caption{Pictorial explanation of the bifurcation method.  
Background level in Region ``A'' can be estimated from the observed  
number of events in the other regions, if CUT1 and CUT2 are uncorrelated.} 
\label{fig:explain_bifurcation} 
\end{figure}

The principal method for background evaluation relied on information from 
outside the signal region and involved the application  of two complementary but  
uncorrelated cuts.   
Fig.~\ref{fig:explain_bifurcation} illustrates this  bifurcation  method  
showing  the parameter space 
of two cuts,  ``CUT1" and ``CUT2".  
The number of background events in the signal region 
(i.e., region ``A") was $A$ events. If the two cuts are  
uncorrelated, that is, if the rejection of a cut does not depend  
on the rejection of the other cut, the ratio of the number  
of background events in region ``A"  
to region ``B" must be equal to the ratio in  
region ``C" to region ``D", i.e., $A/B=C/D$.  
Background events in the signal region are therefore obtained 
from the relation $A=BC/D$.  
 
In practice, the present bifurcation analysis was done through two branches. 
A ``normalization branch" analysis was used to obtain  
the number of events in ``B'' region. 
A ``rejection branch" study was done to  
get the ratio of $D/C$. The rejection was defined as  
$R=(C+D)/C$. The background level in the signal region was then estimated as 
$N_{Bkg} = B/(R-1)$.  
For the case of very small statistics, ``CUT1'' was subdivided  
into two cut categories, and $B$ was estimated in the  
same way as the ``first'' bifurcation. 
 
To check if the two bifurcated cuts were  uncorrelated,  
cuts were loosened simultaneously (as described in Section IV A).
The loosening factors  
were controlled by the predicted background functions, in which  
the loosening factors were the inputs and the outputs were the cut  
positions. By design the functions provided the relative background level 
and the acceptance.  
The background level provided by the functions should agree with the 
observed number of events within 
the newly defined 
regions if CUT1 and CUT2 were  uncorrelated.  
This method  thus provided   
input to  the evaluation of systematic uncertainties (Section~\ref{sec:syst}). 
 
\subsubsection{Analysis Strategy} 
 
The data analysis used the following key steps to determine 
the selection criteria, evaluate the background level, 
investigate the systematic uncertainty,  
measure the acceptance and finally obtain the branching ratio. 
\begin{itemize} 
\item Data were first reconstructed and processed with a number of   
selection criteria to remove  obviously bad events.  
Then the surviving events were divided into 1/3 and 2/3  
portions, in which three sub-samples were also skimmed out  
according to different background features  
(Section~\ref{sec:data_pass}).  
 
\item Blind analysis was adopted in designing, calibrating and tuning 
the selection criteria. Signal-like and background samples were taken  from 
the monitor trigger samples when  applicable. 
When the three skimmed sub-samples were used to have 
the same experimental features of the signal,  
at least one critical selection criterion  
was inverted to avoid examining  the signal  
region (Section~\ref{sec:cuts}).  
 
\item The background level was initially evaluated by applying the 
bifurcation method to the 1/3 data and 
controlled to be much less than one event by tuning the selection 
criteria. At least two uncorrelated cuts  
with large background rejections were chosen.
data samples (Section~\ref{sec:bkg_kp2}-\ref{sec:cex}).  
 
\item The final cut positions were  optimized with respect to  
the signal to background level estimated from the 1/3 data.
This was achieved using the predicted background functions 
estimated in the bifurcation analysis 
(Section~\ref{sec:total_bkg}-\ref{sec:extended}).  
 
\item Correlations between the bifurcated cuts were checked by  
conducting a series of re-evaluations of  the background levels  
outside the signal region. A study of single cut failure was also 
conducted to investigate  a possible flaw in the technique. 
To avoid  the potential bias of the 1/3 background study,  the 
final background evaluations were taken from the 2/3  
portion (Section~\ref{sec:outside_the_box}-\ref{sec:bkg_level}).  
 
\item Acceptances were  measured with the monitor  
trigger samples except for the signal 
phase space, the trigger efficiency and those that could not 
be extracted from the monitor trigger data. These  
were obtained from Monte Carlo. The branching ratio of   
$K_{\pi2}$ was used to validate the acceptance measurement. 
Single event sensitivity was obtained    
from the total $K^+$ exposure and  
the acceptance. (Section~\ref{chap:acc_and_sen}).  
 
\item The signal region was examined. Events observed  
in this region were all considered as the signal candidates 
(Section~\ref{sec:candidate}). 
 
\item The branching ratio was  obtained from a likelihood  
analysis incorporating  the predicted background 
rate and acceptance within the  
signal region (Section~\ref{chap:results}).  
 
\end{itemize} 
 
\subsection{Track Reconstruction} 
 
Throughout this analysis, the events were reconstructed under the  
assumption that they were $K^+\to\pi^+\nu\bar{\nu}$ events with  
only a single $\pi^+$ track in the detector.  
 
\subsubsection{Beam Time Measurements} 
 
To fully reconstruct an event, the initial time of the beam  
particle was required. 
The beam instruments provided several beam times 
from  TDC and CCD measurements on 
the $C_K$, the $C_\pi$, the BWPC and the B4.  
For the TDC measurements, offline analysis treated  
all PMT hits coincident with each other as a cluster.   
The average time of the TDC hits in each cluster gave $t_{C_K}$, 
$t_{C_\pi}$, $t_{BW}$ and $t_{BM}$, respectively. 
The CCD  measurements were used to discriminate  
 cases with more than one particle in a beam (referred to as  
pileup).

\subsubsection{Clustering in the RS} 
 
The track reconstruction routine started by finding clusters in the RS. 
The hit counters of a positively-charged track  
(the track counters)  
were searched for using the TDC timing information in the RS.  
A good $T\cdot2$ sector was found as that closest in time to the 
$DS$. From this $T\cdot2$ sector, adjacent counters in subsequent layers 
within 10~ns of the $T\cdot2$ time and with energy  
greater than 0.5~MeV were selected in the outgoing and clockwise  
direction to find the track counters. 
A good cluster involved at least six consecutive 
layers from inside to outside.  
Once an RS track cluster was identified,  
the track time ($t_{rs}$) was  computed by averaging all the  
time measurements of the track  
counters. The stopping counter was defined as the one in  
the outermost layer  which was   in the most  clockwise direction. 
The $T\cdot2$ sector in a cluster served as a guide 
for tracking in UTC. 
 
\subsubsection{Tracking in the UTC} 
 
When an RS cluster was established, the UTC tracking routine  
started searching for  
clusters of hit anode wires in the $x-y$ plane~\cite{utc}.  
Hit wires in each super-layer were grouped into clusters based  
on their spatial proximity to one another. A straight line  
fit provided a crude vector in each super-layer.  
The solutions due to the left-right ambiguity were included at this stage. 
These vectors were then 
linked to form a track segment in the UTC.  
A circle fit was performed with a  
set of drift distances with left-right ambiguity resolution.    
The radius of the circle gave the measurement of transverse momentum.
 
If a UTC track in the $x-y$ plane was found, the corresponding  
track projection on the $\phi-z$ plane ($\phi$ was
defined as the revolution angle with respect to the closet 
approach to the vertex)
was then sought in the    
clusters of hit strips on the UTC cathode foils. A time window of     
$\pm15$ ns was used to reduce the accidental hits. 
The calculation of the $z$ position 
for a cluster adopted the ratio method suggested  
by Ref.~\cite{khovansky}, which used the three strips with the highest  
ADC counts to derive the centroid and to reduce the bias  
on the $z$ position measurement. 
A straight line fit was performed in $\phi-z$ plane if the $z$ hits were 
found in at least 3 foils, thus determining the slope and intercept.  
The slope was then used to 
convert the measured transverse momentum to the total momentum. 
 
In case of more than one track pointing to the same $T\cdot2$ sector,   
a good UTC track was defined as the one closest to the first  
RS sector crossing point 
or to the clock-wise edge of the stopping counter otherwise. 
The UTC tracking efficiency was checked with the $K\mu2$ monitor data 
and was measured to be better than 95\%.  
The inefficiency was due to the fact that for the  high rate environment  
accidental hits in the UTC caused  problems in  pattern recognition.  
 
\subsubsection{Tracking in the Target and B4} 
 
Traveling almost parallel to the target fibers,   
the incident $K^+$ deposited significant energy 
in each fiber  (usually $>$ 4~MeV)  
and was in coincidence with the $BS$. 
The daughter $\pi^+$ traveled   
nearly perpendicular to the target fibers, and thus left less energy 
in each fiber (1~MeV on average) and was in coincidence with the  
$t_{rs}$. The $BS$, $t_{rs}$ and fiber times, positions and energies were  
the key elements to identify the $K^+$ and $\pi^+$ fibers.  
All the $K^+$ and $\pi^+$ fibers were linked to form a $K^+$ 
cluster and a $\pi^+$ cluster. A good event should only have  
one $K^+$ cluster and one $\pi^+$ cluster.  
 
After a UTC track was found, the target pattern recognition routine  
started to look for the fibers belonging to the $K^+$ path and  
the $\pi^+$ path separately on a 1~cm  
wide strip along the UTC track extrapolated into the target.  
The corresponding energies  
for $K^+$ and $\pi^+$ ($E_K$ and $E_{tg}$)  
and times ($t_K$ and $t_\pi$) were calculated from the sum and  
average in the clusters, respectively. 
The range of $\pi^+$ in the target ($R_{tg}$) was  
calculated as the helix traversed by the $\pi^+$ from  
the $K^+$ decay position to the inner surface of the IC with 
the polar angle $\theta$ correction. 
 
Because of  ambiguity in the entry and stopping ends  
in the pattern of the incoming beam,  
the target reconstruction routine used the  
B4 measurement on the $K^+$ entrance point, which  
was reconstructed by clustering 
the hits in the two B4 layers to determine the beam time $t_{BM}$,  
the energy-weighted beam position and the energy loss in B4. 
The cluster with $t_{BM}$ closest to the $t_K$ was chosen as  
the one caused by the $K^+$ beam. 
A 0.36~cm position  
precision was obtained by the B4 hodoscope in the $x-y$ plane. 
With the B4 position measurement, the target reconstruction was 
repeated to give a better pattern recognition. 
The $K^+$ decay vertex in the $x-y$ plane was determined by  
the $K^+$ fiber closest to the UTC track but furthest from  
the $K^+$ entrance point determined by the B4, while  
the $z$ position was calculated from the UTC track extrapolation  
in the $\phi-z$ plane. 
 
Target CCD information improved the pattern recognition and  
$\pi^+$ energy measurement when pileup occurred in a $K^+$ fiber. 
When $t_K$ and $t_\pi$ were separated by more than 2 ns,  
the CCD pulses in all of the $K^+$ fibers were studied to identify  
 any hidden $\pi^+$ energy.  
 
Isolated hit fibers outside the 1~cm search strip and within  
the time window of $t_{rs}\pm5$ ns  
caused by possible photon(s) or  
other beam particles  were classified into ``photon'' fibers if 
their energies were greater than 0.1~MeV. 
 
Once the first round of target reconstruction was finished,   
the $\pi^+$ passage and $K^+$ decay vertex were determined in the target.  
The procedure of UTC track fitting was repeated with this  
additional  information.  
This aided the resolution of the left-right ambiguity in the UTC track 
reconstruction. 
Iteration of the target reconstruction  
was also performed with the improved UTC track.  
 
\subsubsection{Track Passage in the IC} 
 
An allowed IC hit pattern was either one or two adjacent sectors per event. 
The IC provided energy loss ($E_{IC}$) and time ($t_{IC}$) measurements. 
If there was an IC sector crossing, the $E_{IC}$ was from the sum of two  
IC sectors and the $t_{IC}$ was from the energy-weighted average. 
It was observed that 
1\% of tracks had extra hits in the IC's, confounding the measurements 
of energy and time in IC's. This was resolved by using  TD information in 
addition to the TDC and ADC information. 
The $t_{IC}$ was always the one closest to the $t_{rs}$ and, 
the corresponding energy was taken as $E_{IC}$.  
The range $R_{IC}$ was computed as the length of the extrapolated 
UTC track from the inner to the outer IC radius. 
 
\subsubsection{Tracking in the RS and RSSC} 
 
With a charged track reconstructed in UTC and target,  
the tracking in RS started from the previously found RS cluster. 
The stopping counter was first analyzed by fitting the 
TD information with a double pulse assumption to find a $\pi^+\to\mu^+$ decay  
signature. This also  
determined the $\mu^+$ energy ($E_\mu$)  
from the $\pi^+$ decay at rest. 
 
\begin{figure} 
\centering 
\epsffile{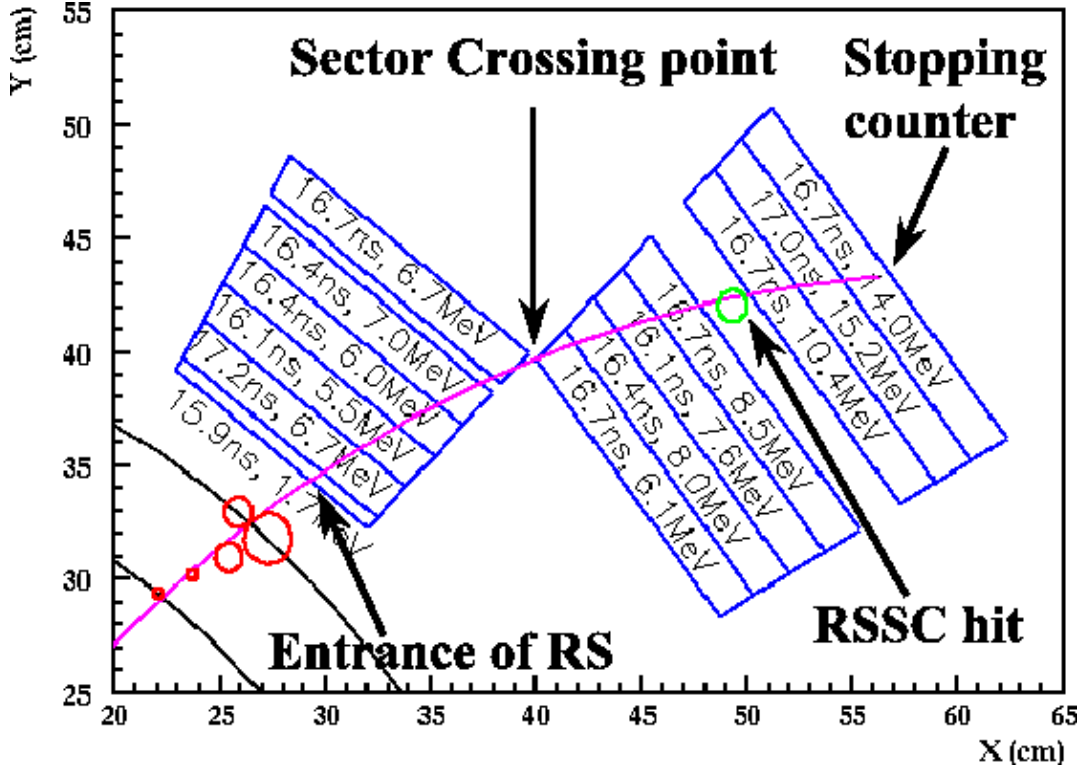} 
\caption{Illustration of the Range Stack track fitting in $x-y$ 
plane. All the relevant definitions are given in this plot.  
The arc represents the extrapolation of the fitted UTC track.} 
\label{fig:trkrng} 
\end{figure}

A sector crossing point as illustrated in  
Fig.~\ref{fig:trkrng} was searched for in the RS. 
A \KPPNN~candidate should not have  more  
than 2 sector crossings. Precise  
position measurement in $x-y$ plane was obtained  
from the sector crossing points. 
The $z$ positions were determined by using  
the end-to-end time differences in the hit counters except the  
T-counter. The average $z$ position resolution was observed to be  
about 4 to 5~cm.  
 
Another precise position measurement was provided by the RSSC's.  
All the adjacent hit straw chambers in the RSSC of the hit sector  
were grouped to form a cluster.  
The candidate cluster in each sub-layer of RSSC  
was defined as the one  
closest to a series of arcs drawn from the UTC track  
extrapolation to the T counter through the sector crossing point(s). 
The $x-y$ position was obtained from the average of two sub-layers  
in the same RSSC.  
To minimize the effect of cross-talk,  
the earliest hit in one sub-layer was chosen as the true hit.       
The $z$ measurement was given by the time difference, and the  
precision was about 1.5~cm.  
 
An RS track fit in the $x-y$ plane used the entrance point  
provided by the UTC track extrapolation, the sector crossing point(s),  
the RSSC hit position(s) and the expected path length predicted  
by the energy losses in the RS layers, taking into account    
$\pi^+$ track propagation with energy loss given by  
the Bethe-Bloch equation in  a 1 Tesla magnetic field.  
The $\chi^2$ of the fitted track was minimized by changing the incident  
momentum and the angle at the entrance to the RS  
(Fig.~\ref{fig:trkrng}).  
 
The total energy loss of the track in the RS ($E_{rs}$)  
was obtained by summing up all the energy losses in the track counters,  
with $E_\mu$ from the fit to pulse shape subtracted from the energy  
in the stopping counter.  
The range in the RS ($R_{rs}$)  
was calculated from the path length of the  
fitted track with the polar angle correction.  
The range in the stopping counter was estimated from the $\pi^+$ energy  
loss.

\subsubsection{Kinematic Measurements of a Track} 
 
The total range $R$ and energy $E$ of the track were calculated as 
$R = R_{tg} + R_{IC} + R_{rs}$ and $E = E_{tg} + E_{IC} + E_{rs}$,  
respectively. 
The total momentum $P$ of the charged track was obtained from the UTC with  
corrections due to energy loss in the target and IC.  
Since the momentum reduction  
in both the target and IC were calculated from the range,  
some correlation between  
$P$ and $R$ was expected, and thus they should not be treated as uncorrelated  
in the bifurcation analysis. 
All of these three kinematic 
measurements also included tiny contributions from 
the inactive  materials in  the UTC.  
In this analysis, the momentum, energy and range resolutions  
were measured to be 1.1\%, 2.8\% and 2.9\% (RMS), respectively,  
from a study of \KPITWO~events (Table~\ref{tab:res_pre}).  
 
\subsection{Monte Carlo Simulation} 
\label{sec:umc} 
 
Detector responses 
were modeled by a Monte Carlo simulation package,  
which was developed for the E787 experiment and maintained or improved  
for E949. The package used several subroutines from the  
electromagnetic-shower simulation package EGS4~\cite{umc3} 
and a number of routines written specially for the experiment, 
including all of the detector 
elements, except for the beam instrumentation upstream of  
the target. Simulation samples were generated with the same  
format as the data except for omission of the  
pulse-shape information and most of the beam counter information. 
 
\subsubsection{Simulation of $K^+$ Propagation} 
 
The simulation of $K^+$ propagation started from a beam file,  
which contained $K^+$ events with 
a list of measured parameters: the $K^+$ stopping position,  
the $t_K$, the number of $K^+$ fiber hits, the number of  
accidental fiber hits, the B4 hit position, the stopping target fiber  
element, the time, energy, fiber element for each $K^+$ and accidental  
hit in the target. This file was obtained from an analysis of  
the stopping distribution of the $K\mu2$ monitor data.  
With this beam file and the corrections described in Section~\ref{sec:mc_trigger}, 
the $K^+$ propagation had exactly the same target  
patten of the data in the simulation. Every $K^+$ decay  
started from the stopping fiber at the given $K^+$ stopping position.  
The $K\mu2$ monitor data had sufficient statistics for all the  
simulation studies in this analysis.  
 
\subsubsection{Simulation of $K^+$ Decay Product}  
 
The $K^+\to\pi^+\nu\bar{\nu}$ decay was generated  
with the matrix element of semileptonic $K^+_{\ell 3}$ decay  
via the V-A interaction, while the $K_{\mu2}$ decay and the $K_{\pi2}$ 
decay were generated via pure phase space. 
Among the $K^+$ decay products,  
photon and electron interactions and their energy deposits 
were calculated using the routines from EGS4. 
For charged particles, 
the energy deposits were calculated by adding the energy losses of  
each ionization along the steps taken by the  
particles. The number of ionization and excitation events was determined  
by dividing the total average energy deposited along the step, 
obtained using the Bethe-Bloch formula, by the minimum energy 
that a particle lost in a collision. 
Multiple Coulomb scatterings of charged particles with various nuclei in the 
detector were calculated according to the theory of Moliere~\cite{umc},  
with corrections for the spin of the scattered particle and  
the form factor of the nucleus~\cite{umc1}. 
Hadronic interactions of positively charged $\pi^+$'s in the plastic  
scintillators were calculated using a combination of data and phenomenological 
models~\cite{umc2}. An option in the simulation package 
allowed users to turn off nuclear absorption reactions 
and decays in flight of $\pi^+$'s. This was useful in the study of the  
acceptance.

\subsubsection{Simulation of Trigger} 
\label{sec:mc_trigger} 
 
All the trigger conditions were simulated except for the  
$DC$, $L1.1$, 
$L1.2$ and those related to the beam instruments.   
Since the beam file generated from the $K\mu2$ monitor data  
included the trigger bias, a correction was needed to  
recover the true beam distribution. 
This was done by introducing a weight function 
derived from a comparison between simulated $K\mu2$ triggers 
and data. In order to get a better precision,  
perpendicular tracks were used to calculate 
the $K\mu2$ trigger acceptance as a function of radial distance  
of the stopping position in the $x-y$ plane.  
 
\subsubsection{Comparison between Data and Simulation} 
 
The performance of the Monte Carlo simulation was 
checked by comparing  
kinematic resolutions between data and simulation as given in 
Table~\ref{tab:res_pre} for the $K_{\pi2}$ decay mode. 
A 0.15~cm deviation was observed in the range resolution, which 
is still not understood. This difference could have affected the acceptance  
estimate and was investigated when performing the acceptance study
(see Section~\ref{sec:acc_mc}). 
It should also be emphasized that 
the main role of Monte Carlo simulation in the E949 experiment 
was to estimate the acceptance factors that could not be obtained from  
real data (e.g. geometrical and relevant trigger acceptances). 
\begin{table} 
\begin{center} 
\begin{tabular}{lccc}\hline\hline 
Experiment& $\sigma_P$ (MeV/$c$) & $\sigma_R$ (cm) & $\sigma_E$ (MeV)\\ \hline 
Data      & 2.299$\pm$ 0.006 & 0.866$\pm$0.002& 2.976$\pm$0.005 \\ 
MC       & 2.399$\pm$ 0.029 & 1.018$\pm$0.008& 3.018$\pm$0.025 \\ 
\hline\hline 
\end{tabular} 
\caption{Comparison of the momentum, range and energy resolutions 
between data and simulation for the $K_{\pi2}$ peak.} 
\label{tab:res_pre} 
\end{center} 
\end{table}

\subsection{Data Processing and Pre-selection} 
\label{sec:data_pass} 
 
Data were stored on the DLT tapes with a total data size  
of  7 Tera bytes. Two steps of data processing (``Pass 1'' and 
``Pass 2'') were taken to reduce and skim 
the data to a reasonable volume.

\subsubsection{Pass 1} 
 
Pass 1  involved  filtering cuts, 
which consisted of  
event reconstruction quality cuts and loose $\mu^+$ background  
rejection cuts. Runs with trigger or hardware problems  
that could not be corrected offline were removed from the data 
analysis. Tracks were required to be successfully reconstructed, 
not to stop in those detector elements which  
had known  hardware problems and to have momentum  $\leq 280$~MeV/$c$. 
 
A $\pi^+\to\mu^+$ double pulse was required to be found by the 
fit to the TD pulse in the stopping counter and no extra hits were found 
in the other 3 sectors associated with the stopping counter TD channel.
Since  photon activity around the stopping counter could 
cause  confusion in the energy measurement, events with sector  
crossing in the stopping layer were rejected. Also rejected were 
the events with a charged track that came to rest in the support materials  
for the second RSSC layer embedded between the 
$14^{\rm th}$ and $15^{\rm th}$  
RS layers.  Pass 1 reduced the data volume by   
a factor of two.  
 
\subsubsection{Pass 2} 
 
Pass 2 involved cuts 
that were applied to  the sample of events surviving Pass 1 
to skim the data into three categories according to the event features.  
Each category was uniformly divided into a 1/3 portion  
and 2/3 portion.  
Pass 2 consisted of five data skimming criteria: quality of the 
target reconstruction, loose photon veto,  
quality of the $\pi^+\rightarrow\mu^+$ double pulse fitting result,  
single beam $K^+$ requirement and delayed coincidence cut. These are  
as discussed  below and listed in Table~\ref{tab:pass2cut}. 
\begin{itemize} 
\item The target reconstruction required that the $K^+$ decay vertex  
       be inside the 
      target volume,  
      $|t_K-t_{BM}|<4$ ns,  
      $|t_\pi-t_{IC}|<5$ ns and  
      $E_{IC}$ be consistent with  
      that expected from the calculated $\pi^+$  
      range.  
\item The loose photon veto rejected events for    
      $|t-t_{rs}|<2$ ns with an energy in the BV greater than 1.5~MeV,  
      $|t-t_{rs}|<1.5$ ns with an energy in the EC greater than 3.5~MeV, 
      $|t-t_{rs}|<1.5$ ns with $E_{RS}>3.0$~MeV, 
      or $|t-t_{rs}|<1$ ns with $E_{tg}>5.0$~MeV, where  
      $t$ was the time measurement in each photon veto counter. 
\item The $\pi^+\rightarrow\mu^+$ sequence required a  
      $\mu^+$  decay pulse in the stopping counter and the absence of  
      hits within $\pm2.5$ ns of the $\mu^+$ time  
      around the stopping counter. 
\item The beam requirements for signal events were such that the
      energy loss in the B4 was greater than 1~MeV, the $t_{BM}$ 
      differed from $t_{rs}$ by more than 1.5 ns, and 
      the number of $C_\pi$ hits was  
      less than 4 with $|t_{C_\pi}-t_{rs}|<1.5$ ns. 
\item The delayed coincidence required $t_\pi-t_K>1$ ns.   
\end{itemize} 
 
As defined in Table~\ref{tab:pass2cut}, Skim 1(4), 2(5) and 3(6) were enhanced 
in \KPITWO, $\mu^+$ and beam backgrounds, respectively,  
for the 2/3(1/3) portions.  
These six skims of Pass 2 output data facilitated the
studies to develop the final selection criteria and
evaluate the backgrounds. Signal candidates, if any,  
remained in the six streams. Study of selection criteria using  
these samples was always done with at least one cut inverted to  
ensure that a blind analysis was  conducted.  
\begin{table} 
\centering 
\begin{tabular}{l c c c}\hline\hline 
Cuts                       & Skim 1 \& 4 & Skim 2 \& 5 & Skim 3 \& 6  
\\ \hline 
Target reconstruction      & $\sqrt{}$ & $\sqrt{}$ & $\sqrt{}$ \\  
Photon veto                & & $\sqrt{}$ & $\sqrt{}$ \\  
$\pi^+\to\mu^+$ sequence   & $\sqrt{}$ & & $\sqrt{}$ \\  
Beam $K^+$ requirement     & $\sqrt{}$ & $\sqrt{}$ & \\  
Delayed coincidence        & & $\sqrt{}$ & \\ \hline\hline 
\end{tabular} 
\caption{The cuts used in Pass 2 for selecting the data streams.  
Skim 1(4) , 2(5) and 3(6) was used for studying the  
$K_{\pi2}$ , $\mu^+$  and  
beam background, respectively.  
See text for more details.} 
\label{tab:pass2cut} 
\end{table} 
\subsection{Selection Criteria of post Pass 1 and Pass 2} 
\label{sec:cuts} 
 
Further selection criteria were designed and applied in order to  
gain more background reduction. According to their characteristics, 
the selection criteria were classified into four categories: 
single beam $K^+$ selection criteria, kinematic reconstruction, 
$\pi^+$ identification and photon veto. All the cuts were  
selected  to optimize the background rejection and the signal acceptance.  
Data samples used for this study came from either the 1/3 data  
(from Skim 4 to Skim 6) or 
the monitor trigger events defined in Section~\ref{sec:moni},  
depending on the nature of the selection criteria that were studied.

\subsubsection{Single Beam $K^+$ Requirements} 
\label{sec:beam_cuts} 
 
The beam cuts were used  to identify beam particles scattering  
either in the beam instruments or in the target and to 
ensure a single beam $K^+$ particle,  
making full use of time measurements from various sub-detectors,  
energy loss measurements and pattern 
recognition information in both the B4 and the target as discussed  below.

\begin{figure} 
\centering 
\epsfxsize 0.9\linewidth 
\epsffile{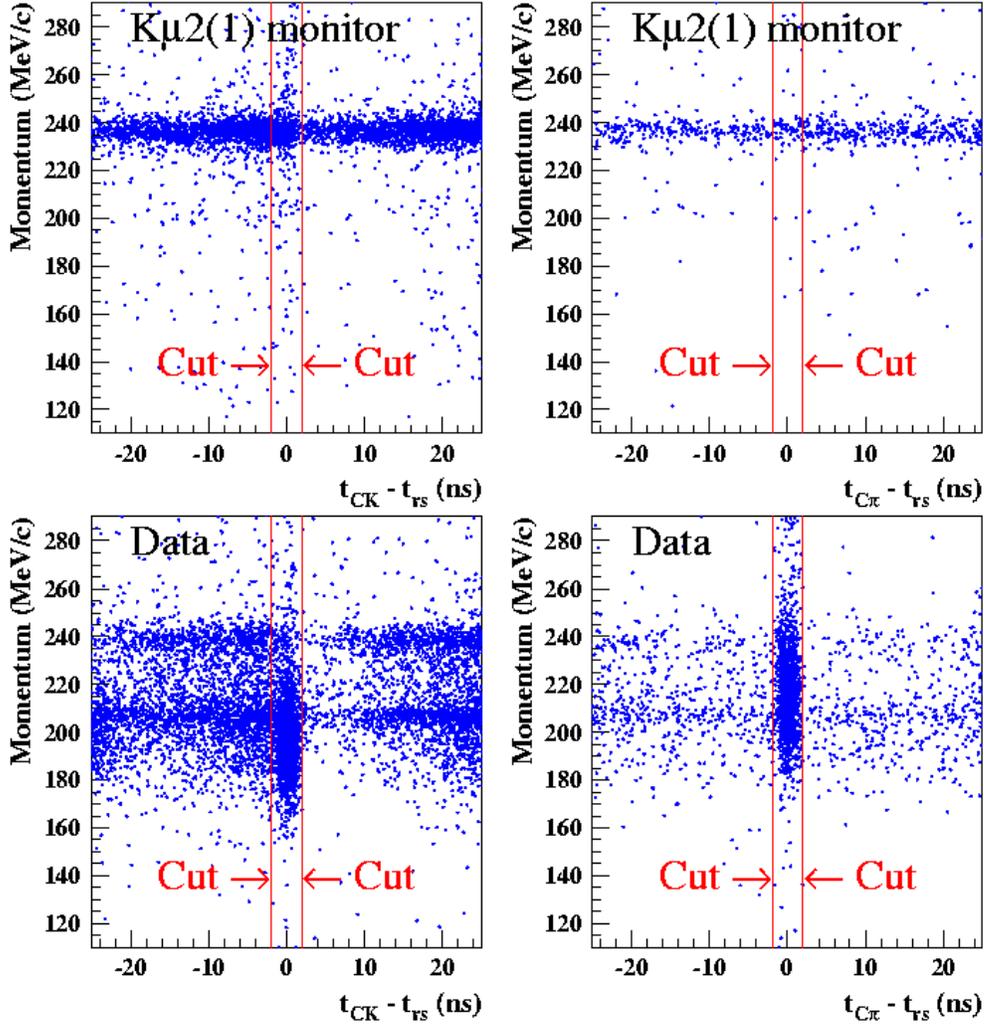} 
\caption{The momentum of charged particles versus $t_{C_K}-t_{rs}$ and  
$t_{C_\pi}-t_{rs}$. 
The $K\mu2$ monitor was used to  represent the single beam $K^+$ events.  
The data plots indicate beam background contamination  
at  beam time ($t_{C_K}$ or $t_{C_\pi}$) close to $t_{rs}$ 
in the $\pi\nu\bar{\nu}(1)$ trigger sample.  
The arrows indicate the rejected timing regions.  
The statistics in these plots account 
for about 0.3\% of $N_K$.} 
\label{ck_cpi_trs} 
\end{figure}

{\it Beam~Times:} 
The $t_{C_K}$, $t_{C_\pi}$, 
$t_{BW}$ and $t_{BM}$ cuts were 
used to reject extra beam particles scattering in the target when one of them 
agreed with $t_{rs}$ within $\pm2$ ns. 
Since the incoming $K^+$ beam intensity was high discriminator dead time 
was important and the time measurements used the CCD information in 
addition to the TDC information  from the $C_K$, $C_\pi$ and B4 
hodoscope. Fig.~\ref{ck_cpi_trs} shows the single beam $K^+$  
signal indicated by the $K\mu2$ monitor and the beam background 
other than the $K_{\mu2}$ or $K_{\pi2}$ peak events rejected 
in the $\pi\nu\bar{\nu}(1)$ momentum distribution.  

\begin{figure} 
\centering  
\epsfxsize 0.8\linewidth 
\epsffile{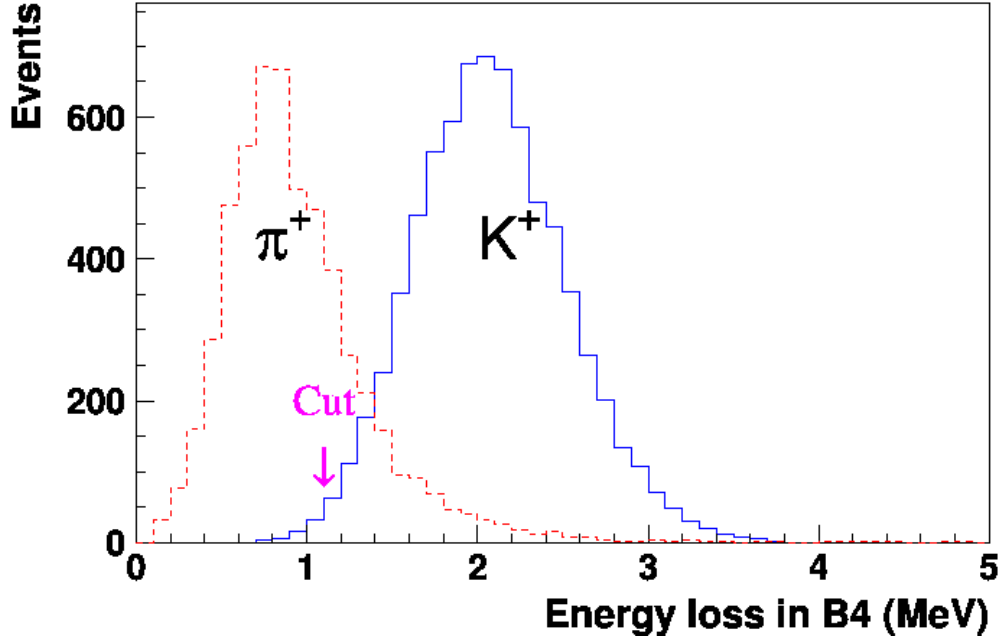}  
\caption{The energy loss in the B4 hodoscope for  beam $K^+$ (solid)  
and beam $\pi^+$ (dashed). 
The $K^+$'s were the $K_{\mu2}$ peak events in the $K\mu2$ monitor sample,  
while the $\pi^+$'s were the scattered $\pi^+$'s  
in the Skim 6 sample. 
The arrow indicates the cut position.} 
\label{fig:b4dedx}  
\end{figure}

{\it Energy~Loss~in~B4:} 
The tuning of the energy loss cut  
in the B4 hodoscope 
used the acceptance sample from the $K\mu2$ 
monitor sample and the rejection sample from the scattered $\pi^+$'s 
in the Skim 6 sample.  
This cut required that the energy loss for an incoming beam 
particle should be consistent with a $K^+$ 
($>1.1$~MeV) as shown in Fig.~\ref{fig:b4dedx}. 

\begin{figure} 
\centering  
\epsfxsize 0.8\linewidth 
\epsffile{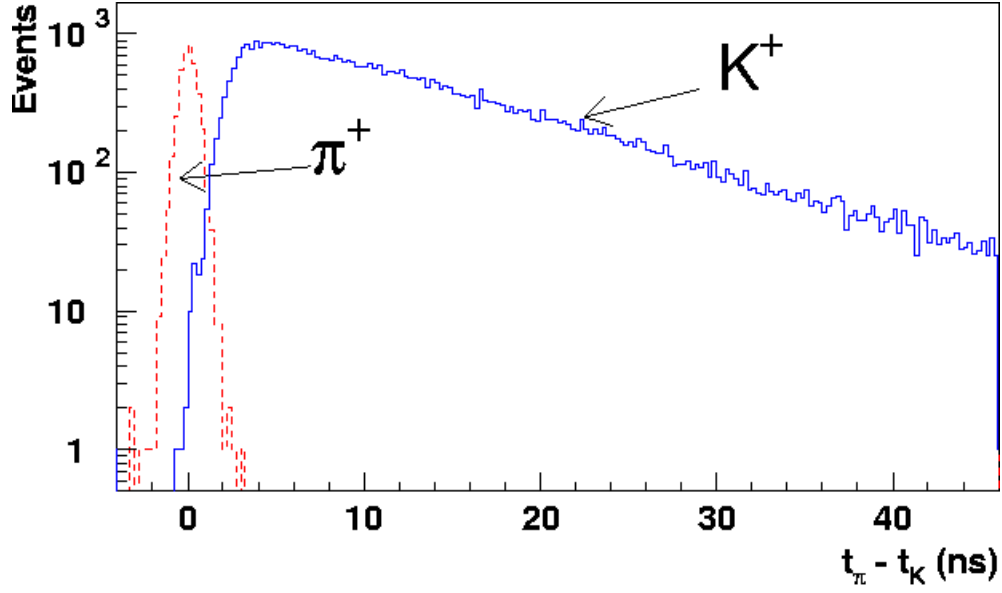} 
\caption{Distribution of the time difference, $t_{\pi} - t_K$ 
measured by the target,  
for the $K_{\mu2}$ peak events (solid) from the $K\mu2$ monitor sample 
and the scattered $\pi^+$'s (dashed) from the Skim 6 sample. 
Beam $\pi^+$'s can be assigned a time $t_K$ by the target fiber  
reconstruction due 
to mis-identification.}  
\label{fig:delco} 
\end{figure}

{\it Delayed~Coincidence:} 
The acceptance samples were taken from the  
$K_{\mu2}$ peak events in the $K\mu2$ monitor sample,  
while the rejection sample  
was from the scattered $\pi^+$'s in the Skim 6 sample.  
As shown in Fig.~\ref{fig:delco}, 
the distribution of $t_{\pi} - t_K$ for $K^+$ decays at rest  
was an exponential, as expected, consistent with  the  
known $K^+$ lifetime~\cite{Yao:2006px}. 
The distribution for the scattered $\pi^+$'s shows a prompt peak around 0 ns.  
The delayed coincidence cut required 
$t_\pi - t_K>2$~ns.  
When not all sub-detector time measurements were available, the time 
resolution was degraded; the delayed coincidence cut was adjusted 
up to 6 ns delay to take into account the resolution. 
It was expected that this delayed coincidence cut had the same rejection 
power to the background from $K^+$'s decaying in flight.  

\begin{figure} 
\centering 
\epsfxsize 0.8\linewidth 
\epsffile{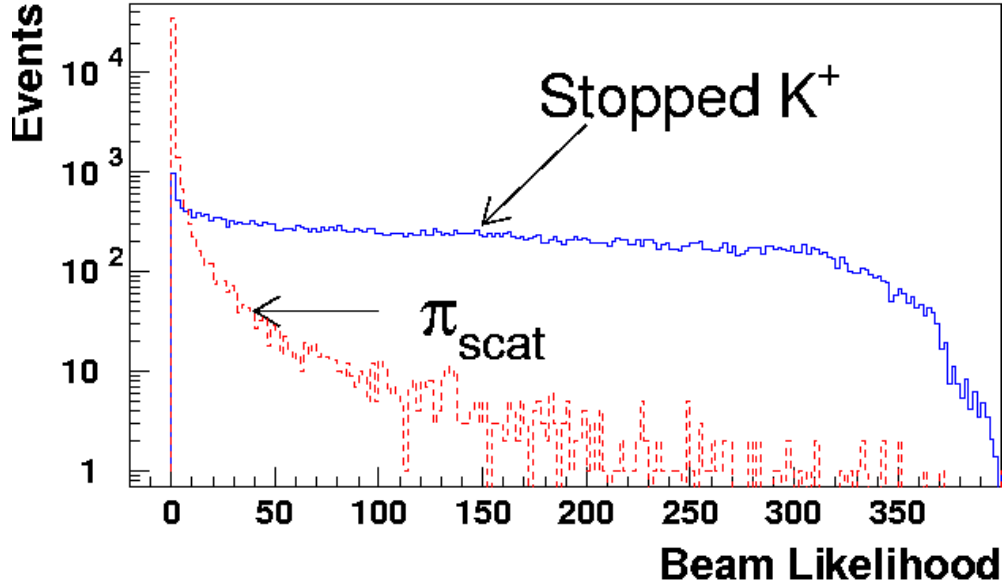} 
\caption{Beam likelihood for $K^+$ decays at rest (solid)  
from the $K\mu2$ monitor sample and scattered $\pi^+$'s (dashed)  
from the Skim 6 sample. 
The energy loss of $K^+$ in  
the target was required to be greater 25~MeV and  
the number of $K^+$ fibers was required to be  
greater than 2 before evaluating the beam likelihood. 
} 
\label{fig:beam_like} 
\end{figure}

{\it Beam~Likelihood:} 
The $K^+$ stops in the target 
were required to have    
energy loss in the B4 and target consistent with that expected 
for the measured $K^+$ stopping position. 
These conditions helped to eliminate  single  
beam backgrounds with a scattered $\pi^+$ which did 
not have a consistent path length in the target.  
The three quantities were combined into a likelihood 
function. Fig.~\ref{fig:beam_like} shows  
the likelihood distributions for signal and background using  
the $K_{\mu2}$ peak events in the $K\mu2$ monitor sample  
and the scattered $\pi^+$'s in the Skim 6 sample. 

{\it Pileup~cut:} 
The pulse shape recorded 
with the target CCD's provided pileup information.   
The time development of the 
output signal was fitted with both single- and double-pulse 
assumptions. If the fitted pulse was more likely to be a double pulse 
and the time of the second pulse was coincident with the $K^+$ decay 
time, the event was rejected.

{\it Pathology~Cuts:} 
In the target scintillator, charged particles sometimes  
underwent nuclear interactions or photons had electromagnetic  
showers. Complicated target patterns could shift  
background events into the signal region. 
Some flaws in the selection criteria were identified by inspecting  
the background events that showed up  
outside the signal region to be described in  
Section~\ref{sec:outside_the_box}.    
Below are the situations addressed by   
pathology cuts.  
\begin{itemize} 
\item Target $\pi^+$ cluster with an identified kink.  
      This was an indication of a hard scattering process, which 
      might lead to an incorrect measurement of kinematic quantities. 
\item In the B4 the measured energies derived from the ADC and  
      CCD were required to be consistent within 1.5~MeV and the  
      measured times derived from the TDC and CCD 
      were required to be consistent within 2 ns.  
      Inconsistency in either the times 
      or energies was likely due to a second beam particle. 
\item Target $\pi^+$ fiber next to the decay vertex  
      with energy greater than 3.5~MeV. This might indicate a 
      $K^+$ fiber being mis-identified as a  $\pi^+$ fiber, causing  
      an incorrect measurement of kinematic quantities. 
\item Target $\pi^+$ fiber with energy greater than 5~MeV. This might  
      imply a $K_{\pi2}$ or a radiative $K_{\mu2}$ event with a photon  
      hiding in the $\pi^+$ fiber. 
\item If a target edge fiber was identified as a $K^+$ fiber with a 
  time within 3 ns of the nearby IC, the event was rejected. 
      This feature was often an indication of a double beam  
      background. 
\item Target fibers on the opposite side of  
      the $\pi^+$ track with respect to the vertex  
      having energy greater than 2~MeV within $\pm$3.0 ns of $t_\pi$. 
      This usually indicated a $K_{\pi2}$ event with  
      a photon conversion 
      opposite to the $\pi^+$.     
\item Target $\pi^+$ track on the opposite side of the $\pi^+$ track  
      with respect to  
      the $K^+$ decay vertex. This usually indicated a double  
      beam or a cosmic ray background in addition to  the  
      first $K^+$ beam particle.      
\item Target edge fiber greater than 5~MeV within $\pm$4 ns of $t_{rs}$. 
      This suggested some photon 
      activity in the edge fibers, or a double beam event with  
      the second beam hiding in the track. 
\item Any activity found in either the UPV or RV within $\pm$ 4 ns 
      of $t_{rs}$. 
\end{itemize} 
Most pathology cuts were related to target 
pattern recognition. These backgrounds could not  
be tagged  with  the usual procedure of background analysis until  
the outside-the-box study given in \ref{sec:outside_the_box}.

\subsubsection{Decay $\pi^+$ Kinematic Requirements} 
\label{sec:kincuts} 
 
Once the  $\pi^+$ kinematic reconstruction routines  
produced the   range-, energy- and 
momentum-related quantities,  cuts on the kinematic 
values (referred to as KIN cuts), were identified and  divided into  
several sub-groups  which are described below.

{\it Fiducial~Cuts:} 
The $\pi^+$ stopping layer was required to be  
RS layers 11-18. No charged track was allowed to stop in the 
RSSC layer embedded between the $14^{\rm th}$ and the $15^{\rm th}$ RS layer.  
Events were also rejected if the stopping layer was 14 with an RSSC 
hit found in the same RS sector or one sector clockwise of the  
stopping counter. The cosine of the polar angle for a charged track  
was required to be within $\pm0.5$.  
The $z$ position from the UTC  
track extrapolation to each RS layer 
was required to be as narrow as $\pm30$~cm to reject  $K_{\mu2}$ 
backgrounds with longer path lengths in RS. 
The effective UTC fiducial volume was defined to be $|z|<25$~cm 
at the UTC outermost layer.

{\it Signal~Phase~Space:} 
The phase space cuts required that the momentum,  
kinetic energy and range of a $\pi^+$ track should be in  
$211 \le P \le 229$~MeV/$c$, $115 \le E \le 135$~MeV, and 
$33 \le R \le 40$~cm.  
As the resolution of kinematic quantities depended on 
the azimuthal and polar angles of the $\pi^+$ track,  
the lower limits of the phase space region were further 
defined by the requirements (referred to as \KPITWO~kinematic cuts) that  
$\mbox{Pdev}=\Delta P/\sigma_P\ge 2.5$,  
$\mbox{Edev}=\Delta E/\sigma_E\ge 2.5$, and 
$\mbox{Rdev}=\Delta R/\sigma_R\ge 2.75$  
where the $\Delta P$, $\Delta E$ and $\Delta R$ were the deviation from  
the \KPITWO~kinematic peak positions, and $\sigma_P$, $\sigma_E$  
and $\sigma_R$ were the corresponding resolutions,  
which were correlated with the azimuthal and polar angles.

{\it Tracking~Quality~in~Target:} 
Good target tracking relied upon a consistent pattern of hits in the   
target. This required 
the nearest target $K^+$ fiber of the track to 
the B4 hit position to be  
no more than 2~cm away. 
The $K^+$ decay vertex was required to be  
 nearest to the extreme tip of the $K^+$ cluster.  
No more than  
one fiber gap was allowed  
between the $K^+$ decay vertex and the  
closest $\pi^+$ fiber.  
A target track could include either $\pi^+$ fibers with 
photons or photon 
fibers mis-identified as the $\pi^+$ fibers, leading to an 
incorrect  energy measurement or event classification if the  
photon veto also failed. The $\pi^+$ fibers were therefore examined  
using likelihoods based on the comparisons of  
the time, energy, and distance to the track between those values expected  
from the simulated \KPPNN~sample and the measured values 
from the $K\pi2(1)$ monitor events.

{\it Tracking~Quality~in~UTC~and~RS:} 
At least 4 hits out of 6 cathode foils were required  
to ensure a good measurement of a track in the UTC.  
Further quality checks were developed with 
respect to the $K_{\mu2}$ momentum peak, taking 
account of all the possible circumstances that could lead to  
an incorrect momentum measurement, such as no hit  
in the outermost foil layer,  
less than 12 layers of  anode wire hits,  
overlapping tracks,  
or too many wire hits in a cluster being excluded from the fit.  
The cuts were adjusted so that the momentum resolution 
effect did not give a significant contribution to the $K_{\pi2}$ and  
$\mu^+$ background estimates (Section~\ref{chap:bkg_estimate}). 
Another good way to ensure a good UTC measurement was to  
require consistency  
among UTC, RS and RSSC's.  
In  the $ x-y$ plane,   
this could be achieved by requiring consistency  between the UTC  
extrapolation and the positions of the sector crossing and the RSSC 
hits. Similarly, in the $\phi-z$ plane consistency was required  
between the UTC extrapolation and the $z$ positions measured by  
RSSC and RS. To establish these requirements, the $\mu^+$'s in Skim 5  
were selected with all the other cuts applied,   
except that the maximum momentum cut was not applied and the \PIMUE~decay  
sequence cuts (see Section~\ref{sec:pimue_id}) were  
inverted, and the $\pi^+$'s in Skim 6 were 
selected with all the cuts applied, except that the cuts  
on single beam $K^+$ were inverted (Section~\ref{sec:beam_cuts}). These  
signal and background samples were also used in the relevant  
studies of energy loss in the RS and range-momentum consistency 
in UTC and RS.

\begin{figure} 
\centering 
\epsfxsize 0.8\linewidth 
\epsffile{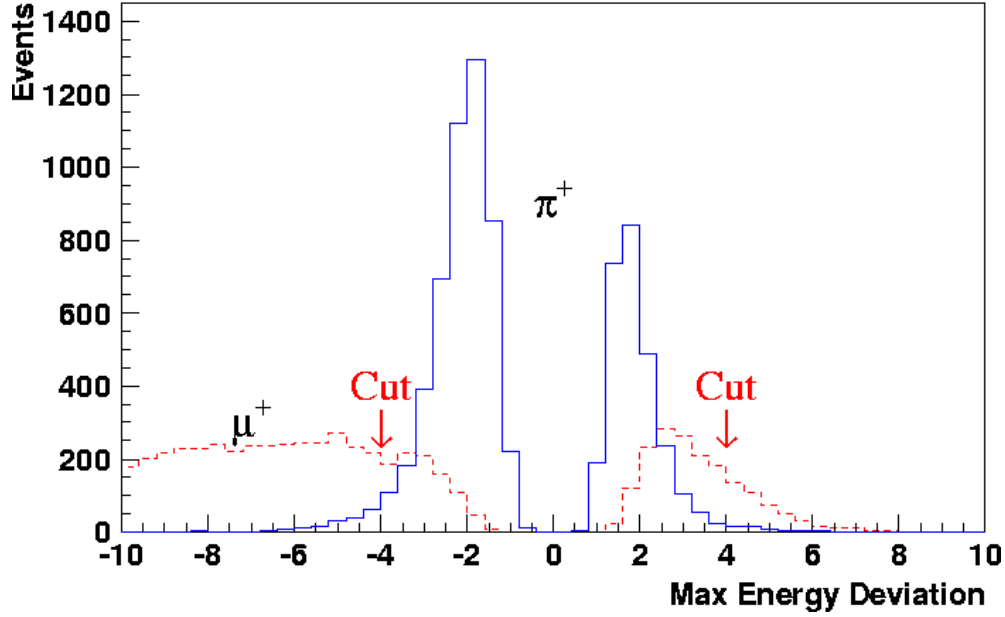} 
\caption{Distributions of the maximum energy  
deviation for $\pi^+$'s (solid) and $\mu^+$'s (dashed) in RS. The  
cut positions are indicated by the arrows.} 
\label{fig:rs_max} 
\end{figure}

\begin{figure} 
\centering 
\epsfxsize 0.8\linewidth 
\epsffile{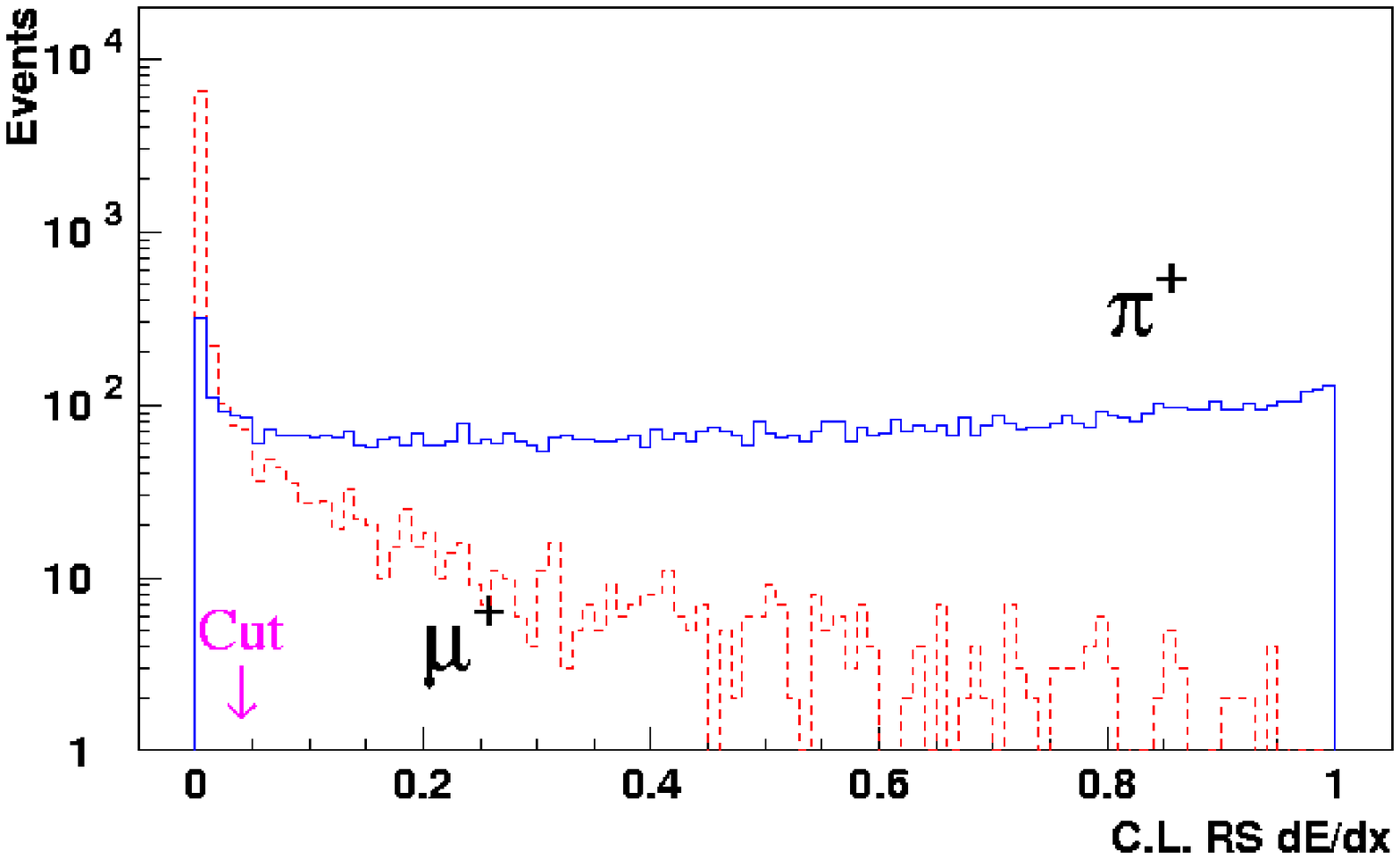} 
\caption{Distributions of the probability of energy loss consistency  
in the RS $dE/dx$ measurement for  
$\pi^+$'s (solid) and $\mu^+$'s (dashed).  
The arrow indicates the cut position. } 
\label{fig:rs_cl} 
\end{figure}

\begin{figure} 
\centering 
\epsfxsize 0.8\linewidth 
\epsffile{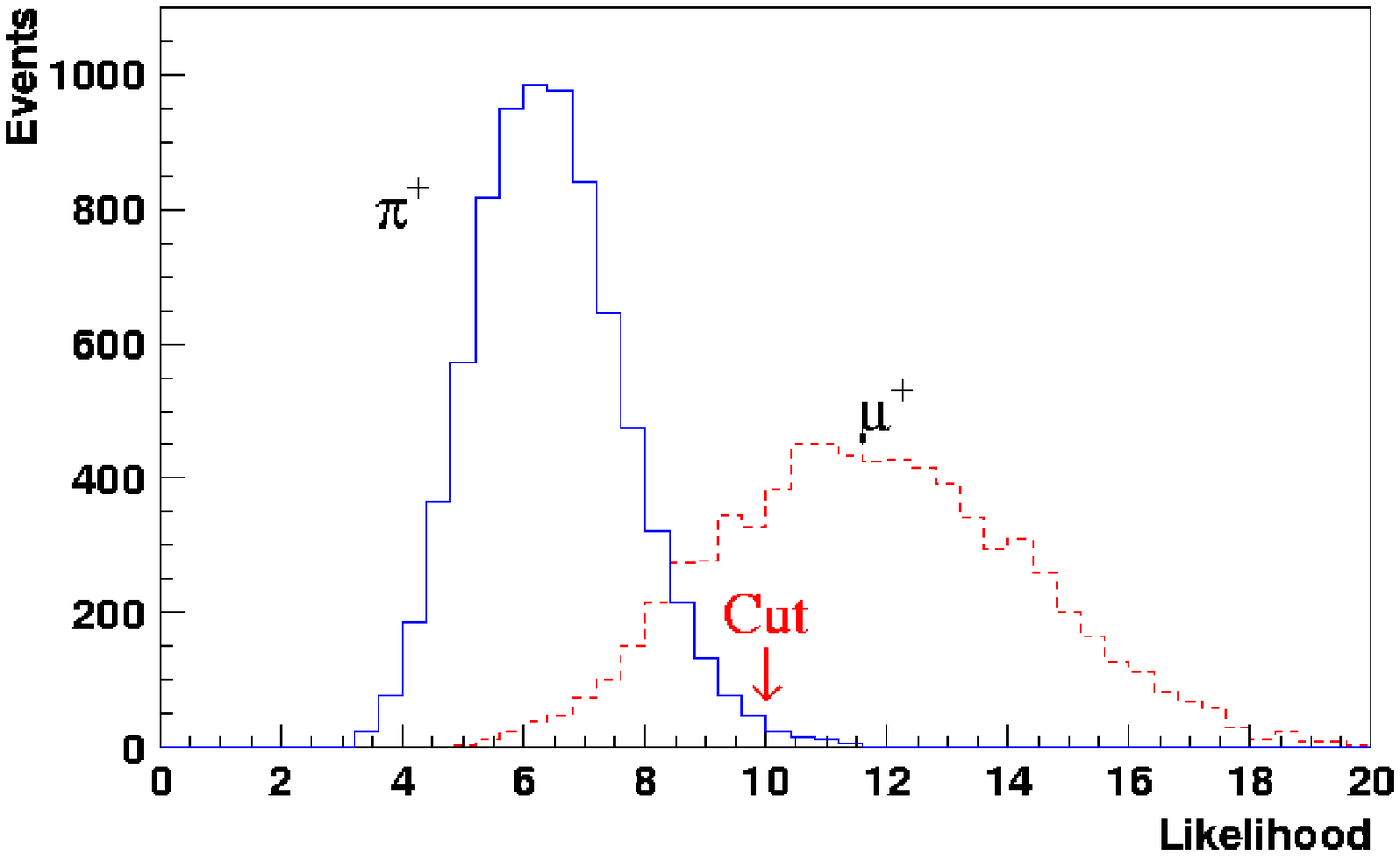} 
\caption{Likelihood distributions of the RS energy  
measurement for $\pi^+$'s (solid) and $\mu^+$'s (dashed).  
The arrow indicates the cut position.} 
\label{fig:rs_like} 
\end{figure}

{\it Energy~Loss~in~RS:} 
It was noted that a $\mu^+$ from the \KMUTWO~ 
might fake a $\pi^+$ due to scattering.  
In addition a $\pi^+$ from the \KPITWO~might fake a signal 
due to either a photon  
or an accidental hit being hidden in the track counters. 
The resulting background was removed by comparing 
 the energy measurement in each RS layer  
 with the expected value from the range.  
The energy deviation was required to be within $\pm4\sigma$'s for  
each RS counter as shown in Fig.~\ref{fig:rs_max}. 
In addition to each layer, a probability for energy loss consistency  
was also calculated  
with all track counters except for the T-counter, the stopping  
counter and the counters with sector crossings. 
As shown in Fig.~\ref{fig:rs_cl}, 
the probability cut provided good separation between 
$\pi^+$'s and $\mu^+$'s.  
Another way to reject the $\mu^+$ background was to calculate  
the likelihood from the ratio of the expected ($E_{expt}^i$)  
and measured ($E^i_{meas}$) energy  
of a track in the $i$-th layer,  
$\Delta E\equiv \log E^i_{expt} - \log E^i_{meas}$,  
taking into account the Landau tail of the expected energy 
distribution. 
This calculation was done up to and including the layer  
prior to the stopping layer.   
Fig.~\ref{fig:rs_like} shows  good $\pi/\mu$ separation using the  
calculated likelihood value. It was noted that this cut on the likelihood 
was correlated with the cuts on the energy deviation and the  
probability given above.   

{\it Range-energy~Consistency~in~IC~and~Target:} 
Despite the poor resolution of energy measurements in the  
IC and target, they could still provide a certain level of  
particle identification in addition to  that from the RS. 
Signal and background samples used in this study were  
from the $K\pi2(1)$ and $K\mu2$ monitors. 
The difference between the $E_{IC}$ and 
the expected value from $R_{IC}$ was required to be between $-5$ to 1.75~MeV.  
The target range ($R_{tg}$) in cm and energy ($E_{tg}$) in MeV  
were used to cut events in which  
$R_{tg}>12$~cm,  
$E_{tg}>28$~MeV,  
$9.5\times E_{tg}>28\times R_{tg}$  
or  
$10\times E_{tg}<21.5\times (R_{tg}-2)$ 
to  
reject background with a photon hiding along  
the $\pi^+$ track in the target.

\begin{figure} 
\centering 
\epsfxsize 0.8\linewidth 
\epsffile{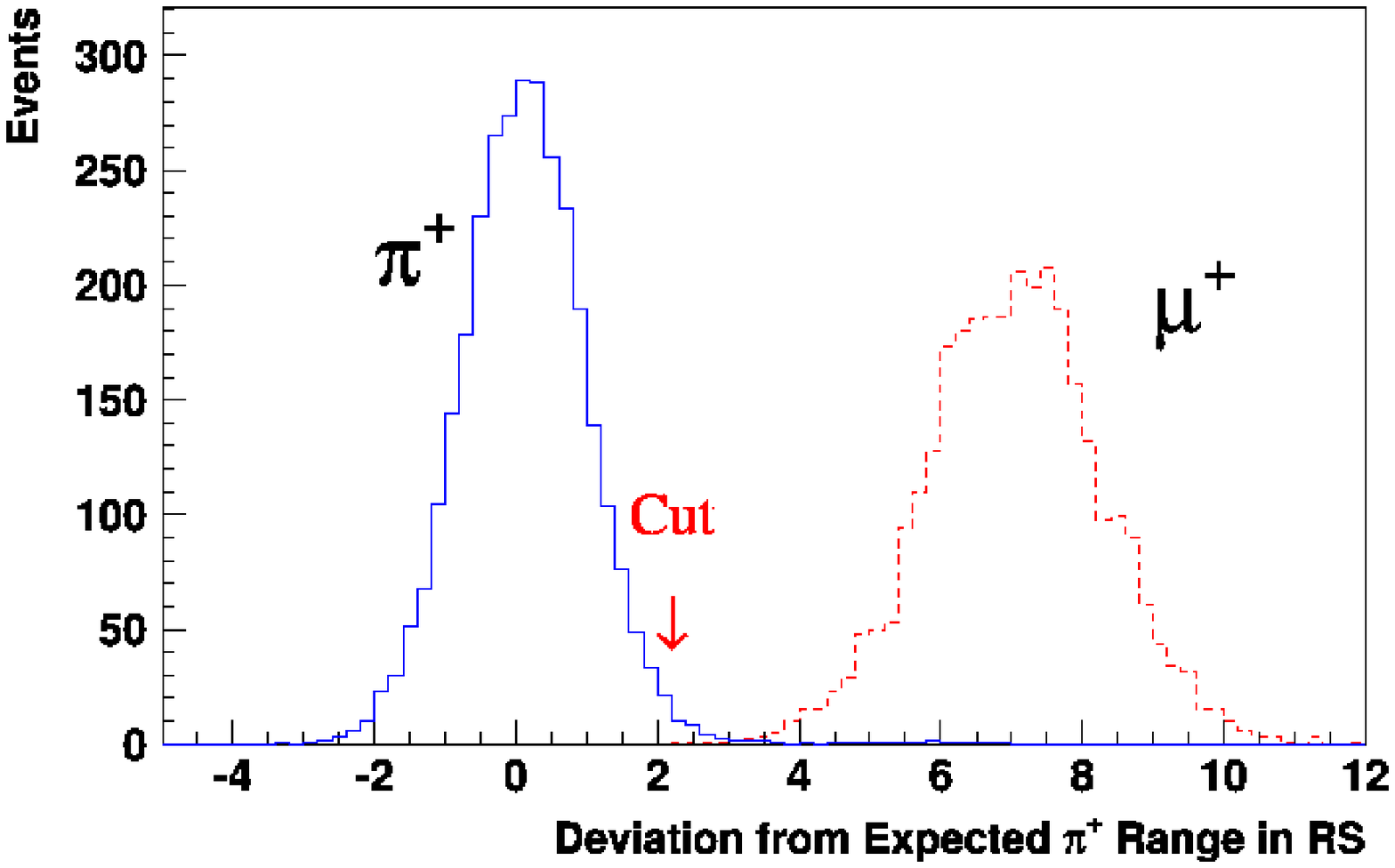} 
\caption{Distributions of the range deviation in  
the RS for $\pi^+$'s (solid) $\mu^+$'s (dashed) tracks.  
The arrow indicates the cut position.} 
\label{fig:rngmom} 
\end{figure}

{\it Range-momentum~Consistency~in~UTC~and~RS:} 
This cut was used to check whether the range of  
the charged track was consistent with that for a $\pi^+$. 
The range deviation in RS was defined as 
$\chi_{rm}=(R_{rs} - R_{utc})/\sigma_R$,  
where $R_{utc}$ was the expected range  
calculated from the momentum measured by the UTC 
with a $\pi^+$ hypothesis, and $\sigma_R$ was the uncertainty of  
the measured range as a function of the momentum. 
The $\mu^+$'s and $\pi^+$'s  
were selected from Skim 5 and Skim 6 samples, respectively, as used  
in the study of energy loss in the RS, except that 
the maximum momentum requirement was also applied in Skim 5 to  
remove the $K_{\mu2}$ range tail events. 
The distributions of the range deviation for  
$\pi^+$'s and $\mu^+$'s were shown in Fig.~\ref{fig:rngmom}.   
Good $\pi/\mu$ separation was observed.  
 
\subsubsection{\PIMUE~Decay Sequence} 
\label{sec:pimue_id} 
 
All cuts for identifying the \PIMUE~decay sequence were  
put into a special group (referred to as TD cuts).  
The TD information recorded the pulse shape, providing  
a tool to recognize the decay sequence. As can be seen below,  
the TD cuts were independent of the kinematic reconstruction  
and photon veto and could be used in the bifurcation study.  
The signature for this decay sequence was: 
\begin{itemize} 
    \item Three energy deposits  
      (pulses)  
      corresponding to the $\pi^+$ kinetic energy,  
      $\pi^+\to\mu^+\nu_\mu$  
      and $\mu^+\to e^+\nu_e\bar{\nu}_\mu$ decays  
      were found in the stopping counter. 
    \item The kinetic energy of the $\mu^+$ from  
      $\pi^+\to\mu^+\nu_\mu$ decay was 4.1~MeV, but due to saturation  
      the observed energy was about 3~MeV~\cite{birks}. 
      Since the path length of the $\mu^+$ was $\sim\!1.4$~mm in RS,    
      the fraction of $\mu^+$ exited the stopping counter  
      without depositing more than 1~MeV  
      was only $\sim\!1\%$ of $\pi^+$ decays. 
    \item The $e^+$ from $\mu^+\to e^+\nu_e\bar{\nu}_\mu$ 
      decay has a kinetic energy of $E_e<53$~MeV.  
      Most of the $e^+$'s exited the stopping  
      counter and deposited energy in the other RS counters. 
\end{itemize} 
The three energy deposits  
from the \PIMUE~decay sequence should be  
observed by the TD's at both ends of the stopping counter.  
 
For the $\mu^+$ background,  only two pulses due to muon kinetic energy 
and decay would be  produced. A $\mu^+$ could fake  
a $\pi^+$ when an extra pulse  
was detected in addition to the expected two pulses.  
To suppress the $\mu^+$ background, two stages of cuts were imposed. 
 
\begin{figure} 
\centering 
\epsfxsize 0.9\linewidth 
\epsffile{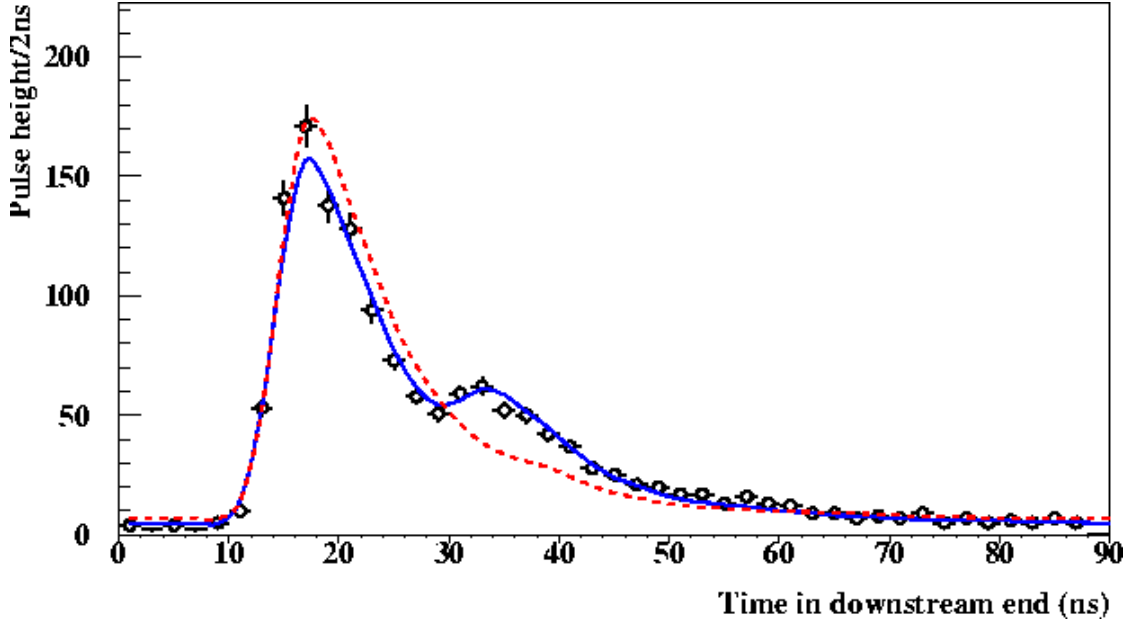} 
\caption{Results of fits with the single- and double-pulse hypotheses 
to the TD pulse shapes in the stopping counter.  
The solid curve is for the double-pulse hypothesis 
and the dashed curve is for the single-pulse hypothesis. 
} 
\label{fig:fitpi4} 
\end{figure}

At the first stage evidence for the \PIMU~decay was sought. 
The pulse development in the stopping counter as recorded by the TD's 
was fitted with a single- and double-pulse hypothesis in an interval 
of $\sim\!4$ $\pi^+$ lifetimes (typically 104 ns). The template shapes 
used in the fit were derived from the average of measured pulses from 
$\mu^+$ traversal for each end of each RS counter. In addition, a correction 
was applied to the template shape to take into account 
the change in shape due to propagation along the counter.  
The parameters of the single-pulse fit were the time, the total 
area of the pulse and a constant corresponding to a pedestal 
of typically 3 TD counts. The parameters of the double-pulse fit 
were the time of the first pulse, the time difference of the two 
pulses, the total pulse area, the fractional area of the second pulse 
and the pedestal. A fit to a triple-pulse 
hypothesis was attempted if evidence for a third pulse was found 
based on a rudimentary analysis of the TD information or 
if evidence for the \PIMU~decay from the double-pulse fit was lacking. 
The two additional parameters in the fit 
were the time difference of the third pulse with respect to the first 
and the fractional area of the third pulse. The results of the single- 
and double-pulse fit hypotheses for \PIMU~decay were 
shown in Fig.~\ref{fig:fitpi4}. 
Loose requirements were first 
applied on the observed $\mu^+$ energy $1<E_{\mu}<14\ {\rm MeV}$ 
and on the relative quality of the results  
of the single- and double-pulse fits, 
$R(1)>1$ or $R(2)>2$,  and $R(1)\times R(2)>1$ where  
the quantities are defined in 
Table~\ref{tab:fitpi_quantities}. 
 
\begin{table} 
\centering 
\begin{tabular}{l l} 
\hline\hline 
Quantity                    & Definition or use\\ 
\hline 
$E_{\mu}$                   & Energy of the second pulse \\ 
$T_{\mu}$                   & Time of the second pulse\\ 
$\chi^2_n(i)$               & $\chi^2$ for the $n$-pulse hypothesis  
                              for the counter\\ 
                            & end $i$ ($i=1,~2=$upstream, downstream)\\ 
$R(i)$                      & $\chi^2_1(i)/\chi^2_2(i)$ \\ 
$\log_{10}(R(1)\cdot R(2))$ & Neural net input \\ 
$\log_{10}(\chi^2_1(1)\cdot\chi^2_1(2))$ & Neural net input \\ 
$dz$                        & $=z_\pi - z_\mu$, neural net input component \\ 
 
$dt$                        & Time difference between both counter ends \\ 
                            & for the second pulse, neural net input\\  
\hline\hline 
\end{tabular} 
\caption{\label{tab:fitpi_quantities} Definitions of quantities determined 
by the pulse-fitting in the stopping counter. The $z$ positions were 
determined from the energy ratio between the two ends of the stopping counter. 
The $z$ position of the nominal $\pi$ and $\mu$  
pulse was $z_\pi$ or $z_\mu$, respectively. 
The term ``second'' pulse identifies the $\mu^+$-candidate pulse.  
The neural net is described in the second stage of cuts.} 
\end{table}

At the second stage, five cuts listed in Table~\ref{tab:tdcut_table} 
were then applied to 
suppress the following four mechanisms when a $\mu^+\to e^+$  
decay could fake the three-pulse \PIMUE~decay sequence. 
\begin{itemize} 
   \item $\pi^+$ time accidental: 
     Accidental activity produced the first pulse, while  
     the $\mu^+\to e^+$ decay gave the second and  
     third pulses in the timing sequence. 
   \item Early $\mu^+$ decay: The $\mu^+\to e^+$ decay  
     occurred at an early time 
     ($\le 100$~ns), producing the second pulse,  
     and accidental activity was identified as 
     the third pulse.  
   \item $\mu^+$ time accidental: The  
     $\mu^+\to e^+$ decay made the third pulse,  
     while the second pulse was produced by  accidental activity 
     occurring between the $\mu^+$ stop and 
     decay. 
   \item Tail fluctuation: A fluctuation in the falling edge  
     of the first pulse was identified as the second pulse.  
     The decay positron from the $\mu^+$ decay made the  
     third pulse. 
\end{itemize} 
 
\begin{table} 
\centering 
\begin{tabular}{l c c c c}\hline\hline 
\multirow{2}{*}{Cut}& $\pi^+$ time & Early $\mu^+$ & $\mu^+$ time   & Tail \\  
                    &   accidental & decay         & accidental     & fluctuation \\ \hline 
$\pi^+$ time consistency Cut & $\sqrt{}$ & & & \\  
$\mu^+\to e^+$ decay requirement & & $\sqrt{}$ & & \\  
Cut on $\mu^+$ time accidental & & & $\sqrt{}$ & \\  
Cut on $\mu^+$ time accidental & & &\multirow{2}{*}{$\sqrt{}$} & \\ 
in the track counters       & & & & \\  
Neural net $\pi^+\to\mu^+$ decay cut & & & $\sqrt{}$ & $\sqrt{}$ \\ \hline\hline 
\end{tabular} 
\caption{List of the backgrounds targeted by \PIMUE~decay sequence cuts.} 
\label{tab:tdcut_table} 
\end{table}

{\it $\pi^+$~Time~Consistency~Cut:} 
This cut suppressed the $\pi^+$ time accidental background. 
When accidental activity made the first pulse 
and a charged track made the second pulse in the stopping counter, 
the timing of the first pulse obtained by the TD ($t_{\pi,TD}$)  
was not coincident with $t_{RS}$ obtained from the 
other RS counters along the track. Events were rejected if  
$|t_{\pi,TD} - t_{RS}| > 2.5~{\rm ns}$.

\begin{figure} 
\centering 
\begin{minipage}{0.49\linewidth} 
\centering 
\epsfxsize 0.8\linewidth 
\epsfysize 4cm 
\epsffile{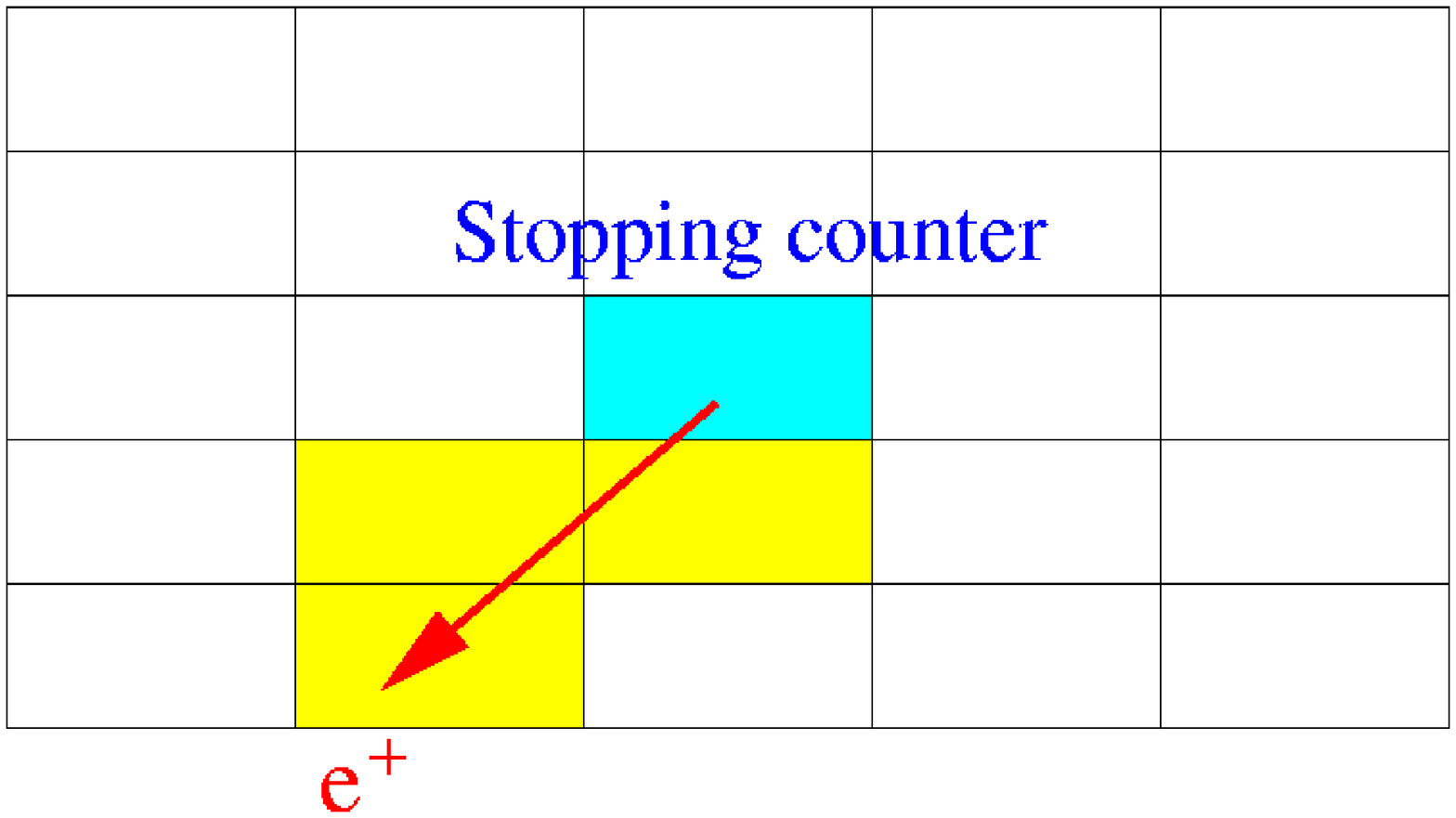} 
\end{minipage}\hfill 
\begin{minipage}{0.49\linewidth} 
\centering 
\epsfxsize 0.8\linewidth 
\epsfysize 4cm 
\epsffile{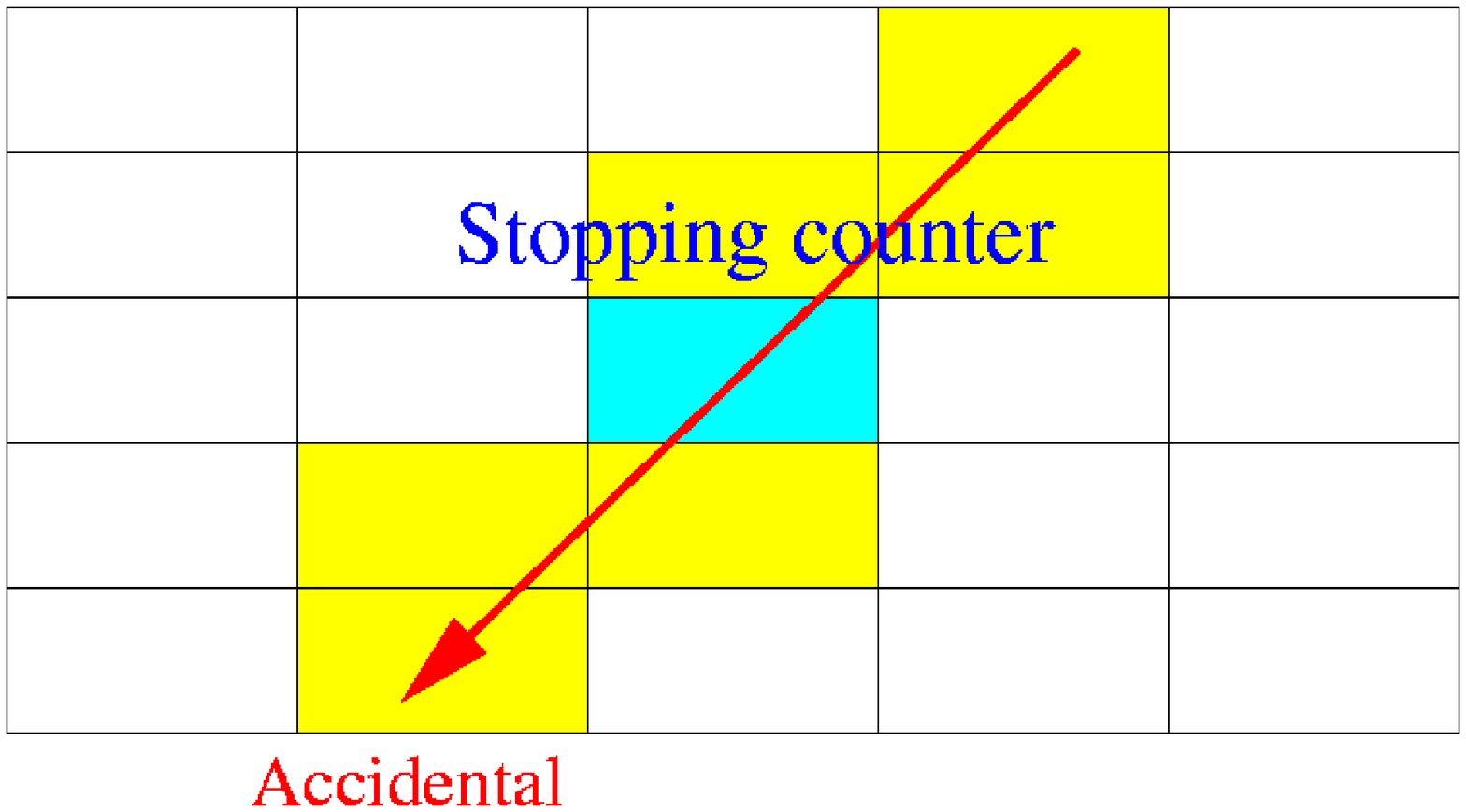} 
\end{minipage} 
\caption{Schematic  view of signal (left) and background (right) of  
the $\mu^+\to e^+$ decay. 
Each rectangle represents an RS counter.  
The central counter represents the stopping counter and the 
shaded rectangles represent hit counters. The arrow indicates  
the possible direction of the positron or charged track producing the hits.} 
\label{fig:ev5fig} 
\end{figure}

{\it $\mu^+\to e^+$~Decay~Requirement:} 
The positron from the $\mu^+\to e^+$ decay  
(the third pulse) generally deposited energy in  
the stopping counter and other neighboring counters  
as depicted in Fig.~\ref{fig:ev5fig}. 
The positron finding started by looking for a  
cluster of TDC hits in the  RS counters in 
the region within $\pm 1$~sector and $\pm 2$~layers of the stopping counter. 
The cluster should include the  
stopping counter and an additional hit in the same 
sector as the stopping counter. 
Candidates for the positron were found when  
the average time of the hits in the cluster 
was within $\pm2.4$~ns of the TDC time (the third pulse) in  
the stopping counter. 
The $z$ positions of the hits in the cluster, obtained  
from the end-to-end time differences of the hits, 
were also required to be consistent with the $z$  
position in the stopping counter.  
If the candidate was due to a track that passed through the 
stopping counter, then hits might be found  
on both sides of the stopping counter (Fig.~\ref{fig:ev5fig}).  
The early $\mu^+$ decay background was removed by requiring that  
the second pulse from 
the TD pulse fitting agreed with the time of the cluster. 

{\it Cut~on~Accidental~Activity:} 
Accidental activity in the stopping counter was frequently  
associated with activity in other RS counters as well as the BV, BVL and EC. 
Hence, events with activity coincident with the second pulse in the stopping 
counter were targeted for rejection. 
The time windows and energy thresholds for  
the various subsystems in RS, BV, BVL and EC were  
optimized in order to have the 
highest rejection power at a given acceptance value of 94\% for this cut.  
Events were rejected 
if the energy sum of the hits within a time window in any of the subsystems  
was greater than the 
threshold.  
There was also a kind of accidental activity  
that overlapped the charged track 
and made a second pulse in the stopping counter.  
To reject this accidental background,  
fits were performed to a double-pulse hypothesis  
in the two RS counters along the track  
prior to the stopping counter. If the time of the fitted second pulse  
was within $\pm5$~ns of the second pulse in the stopping counter, 
the fitted energy of the second pulse was greater than 1~MeV 
and the $\chi^2$ ratios of the single- to the double-pulse fit 
hypotheses were greater than 4, then the event was rejected.

\begin{figure} 
\centering 
\epsfxsize 0.85\linewidth 
\epsffile{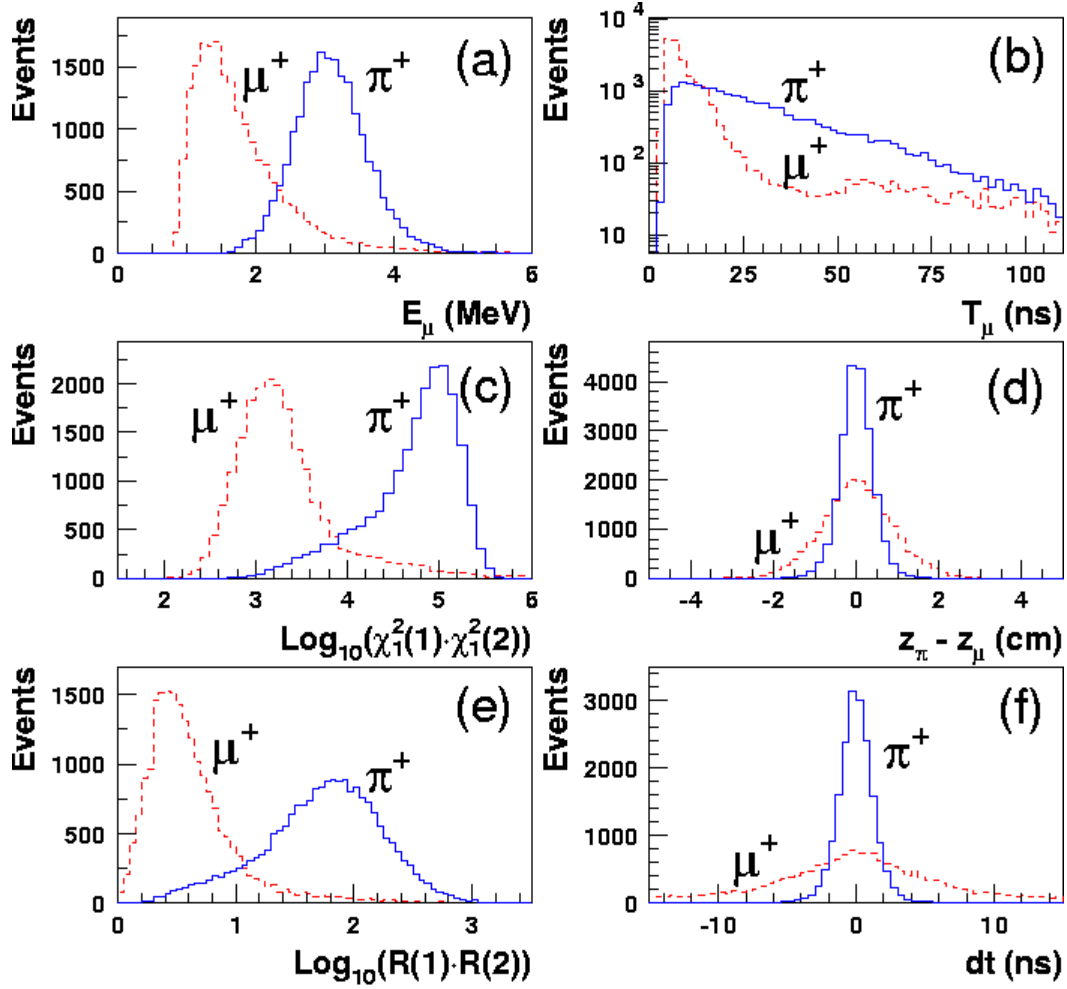} 
\caption{Distributions of the input variables for  
the Neural Net function in \PIMU~induced  (solid) 
and \MUE~induced  (dashed) decays. 
The energy of the second pulse (a), 
time of the second pulse (b),  
log of the product of the single-pulse fit $\chi^2$'s  
for both ends (c), 
$z$ position difference between the first and second  
pulses obtained from the energy ratio of  
both ends (d), 
log of the product of the $\chi^2$ ratios of  
single- to double-pulse fits for both ends (e) and  
time difference  
of the second pulses in both ends (f).} 
\label{fig:nninp} 
\end{figure}

\begin{figure} 
\centering 
\epsfxsize 0.8\linewidth 
\epsffile{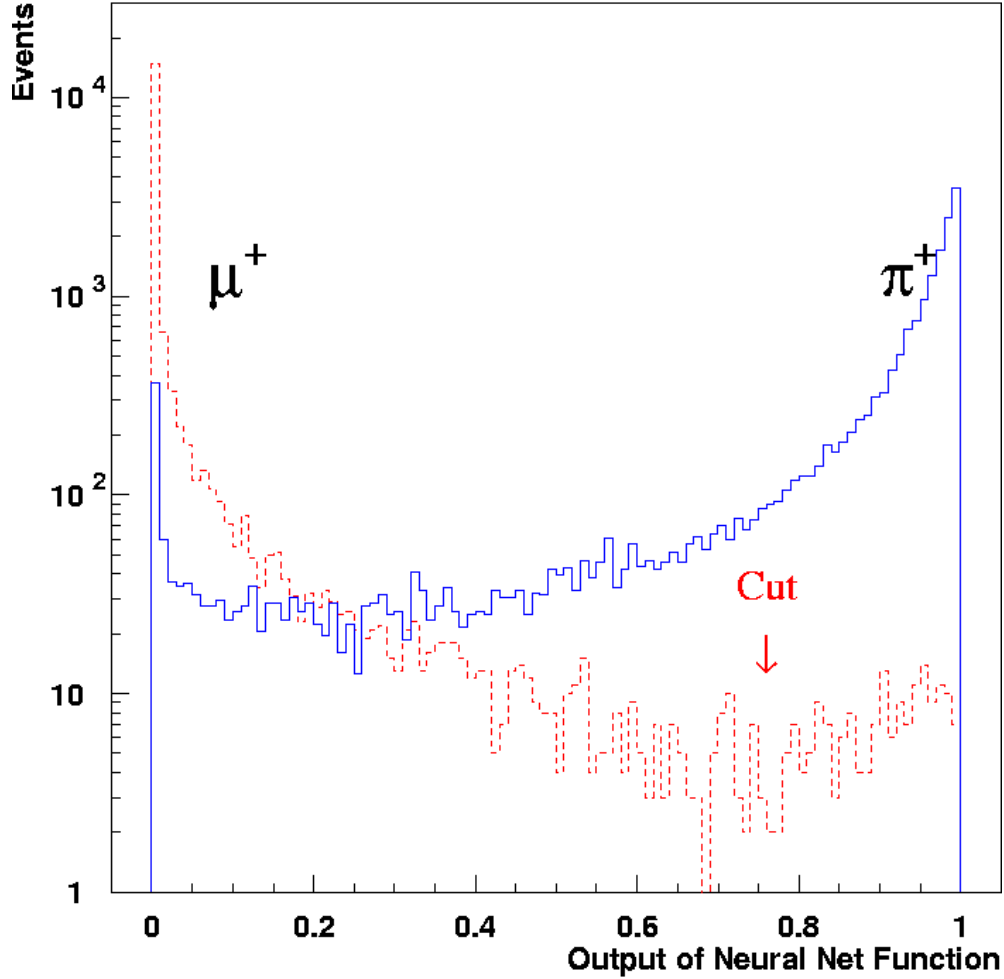} 
\caption{Distributions of the outputs of the NN  
function for $\pi^+$ (solid) and $\mu^+$ (dashed) events. 
Events with an output of the NN function less than 0.76 were rejected. 
} 
\label{fig:nn_cut} 
\end{figure}

\begin{figure} 
\centering 
\epsfxsize 0.8\linewidth 
\epsffile{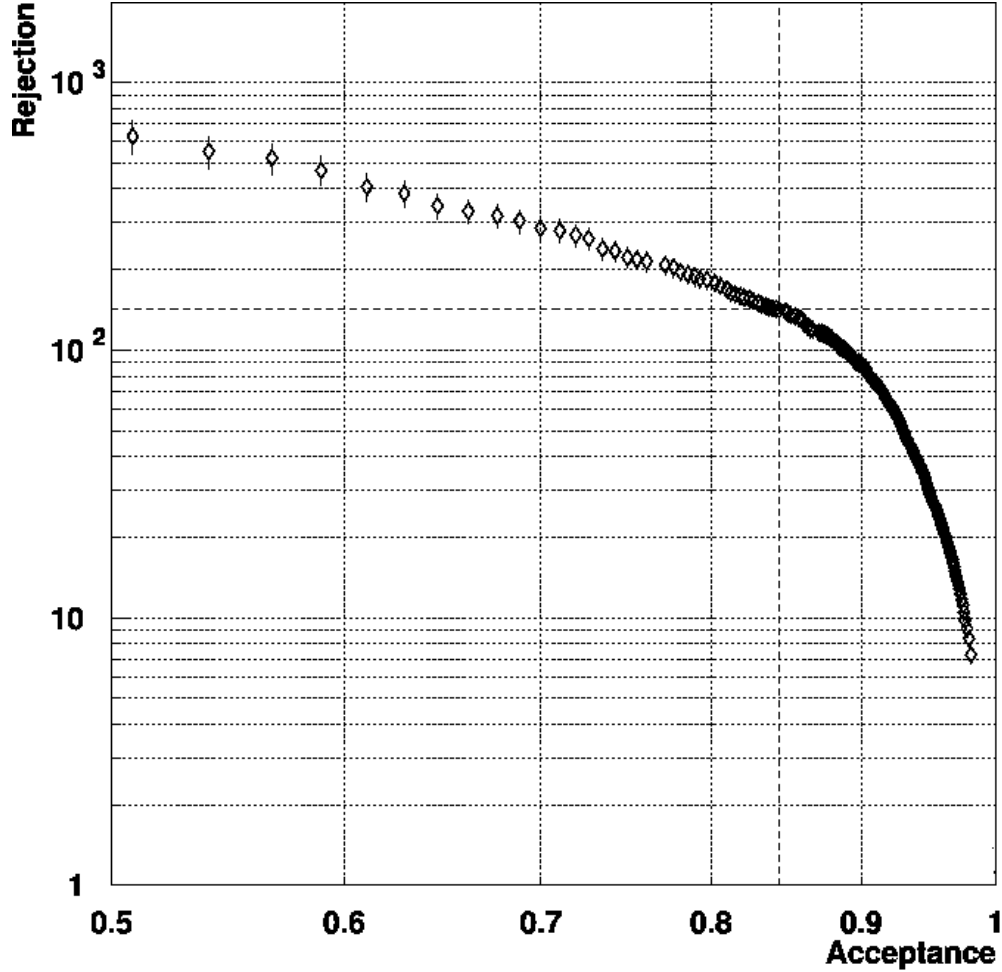} 
\caption{ 
Rejection of the NN $\pi^+\to\mu^+$ decay  
cut as a function of the acceptance. 
The intersection of the vertical and horizontal  
dashed lines shows the rejection and acceptance  
at the nominal cut position. 
} 
\label{fig:nn_func} 
\end{figure}

{\it Neural~Net~$\pi^+\to\mu^+$~Decay~Cut:} 
The tail fluctuation background 
mimicking the 
energy deposit for a $\mu^+$ at the falling edge of the $\pi^+$-induced 
pulse was characterized by a small decay time and  
a low pulse area in the second pulse.  
The variables shown in Fig.~\ref{fig:nninp} and 
described in Table~\ref{tab:fitpi_quantities}  
differed for events induced by \PIMU~ and \MUE~ decays. 
Application of a fixed cut to each variable would cause  
a non-negligible acceptance loss. 
In order to achieve a higher acceptance at the same  
rejection as the fixed cuts, 
a Neural Network (NN) technique was adopted.  
The NN function was derived via a Multi-Layer Perception 
program incorporated  
in the library of Physics Analysis Workstation (PAW)~\cite{pawmlp}. 
To create the NN function, the scattered $\pi^+$'s and  
\KMUTWO~range tail events in the 
$\pi\nu\bar{\nu}(1)$ trigger, which passed  
all other \PIMUE~decay sequence cuts, 
were used as signal and background samples, respectively.  
A 5-variable NN function was obtained using the six variables  
shown in Fig.~\ref{fig:nninp} with the differences in $z$ position 
and time combined to create a single input variable, 
\begin{equation} 
\chi^2(z,t) \equiv  
   \left({dz}/{\sigma_{dz}}\right)^2 + \left({dt}/{\sigma_{dt}}\right)^2, 
\end{equation} 
where the $z$ position and the time were obtained from the energy ratio of  
both ends, and time difference  
of the second pulses in both ends. 
The resolution of  $dz$ ($dt$) was 
$\sigma_{dz}$ ($\sigma_{dt}$). Fig.~\ref{fig:nn_cut} shows 
the distributions of the output of  
the NN function for $\pi^+$'s and $\mu^+$'s. 
The rejection of the NN $\pi^+\to\mu^+$ decay  
cut as a function of the acceptance is shown in Fig.~\ref{fig:nn_func} 

\subsubsection{Photon Veto} 
\label{sec:pv} 
 
To achieve the background level much less than one event,  
the total $\pi^0$ rejection  
was required to be of order of $10^6$. A rejection factor of  
$\sim\!10^4$ was already achieved  online, leaving a further  
$\sim\!10^2$ rejection factor to be achieved by  
the offline analysis. The corresponding photon veto cuts  
(referred to as PV cuts) were used to identify the  
photon activities detected by all the PV counters.  
 
\begin{figure} 
\centering 
\epsfxsize 0.9\linewidth 
\epsffile{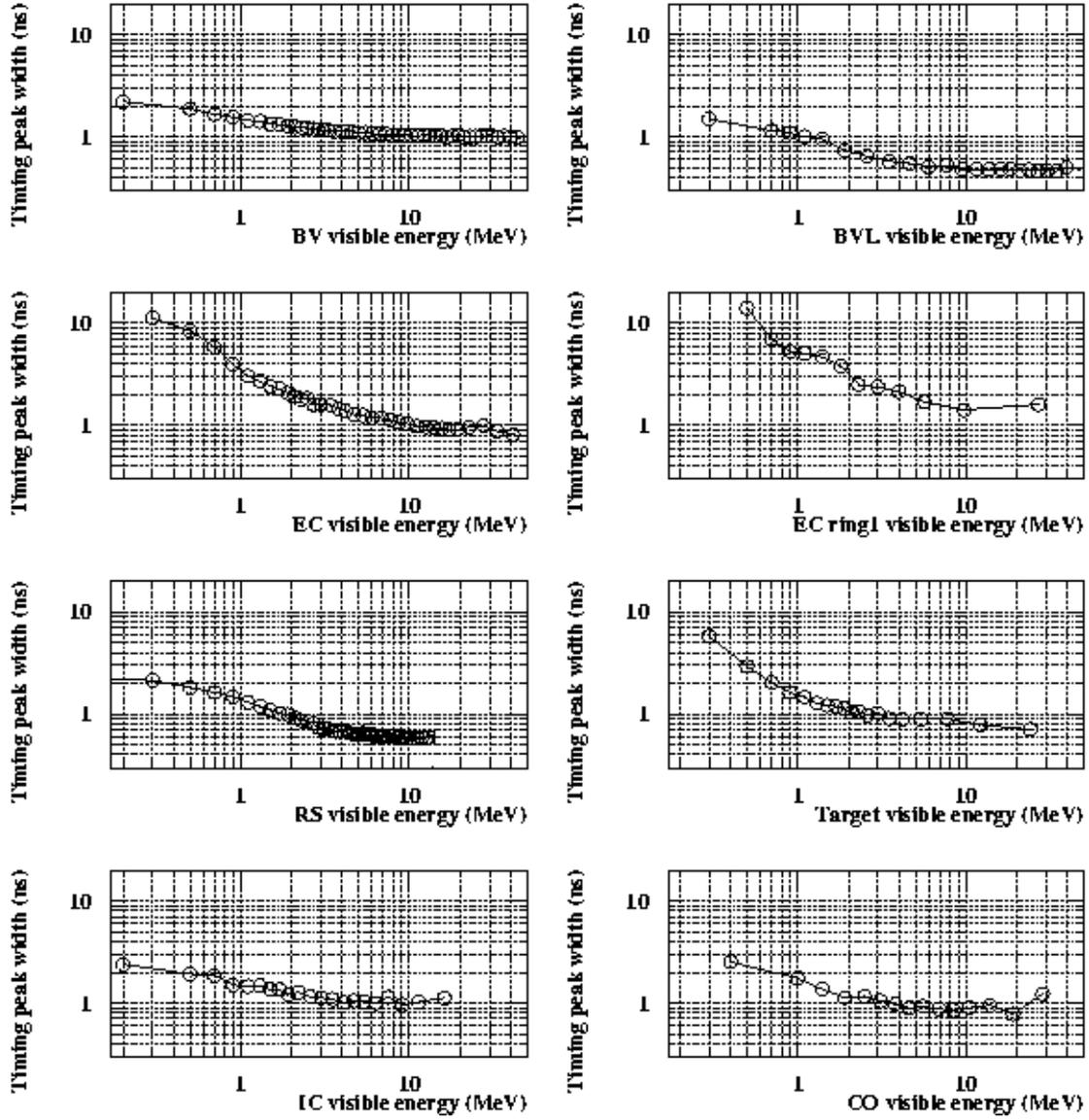} 
\caption{The timing resolution as a function of visible energy 
in various PV counters.} 
\label{fig:pv_twin} 
\end{figure}

A search for the photons coincident with the track time was performed  
in the subsystems of the BV, BVL, RS, EC, 
target, IC, VC, CO and $\mu$CO.  
The timing resolution of  
each photon detection system was a key ingredient in determining  
the time window for the PV due to the false 
veto rate in the high rate experimental environment.  
Fig.~\ref{fig:pv_twin} shows the measurements of  
the timing resolution for each system as a function of the  
visible energy. Since the EC consisted of 4 rings and the inner ring (ring1) 
was exposed to high accidental rates from the beam, the corresponding  
time resolution was worse.  
The time window and energy thresholds in each subsystem  were optimized  
by adjusting  the cut positions to maximize rejection for a given acceptance. 
The rejection sample was from Skim 4 with   
$K_{\pi2}$ events, while the acceptance  
sample was from the $K\mu2$ monitor. To ensure the events were 
from the \KPITWO~ peak, the measured range, energy and momentum  
of the charged track were required to be within three standard deviations 
of the nominal values. 
For the $K_{\mu2}$, the momentum of the charged track was 
required to be within three standard deviations 
of the nominal value, and the range was required to be  
longer than 37~cm to remove any event which could contain photon(s).  
The selected \KPITWO~and $K_{\mu2}$ events were also required to  
pass the KIN cuts and the beam cuts. 
 
\begin{figure} 
\centering 
\epsfxsize 0.8\linewidth 
\epsffile{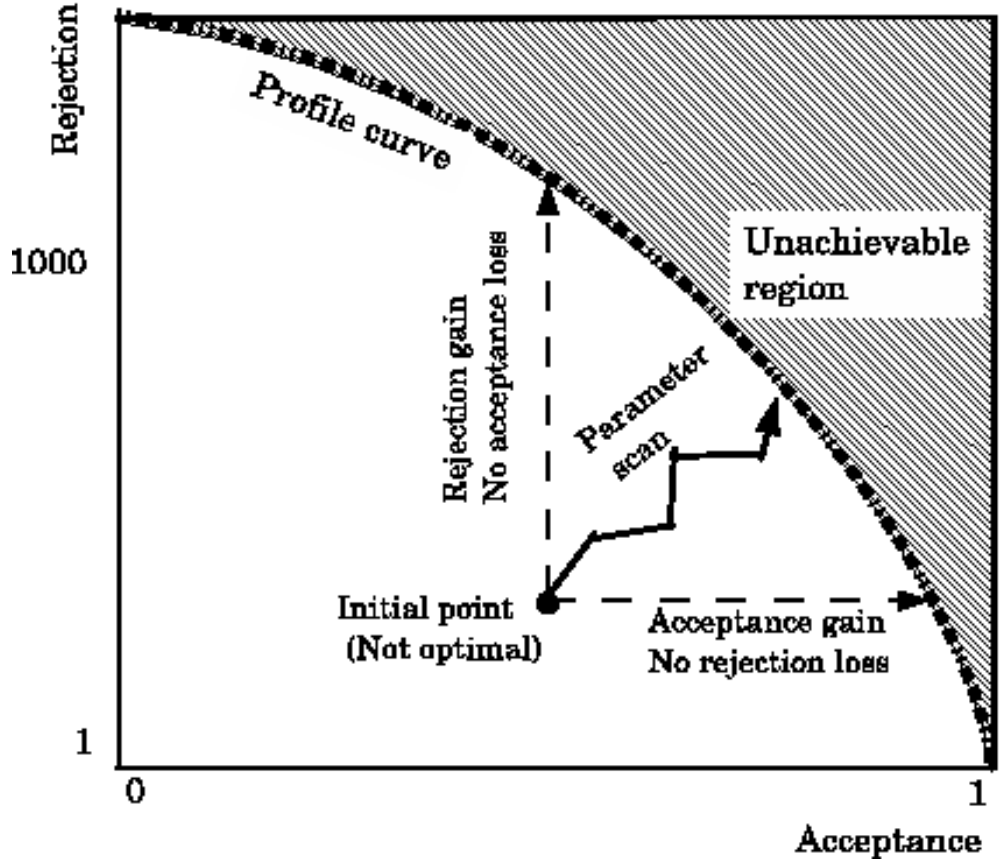} 
\caption{Illustration of the optimization process to determine 
the photon veto parameters.} 
\label{fig:pv_tune} 
\end{figure}

The optimization process~\cite{photon_veto} started from the initial set of cut 
parameters. For a new set 
of parameters, the rejection and acceptance were 
re-measured. Only one subsystem's cut parameters were varied at a time. 
If the rejection  
increased without losing acceptance or the acceptance increased 
without losing rejection, the set of parameters was regarded as a good input  
for the  iteration. More preferable cases occurred when both the  rejection and 
acceptance were improved. Fig.~\ref{fig:pv_tune} illustrates this  
optimization  process for the photon veto. The optimization process continued 
until no more gain was obtained in  rejection without losing  
acceptance. The boundary point was measured at every given 
acceptance position. As a result, a profile curve, which gave  
the maximum achievable  rejection, was obtained.  
 
\begin{figure} 
\centering 
\epsfxsize 0.8\linewidth 
\epsffile{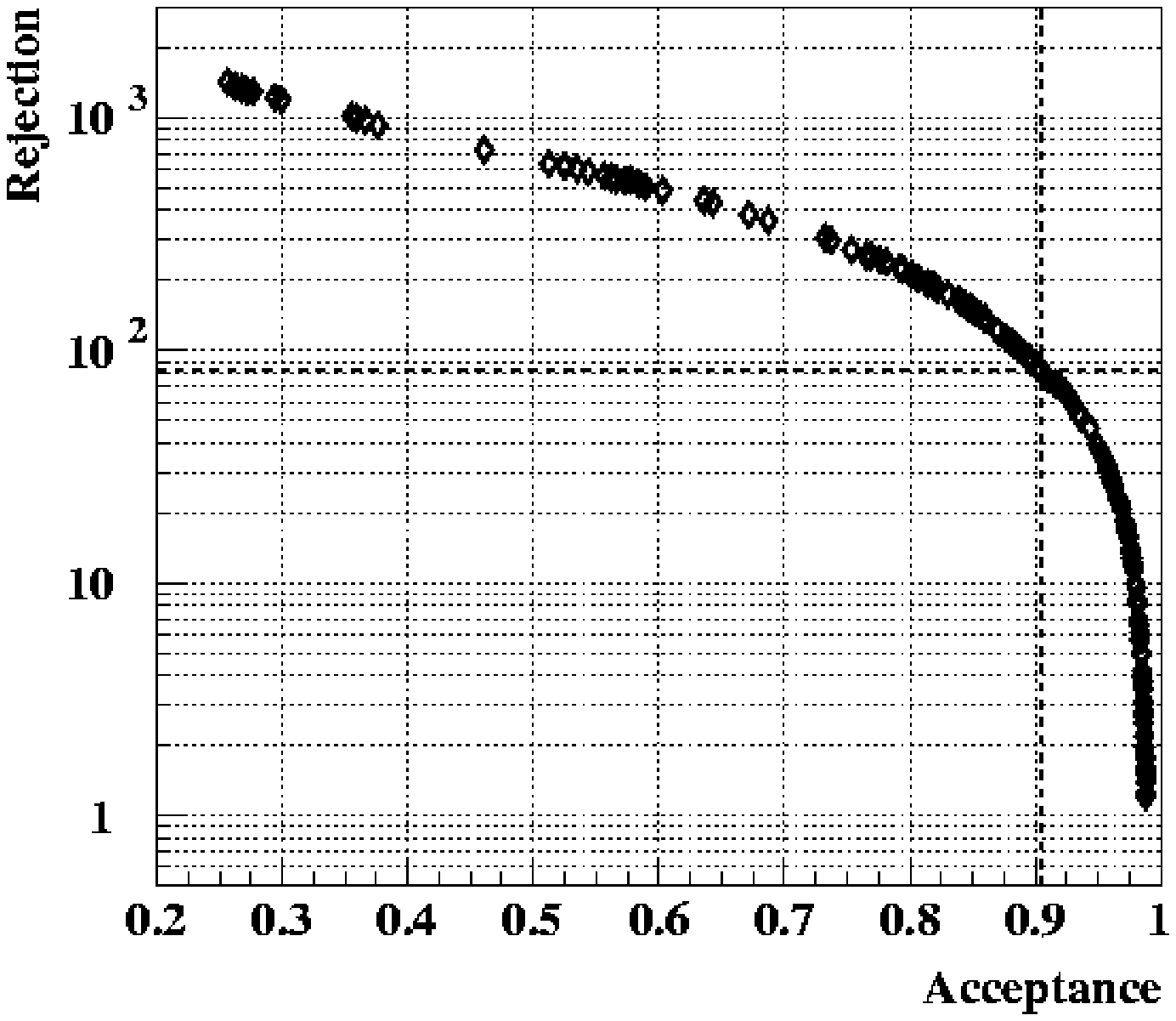} 
\caption{ 
Offline rejection of the photon veto cuts against the  
\KPITWO~background as a function of the acceptance. 
The crossing point of the vertical and horizontal lines  
shows the rejection and acceptance 
at the cut position.} 
\label{fig:pv_tune_rva} 
\end{figure}

Fig.~\ref{fig:pv_tune_rva} shows the offline rejection  
of the photon veto cuts  
against the \KPITWO~background as a function of acceptance. 
The time window and energy threshold for each category are tabulated  
in Table~\ref{tab:pv_def}. As a reminder, the total offline photon veto  
is not the  simple product  
of the rejections listed in Table~\ref{tab:pv_def} due to 
mutual correlations.  
\begin{table} 
\centering 
\begin{tabular}{l c c c c}\hline 
Category& Time window (ns)  
        & Energy threshold (MeV) & Rejection \\ \hline\hline 
BV      & $\pm$4.50 & 0.20 & 14.556 \\  
BVL     & $\pm$2.00 & 0.00 &  1.247 \\  
RS      & $\pm$1.50 & 3.80 &  3.329 \\  
EC      & $\pm$2.25 & 3.80 &  2.296 \\  
Inner EC& $\pm$1.75 & 1.00 &  1.186 \\  
Target  & $\pm$1.30 & 4.79 &  1.493 \\  
IC      & $\pm$2.25 & 0.40 &  1.321 \\  
VC      & $\pm$2.00 & 1.60 &  1.054 \\  
CO      & $\pm$1.25 & 0.80 &  1.052 \\  
$\mu$CO      & $\pm$1.50 & 2.80 &  1.003 \\ \hline\hline 
\end{tabular} 
\caption{Time window and offline energy threshold for each category of  
the photon veto cuts. Also listed are individual rejection values  
contributed by each sub-system.  
} 
\label{tab:pv_def} 
\end{table} 
 
\subsection{Background Evaluation} 
\label{chap:bkg_estimate} 
 
To have an unbiased result, the signal region was always masked until all  
the background evaluation studies were completed. As described in  
Section~\ref{sec:bkg_overview}, the stopped $K^+$ decay  
background and the beam background were subdivided into the following 
categories:  
\begin{itemize} 
\item \KPITWO~background, 
\item \KMUTWO~range tail background,  
\item $\mu^+$ band backgrounds, 
\item single beam background, 
\item double beam $K^+-K^+$ background, 
\item double beam $K^+-\pi^+$ background, and 
\item CEX background. 
\end{itemize} 
Except for CEX, all  background levels  were estimated using the  
data by means of the  bifurcation method. Table~\ref{bifur_cuts} gives  
the bifurcation cuts, the data stream categories and the results  
for the 1/3  data sample. The background  
levels given in Table~\ref{bifur_cuts} 
must  be scaled by a factor of 3 to obtain  estimates for 
the full sample. The 1/3 sample was studied first in order to  tune  
and optimize the  cuts. Then 
the 2/3 sample was used to give the final background  
estimates. When the normalization branch 
contained only one or few events  in region B, a  
second bifurcation analysis was performed in this branch to improve the  
statistics.  Details of these procedures are given  
in the following sections. 
\begin{table} 
\centering 
\begin{tabular}{l l l l c c c}\hline\hline 
Bkg.       & CUT1  & CUT2  &Category&$B$& $(C+D)/C$&$BC/D$\\ \hline  
$K_{\pi2}$ & PV    & KIN   &Skim 4&  
$0.39\pm0.11$&$85.2\pm2.5$     &$0.0046\pm0.0013$\\  
$K_{\mu2}$ & TD    & KIN   &Skim 5&  
$1.81\pm0.16$&$(4\pm1)\times 10^2$  &$0.0041\pm0.0011$\\  
$\mu^+$ band   & TD    & KIN   &Skim 5&  
$2.37\pm0.74$&$(4\pm1)\times 10^2$  &$0.0053\pm0.0021$\\  
Single-beam   & DC    & B4    &Skim 6&  
8             &$(7\pm4)\times 10^3$&$0.0011\pm0.0007$\\  
Beam $K^+-K^+$  & BWPC  & B4    &Skim 6&  
$0.04\pm0.04$&$117\pm37$   &$0.0003\pm0.0003$\\  
Beam $K^+-\pi^+$& BWPC  & B4    &Skim 6&  
$0.26\pm0.11$&$(7\pm4)\times 10^3$&$<0.0001$\\ \hline\hline 
\end{tabular} 
\caption{Results of the bifurcation analyses for the backgrounds 
in the 1/3 sample only.  
All the results for the region B were from the second  
bifurcations except for the single-beam background as explained in the text.   
Details for CUT1 and CUT2 are described in the text.  
Errors are statistical only.} 
\label{bifur_cuts} 
\end{table} 
 
\subsubsection{\KPITWO~Background} 
\label{sec:bkg_kp2} 
 
A \KPITWO~decay event should have a charged track with  
a monochromatic momentum, range and energy plus two photons. Experimentally, if 
a \KPITWO~event appeared in the signal region, the photons  
from the $\pi^0$ decay  
must have escaped detection and the \KPITWO~kinematics must have been 
distorted by scattering or resolution effects as well. 
 
In the background study, the two  
bifurcation cuts were chosen as the PV cuts (CUT1)  
and the signal phase space cuts in  
the KIN cuts (CUT2), since both of these  
could independently give  powerful 
rejection of  the \KPITWO~background. In order to remove  
the contamination from $\mu^+$ and beam events, 
the bifurcation analysis sample was selected from the Skim 4 sample 
by applying the TD cuts, the beam cuts and the KIN cuts  
other than the phase space cuts. 
In the rejection branch,  \KPITWO~events were selected by  
inverting the signal phase space cut in  
the KIN cuts ($\overline{\mbox{CUT2}}$),  
giving 95,797 events for the region C+D.  
The PV cuts (CUT1) were then applied to the remaining  
\KPITWO~events, leaving 1,124 events for the region C. 
In the normalization branch,  
the \KPITWO~events with photon activity 
were selected by inverting the PV cut ($\overline{\mbox{CUT1}}$). 
The signal phase space cuts in the KIN cuts (CUT2) were  
applied to the above selected \KPITWO~sample, resulting in no   
events ($B=0$) left in the normalization branch.  
 
\begin{figure} 
\centering 
\epsfxsize 1.0\linewidth 
\epsffile{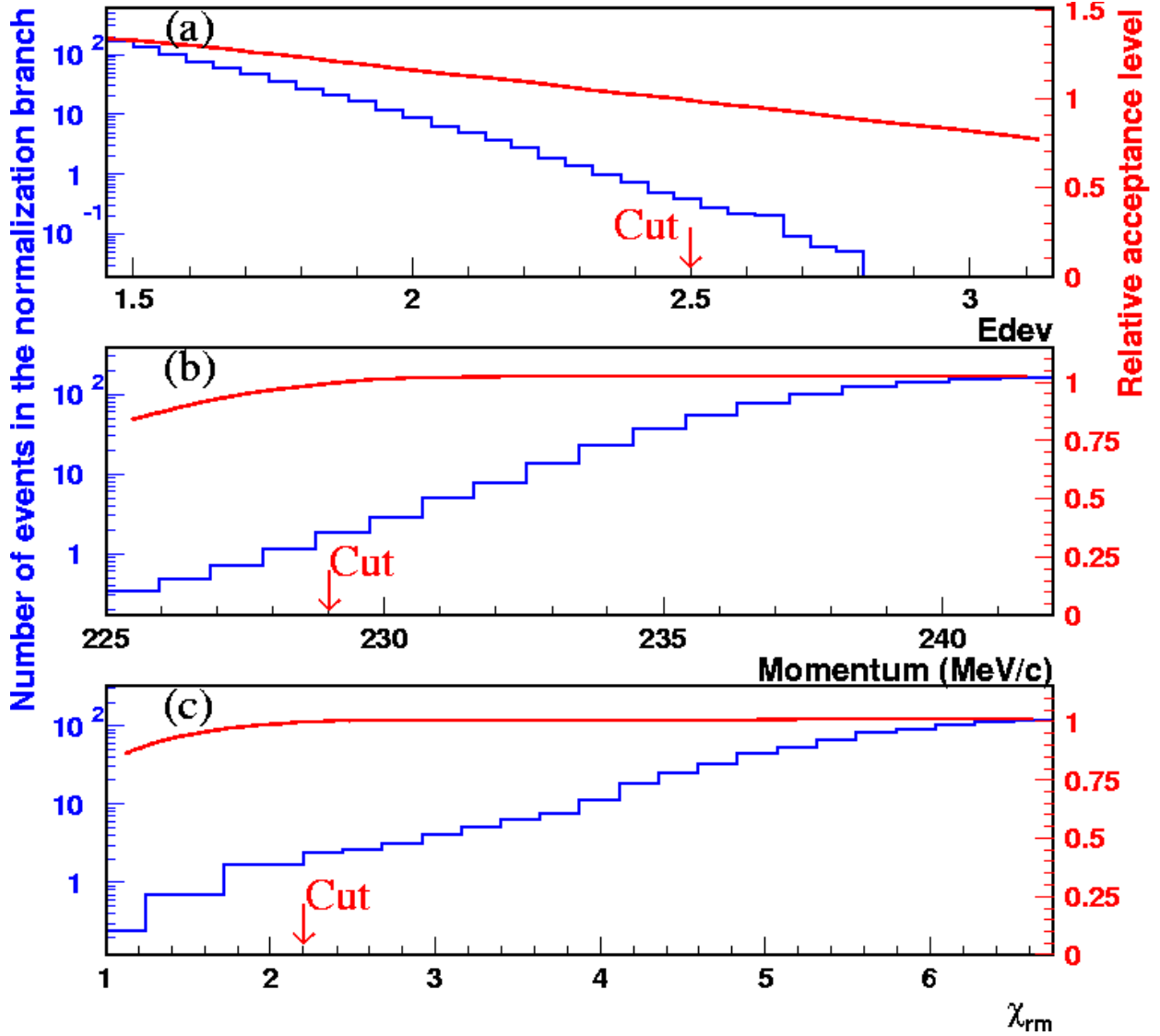} 
\caption{ 
The expected kinematic background events in the 1/3 normalization branch   
(histogram with left axis) as a function of  
$K_{\pi2}$ kinematic cut (Edev) position for the $K_{\pi2}$ background (a),  
maximum momentum cut position for the  
$K_{\mu2}$ range tail background (b) and  
range-momentum cut ($\chi_{rm}$) position for the $\mu^+$ band background (c). 
Also shown is the relative acceptance change as the cuts  
(curve with right axis).}  
\label{fig:kin_bg_func} 
\end{figure}

To deal with the above situation and give a non-zero events in  
the normalization branch, another (second) bifurcation analysis  
was performed by separating 
the $K_{\pi2}$ kinematic cuts into Edev (CUT1) and  
Rdev+Pdev cuts (CUT2), since the $E$ measurement was almost independent  
of $R$ and $P$ measurements.  
These two bifurcation cuts were applied  
sequentially to the selected $K_{\pi2}$ sample  
in the normalization branch. In this second bifurcation  
study, the lower boundary  
cuts on $E$, $R$ and $R$ were removed. Changing the cut  
positions on the Rdev, Pdev and Edev 
gave the number of events in the normalization branch ($N_{RP}$) and  
the rejection branch ($R_E$) in the second bifurcation analysis,  
which were then used to calculate 
the expected number of events in the normalization  
branch by means of $B = N_{RP}/(R_E-1)$. This second 
bifurcation method gave smaller uncertainty, providing a  
way to optimize the phase space cut positions to reject  
the $K_{\pi2}$ background with less acceptance loss.  
The expected number of events from the second bifurcation analysis  
was found to be $\sim\!10\%$ less than  
the observed number from the first bifurcation analysis in the region 
B+D. 
This was due to the small level of correlation between $R$ and $E$  
when estimating the $\pi^+$ range in the stopping layer from the  
measured energy in the stopping layer. The cut positions were 
chosen at $\mbox{Rdev}>2.75$, $\mbox{Edev}>2.5$ and $\mbox{Pdev}>2.5$,  
in order to reduce the expected $K_{\pi2}$ background level to about 
0.01 events level as shown in Table~\ref{bifur_cuts}. 
Fig.~\ref{fig:kin_bg_func} shows the expected $K_{\pi2}$ background  
from the second bifurcation in  
the 1/3 normalization branch as a function of the Edev cut position. 
The acceptance was measured using the simulated $K^+\to\pi^+\nu\bar{\nu}$  
sample.  
 
The expected number of events from the second bifurcation  
was used to give the result for the normalization branch, which 
was then used to give the estimated background level for the  
1/3 sample as shown in Table~\ref{bifur_cuts}.

\subsubsection{$\mu^+$ Background} 
 
The $\mu^+$ background consisted of the \KMUTWO~range tail events and 
the $\mu^+$ band events as indicated in Fig.~\ref{fig:r_vs_p_kp21}. 
These $\mu^+$'s lost energy and eventually came to rest in RS and   
could migrate into the signal region through  
resolution effects if the TD cuts failed.  
 
The two bifurcation cuts with the most powerful rejection  
were the TD cuts (CUT1) and the selected KIN cuts (CUT2),  
which excluded the fiducial cuts,  
the lower boundary in the signal phase space cuts,  
the cuts on the tracking quality in target and the  
cuts on the range-energy consistency in IC and target as detailed in  
Section~\ref{sec:kincuts}. 
The $\mu^+$ background sample used for the bifurcation study  
was selected from the Skim 5 sample 
by applying the PV cuts, the beam cuts and the KIN cuts not  
used in this bifurcation 
study, in order to remove the $K_{\pi2}$ and beam backgrounds. 
In the rejection branch, there were 
7,119 $\mu^+$ events in the region C+D when inverting  
the selected KIN cuts ($\overline{\mbox{CUT2}}$). 
The TD cuts (CUT1) were applied to the $\mu^+$ events,  
leaving 16 events in the region C. 
In the normalization branch,  
the $\mu^+$ events were obtained by inverting the TD cuts 
($\overline{\mbox{CUT1}}$). The selected KIN cuts (CUT2) were applied to 
the above selected $\mu^+$~sample, leaving only one 
event ($B=1$) in the normalization branch.  
 
The result from the bifurcation analysis given above had a large  
statistical uncertainty, since only one events remained in  
the normalization branch. The \KMUTWO~range tail and the  
$\mu^+$ band events were the only two possible $\mu^+$ backgrounds  
in the signal region. The origins of  
these two backgrounds were due to  
the momentum and range resolution effects.  
Once these resolution  
effects were  known  in the signal region, a better estimate of corresponding  
events $B$ in the normalization branch could be obtained. In this analysis,  
the events with $P>225$~MeV/$c$ were regarded as  
the \KMUTWO~range tail background. 
 
The momentum resolution effects could be well described by using    
the momentum distribution from the $K_{\mu2}$ peak events.   
In order to enhance the number of events in  
the normalization branch, the RS energy loss cuts were 
removed from the KIN cuts. Also removed were the upper boundary 
cuts on the $R$, $E$ and $P$. The $K_{\mu2}$ peak 
events were selected by requiring $R>50$~cm. 
The selected $K_{\mu2}$ peak events in the $\pi\nu\bar{\nu}(1)$ trigger 
were found to have longer range in the target, leading to a bias of 0.5 
MeV/$c$ higher momentum measurement when applying a $\pi^+$ hypothesis  
to the contribution from the energy loss in the target. After  
subtracting this bias for each $K_{\mu2}$ peak event, the momentum 
distribution was seen to be in good agreement  
with that from the $K_{\mu2}$ range tail events and with more  
statistics in the signal region defined below 229~MeV/$c$.  
Normalizing the number of $K_{\mu2}$ peak events to that  
of the $K_{\mu2}$ range tail events observed in the region  
B+D gave the expected  
number of events $B$ in the normalization branch. The expected  
$K_{\mu2}$ range tail background is given in Table~\ref{bifur_cuts}. 
Fig.~\ref{fig:kin_bg_func} shows the number of $K_{\mu2}$  
range tail background  
events as a function of the maximum momentum cut position  
in the normalization branch.

The $\mu^+$ band background came from the $RP$ 
resolution effects, which were well described by the  
range deviation ($\chi_{rm}$) in RS.   
Enhancing the number of events in the study of the $RP$ 
resolution effects was achieved by 
removing the PV cuts and the RS energy loss cuts in the normalization branch 
sample, since they were not correlated with the range and momentum 
measurements. Within the statistical uncertainty,  
both distributions were seen to be consistent except that 
the distribution without the above requirements gave higher statistics in  
the normalization branch (Fig.~\ref{fig:kin_bg_func}). 
The estimated background level for  
the $\mu^+$ band is given in Table~\ref{bifur_cuts}.

\subsubsection{Single Beam Background} 
 
In the study of single beam background, 
the two bifurcation cuts chosen were the offline  
delayed coincidence (DC) cuts (CUT1) and the B4 energy loss cut (CUT2). 
The offline DC cuts were from the precise offline time measurements  
from the beam instrumentation,  
the target, the IC and the RS (They were not the ones used in the trigger). 
The events were selected from the Skim 6 sample  
by applying the PV cuts, the KIN cuts, the TD cuts, and  
the beam cuts except for the DC cuts 
and the B4 energy loss cut. 
In the normalization branch,  
the single beam events were selected by inverting the DC cuts 
($\overline{\mbox{CUT1}}$). The the B4 energy loss cut (CUT2)  
was applied to the above selected single beam sample, leaving 8 
events in the region B for the normalization branch.  
In the rejection branch, 29,100 single beam events in the region C+D 
were selected by inverting the B4 energy loss cut  
($\overline{\mbox{CUT2}}$). 
The DC cuts  were applied to these selected events,  
resulting in 4 events in the region C. 
The rejection factor was applied for both the $\pi^+$  
scattering events and the $K^+$ decay-in-flight events, 
since there was no reason to have  different rejections for  
these.  
The background estimate for the single beam background  
is given in Table~\ref{bifur_cuts}.  
 
\subsubsection{Double Beam Background} 
 
As already defined in Section~\ref{sec:bkg_overview},  
the double beam background could be due to a $K^+-K^+$ event 
or a $K^+-\pi^+$ event. 
For this double beam background, the DC cuts  were insufficient to remove the  
double beam background, but the time difference between the  
beam instrumentation 
($C_K$, $C_\pi$, BWPC's and B4) and the $\pi^+$ track 
was a  good indicator. Another independent way was  
to use the target pattern to identify    
extra particles other than the initial $K^+$ hit. 
 
The two bifurcation cuts chosen were the  
BWPC timing cuts (CUT1) and the B4 timing cuts (CUT2). 
The events were selected from the Skim 6 sample  
by applying the PV cuts, the KIN cuts, the TD cuts and  
the beam cuts except for the BWPC  
and the B4 timing cuts. These cuts removed the \KPITWO,  
$\mu^+$ and single beam backgrounds.  
In the normalization branch, there were no events left  if the  
B4 timing cuts (CUT2) were applied. In order to give a more 
precise estimate, the second bifurcation was adopted.  
Time measurements including the trailing edge TDC time from  
the $C_K$ and  
$C_\pi$ were used to select $K^+-K^+$ events and $K^+-\pi^+$,  
separately. The bifurcation analyses were then performed using  
the B4 timing cuts and the target pattern recognition cuts. 
Results are given in Table~\ref{bifur_cuts}. 
In the rejection branch, the double beam background  
events were selected by inverting the B4 timing cuts  
($\overline{\mbox{CUT2}}$). The $K^+-K^+$ and $K^+-\pi^+$ events 
were tagged by the $C_K$ and $C_\pi$, separately; this 
resulted in 1,170 and 22,150 events for both cases.  
Applying the BWPC timing cuts (CUT1) resulted in 10 and 3  
events observed for the $K^+-K^+$ and $K^+-\pi^+$ backgrounds, respectively.  
The resulting rejections and background estimates  
are given in Table~\ref{bifur_cuts}. 
The $K^+-K^+$ background was found to 
dominate the double beam background.  
 
\subsubsection{Charge Exchange Background} 
\label{sec:cex} 
 
Since there was no reliable way to isolate the CEX events from the  
$\pi\nu\bar{\nu}(1)$ trigger data,  
the background study could only rely on the Monte Carlo simulation. 
The CEX simulation needed a number of inputs, such as 
the CEX re-generation rate as a function of $K^0_L$ energy,  
the $K^0_L$ decay vertex and the $K^0_L$ momentum,  
all of which could only be obtained from the real data.  
A special CEX monitor trigger as described in Section~\ref{sec:moni} 
was used for collecting data with two  
charged tracks from the $K^0_S$ decay.  
The $K^0_S$'s were reconstructed in the $\pi^+\pi^-$ decay mode and  
used to measure the $K^0_S$ production rate, momentum spectrum,  
the B4 hit information, decay vertex 
distribution, and pattern of target $K^+$ fibers' time and energy. 
Since a $K^0$ decays approximately equally  to $K^0_L$ and $K^0_S$ states, 
the measured rate of $K^0_S$ decays can be used to obtain the  
$K^0_L$ production rate in the target 
\begin{eqnarray} 
R_{K^0_L}&\equiv&  
                    \frac{N_{K^0_S}}{\epsilon_{K^0_S}\cdot A_{PV}\cdot{\cal B} 
                    (K^0_S\to\pi^+\pi^-)\cdot N_K/PS},\\ 
            &=     &2.73\times 10^{-5}, 
\end{eqnarray} 
where the quantities used in this calculation  
are summarized in Table~\ref{tab:Ks_parameters}. 
\begin{table} 
\centering 
\begin{tabular}{l l c}\hline\hline 
Description & Parameter & Values \\ \hline 
Number of selected $K^0_S$ events        & $N_{K^0_S}$              & 8,086 \\ 
 $K^0_S$ selection efficiency        & $\epsilon_{K^0_S}$       & 0.138 \\ 
       Photon Veto efficiency        & $A_{PV}$                 & 0.680 \\ 
$K^0_S\to\pi^+\pi^-$ Branching Ratio & \BR($K^0_S\to\pi^+\pi^-$)& 0.686 \\ 
Number of $K^+$ Triggers ($10^{12}$)     & $N_K$              & 1.77 \\ 
Prescaling Factor                    & $PS$                     & 384 \\ \hline 
$K^0_L$ Production Rate              & $R_{K^0_L}$ 
& $2.73\times10^{-5}$ \\ \hline\hline 
\end{tabular} 
\caption{$K^0_L$ production rate and the quantities that  
are used to estimate it.} 
\label{tab:Ks_parameters} 
\end{table} 
 
Both $K^0_L\to\pi^+\mu^-\bar{\nu}_\mu$ 
($K^0_{\mu3}$) and $K^0_L\to\pi^+ e^-\bar{\nu}_e$ ($K^0_{e3}$) decays  
were simulated to estimate the CEX background level.  
Each decay mode was generated with the amount of $K^0_L$ decays equivalent to  
$7.604\times 10^{15}$ $K^0_L$'s.   
These Monte Carlo events were passed through all of the selection  
criteria, except for the \PIMUE~decay sequence  
cuts and the beam cuts that were not related to 
 target quantities. There were 56 $K^0_{e3}$ events 
and 21 $K^0_{\mu3}$ events surviving in the signal region. Therefore,  
the expected CEX events can be estimated by means of  
\begin{eqnarray} 
N_{CEX} & = & \left( N^{K^0_{e3}}_{pass} +N^{K^0_{\mu3}}_{pass}  
                 \right)\times\frac{N_K}{N_K^{MC}} 
		 \times F_{acc} \nonumber \\ 
        & = & \left(56 + 21  
                \right)\times\frac{1.77\times 10^{12}}{7.604\times 10^{15}} 
                \times 0.250 \nonumber \\ 
        & = & 0.00448\pm0.00051(stat), 
\label{cex_bkg} 
\end{eqnarray} 
where $N^{K^0_{e3}}_{pass}$ and $N^{K^0_{\mu3}}_{pass}$  
are the numbers of $K^0_{e3}$ and $K^0_{\mu3}$ events  
surviving all the cuts, $N_K^{MC}$ was the total exposure of  
generated $K^+$'s and $F_{acc}$ was the acceptance 
of the TD and beam cuts that were not applied to in the  
simulation.  
 
\subsubsection{Initial Background Evaluated from 1/3 Sample} 
\label{sec:total_bkg} 
 
The initial total background evaluation based on the  
1/3 sample was $0.05\pm0.01_{stat}$ events,  
which came from the results in Table~\ref{bifur_cuts}  
and Equation~(\ref{cex_bkg}).   
Given this relatively low background level, the signal region was expanded  
to gain acceptance at the cost of more background. 
In addition to the total  
background level, the bifurcation analyses  gave  
the corresponding predicted background functions for the TD, the PV and the  
kinematic cuts, with which  the expected  
background level and relative acceptance change for a given 
cut position were obtained. These functions were used in optimizing the  
selection criteria, studying the correlation of two bifurcation 
cuts and determining the branching ratio as well.    
 
\subsubsection{Optimization of Signal Region} 
\label{sec:extended} 
 
The distributions of signal and background in the cut space were 
described well by the predicted background functions and were 
used to enlarge the signal region. 
The cuts to be loosened were the NN $\pi^+\to\mu^+\to e^+$ decay cut  
in the TD cuts (Fig.~\ref{fig:nn_func}), the  
PV cuts (Fig.~\ref{fig:pv_tune_rva}) and the $K_{\pi2}$ kinematic  
cuts (Fig.~\ref{fig:kin_bg_func}).  
Loosening the cuts for the beam backgrounds and the other TD and KIN cuts  
did not provide much acceptance gain.  
Simultaneously loosening the cut positions  
of the NN $\pi^+\to\mu^+\to e^+$ decay, PV and \KPITWO~kinematic 
cuts could increase the background levels to an unacceptable level.  
Instead, when one of the three cuts was loosened, the cut positions  
of the other two were kept unchanged.  
The revised (extended)   
signal region consisted of the standard signal region  
plus three extensions. 
Hereafter, the revised and standard signal regions were 
referred to as the extended signal region 
and the standard region, respectively.   
The regions created by loosening the NN $\pi^+\to\mu^+\to e^+$ decay cut,  
the PV cuts, and the \KPITWO~kinematic cuts 
were referred to as $\pi^+\to\mu^+\to e^+$ extended, PV extended,   
and \KPITWO~kinematic extended regions,  
respectively.  All of which included the standard regions. 
The loosening factors for these three cuts,  
$f_{\pi\mu e}=4.2$, $f_{PV}=4.0$ and $f_{K_{\pi2}}=9.5$,  
meant the corresponding background increased by the same factor. 
The total estimated acceptance gain by enlarging the signal region  
was 31\% (12\% from the NN $\pi^+\to\mu^+\to e^+$ decay cut,  
7\% from the PV cuts,  
and 12\% from the \KPITWO~kinematic cuts).

\subsubsection{Correlation and Single Cut Failure Study} 
\label{sec:outside_the_box} 
 
The bifurcation procedure described above assumed that the  
two bifurcation cuts were not correlated.  
This assumption was tested by comparing the predicted and observed rates  
near but outside the signal region. 
\begin{figure} 
\centering 
\epsfxsize 0.9\linewidth 
\epsffile{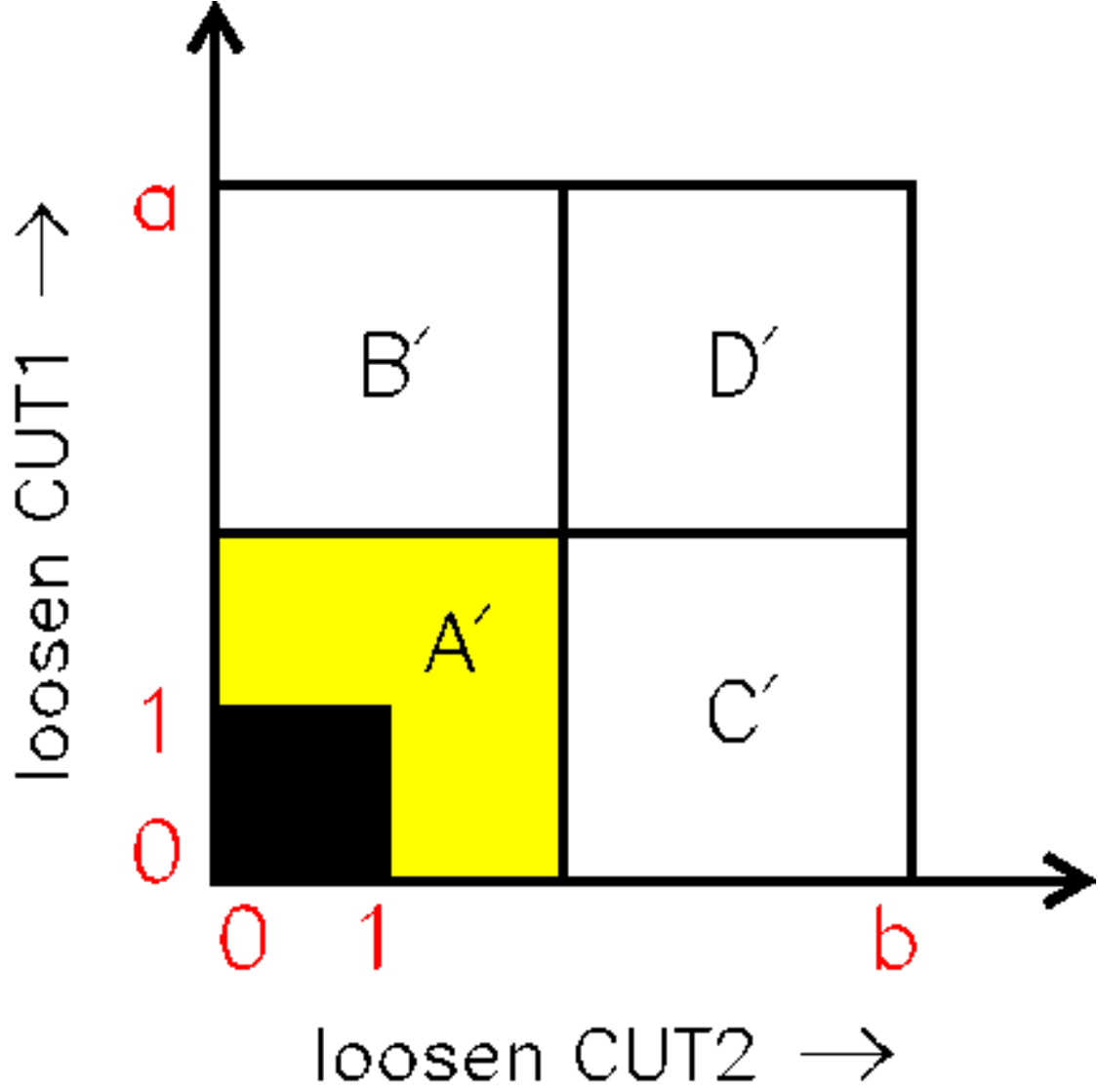} 
\caption{Pictorial explanation of the correlation study using the events near, 
but outside the signal region.  
The vertical axis ``$a$'' and the horizontal axis ``$b$''  
are for the loosening factors, while the $1\times1$ black region  
represents the standard signal region.} 
\label{fig:explain_osb} 
\end{figure} 
 
A schematic representation of the region near, but outside, the  
signal region is shown in Fig.~\ref{fig:explain_osb}. 
A near region outside the signal region  
(${\rm A}'$ without the black region  
in Fig.~\ref{fig:explain_osb}) was defined by 
loosening two bifurcation cuts (CUT1 and CUT2) simultaneously  
by the fixed  factors $a$ and $b$, respectively. 
If the predicted background functions associated with the  
CUT1 and CUT2 were correct, the expected number of 
background events in the near region ($BG^\prime$) was estimated by 
the  bifurcation method as  
\begin{equation} 
BG^\prime = B^\prime C^\prime/D^\prime - BC/D  
\label{eq:bifurcation} 
\end{equation} 
If a deviation was seen between the observed and the predicted numbers 
in the near region, 
then a correlation between the bifurcation cuts could be indicated  
and the background estimate might be unreliable. 
The observation was performed 
in the same way as the bifurcation method used in the  
background estimation in the signal region, 
except that two bifurcation cuts were loosened by factors of $a$ and 
$b$.  
 
The results of the correlation study for the  
\KPITWO, \KMUTWO~range tail and $\mu^+$ band backgrounds are  
summarized in Table~\ref{tab:outside2} for the 2/3 data sample.  
Good agreement was found between the observed and the predicted  
number of events. 
The test results were obtained from the comparisons between  
the observed and the predicted numbers of events using   
the predicted background function method.

In addition to the study of correlation between the cuts used for 
the bifurcation method, events that passed all except for a single cut  
were examined  
to determine if each cut operated as designed for the appropriate background 
mechanism. Such a study  provided a way to discover any new type of 
background or potential analysis flaw.  
In the 1/3 sample, six of the eight  
events that failed a single cut only 
were far from the cut position, while the other two events showed  
potential analysis flaws. The first flaw  
would artificially increase the measured range and 
momentum of $\pi^+$'s from \KPITWO\ decays that exited the  
upstream end of the target through the gap between the front face of 
the target and the B4 hodoscope. 
Additional cuts with minimal  
acceptance loss were devised to eliminate such events.  
The second revealed a possible correlation  
between the PV cuts and the KIN cuts when using the $\pi^+$ 
polar angle ($\theta$) as a reference to exclude the  
accidental hits in the opposite site of the PV counters.    
The corresponding calculation used in the PV cuts was subsequently removed.   
All the cuts designed at this stage  
were referred to as pathology cuts as  described  
in Section~\ref{sec:beam_cuts}.

\subsubsection{Final Background Evaluated from 2/3 Sample} 
\label{sec:bkg_level} 
 
Evaluation of the final background levels came from the 2/3 
sample. To get the values for the extended box, the corresponding  
values were scaled by the loosening  factors given in  
Section~\ref{sec:extended}.  
It was noted that loosening the NN $\pi^+\to\mu^+\to e^+$ decay cut,  
the PV cuts, and the \KPITWO~kinematic cuts could lead to a small change  
in the beam background levels. Therefore, the beam backgrounds were 
re-estimated in the extended signal region.  
The TD rejection  
and PV rejection were measured to be $445\pm111$ and $84.3\pm1.2$ with  
the 2/3 sample in good agreement with  
those obtained from the 1/3 sample given in Table~\ref{bifur_cuts}.  
The final background estimates   
are summarized in Table~\ref{tab:total_bkg_level}. 
 
The total background level in the extended signal region  
was estimated to be $0.30\pm0.03_{stat}$ events, which was dominated by 
the \KPITWO~background contribution. 
As the background distribution was not uniform in the  
signal region, the predicted background functions obtained in  
the background study were exploited to interpret any possible  
candidate events observed and to give a proper branching ratio 
measurement by using the likelihood technique  
described in Section~\ref{chap:results}. 
\begin{table} 
\centering 
\begin{tabular}{l c c}\hline\hline 
Background & Standard        & Extended  \\ \hline 
\KPITWO    & 0.019$\pm$0.004 & 0.216$\pm$0.023\\  
\KMUTWO~range tail     
           & 0.010$\pm$0.001 & 0.044$\pm$0.005\\  
$\mu^+$ band 
           & 0.005$\pm$0.002 & 0.024$\pm$0.010\\  
Single-beam     & 0.004$\pm$0.002 & 0.006$\pm$0.002\\  
Double-beam     & 0.003$\pm$0.002 & 0.003$\pm$0.002\\  
CEX        & 0.004$\pm$0.001 & 0.005$\pm$0.001\\ \hline 
Total      & 0.05$\pm$0.01 & 0.30$\pm$0.03\\ \hline \hline  
\end{tabular} 
\caption{Total background level and the contribution  
from each background source for both standard and extended  
signal regions as estimated from the 2/3 data samples.  
The errors are statistical.} 
\label{tab:total_bkg_level} 
\end{table}

\subsubsection{Systematic Uncertainty} 
\label{sec:syst} 
 
Systematic uncertainty in the background estimates  
arose from the possible correlation 
between the two bifurcation cuts. The correlation   
was investigated using the 2/3 sample and  the results are given  
in Table~\ref{tab:outside2}. In addition,  
the ratios of observations over predictions were used to  
quantify the degree of consistency and  
were found to be consistent with unity within a relative uncertainty of 15\%,  
confirming that the background estimations obtained with the bifurcation  
method were reliable. 
 
\begin{table} 
\centering 
\begin{tabular}{l c c c c c}\hline\hline 
\KPITWO
& \multicolumn{5}{c}{(PV cuts)~$\times$~($K_{\pi2}$ kinematic cuts)}\\ \hline 
Loosening factor    & 10$\times$10   & 20$\times$20    
                 & 20$\times$50   & 50$\times$50   & 50$\times$100 \\ \hline 
Prediction   & 1.1$\pm$0.2    & 4.9$\pm$0.6  
                 &12.4$\pm$1.3    &31.1$\pm$3.1    & 62.4$\pm$5.6 \\ 
Observation      & 3              & 4  
                 & 9              & 22             & 53 \\ \hline\hline 
\KMUTWO~Range Tail 
& \multicolumn{5}{c}{($\pi^+\to\mu^+\to e^+$)~$\times$~(Maximum momentum cut)} \\ \hline 
Loosening factor    & 10$\times$10   & 20$\times$20  
                 & 50$\times$50   & 80$\times$50   & 120$\times$50 \\ \hline 
Prediction   & 0.4$\pm$0.0    & 1.4$\pm$0.1  
                 & 9.1$\pm$0.6    &14.5$\pm$1.0    &21.8$\pm$1.5 \\ 
Observation      & 0              & 1   
                 & 12             & 16             & 25 \\ \hline\hline 
$\mu^+$ Band 
& \multicolumn{5}{c}{($\pi^+\to\mu^+\to e^+$)~$\times$~(Range-momentum cut)} \\ \hline 
Loosening factor    & 10$\times$10   & 20$\times$20  
                 & 50$\times$20   & 80$\times$20   & 80$\times$40 \\ \hline 
Prediction   & 0.3$\pm$0.1    & 1.3$\pm$0.4 
                 & 3.2$\pm$0.9    & 5.2$\pm$1.5    &10.4$\pm$2.8 \\ 
Observation      & 1              & 1 
                 & 4              & 5              &11 \\ \hline\hline 
\end{tabular} 
\caption{Results of a correlation study for the 
\KPITWO~(top), \KMUTWO~range tail (middle) 
and $\mu^+$ band (bottom) backgrounds in the 2/3 sample.  
The errors in the predictions are statistical uncertainties.} 
\label{tab:outside2} 
\end{table} 
 
\subsection{Acceptance and Sensitivity}  
\label{chap:acc_and_sen} 
 
To reduce the estimated background level to less than one event in the  
signal region, this analysis utilized many selection criteria. 
The corresponding acceptances for these selection criteria 
were estimated directly from the data when possible,  
by splitting them into components  
that could be measured separately using the monitor trigger data or the Monte  
Carlo simulation. The latter gave the  
estimates on the decay phase space, the trigger efficiency 
and the nuclear interaction effects.  
 
\subsubsection{Acceptance Factors from \KMUTWO~Events} 
\label{chap:acc_kmu2} 
 
Since the \KMUTWO~events have the same features as the  
signal regarding the $K^+$ beam, the charged track 
and the event topology, the acceptances associated with the  
relevant cuts as listed in Table~\ref{tab:acc_kmu2_all} 
were directly measured using the $K\mu2$ monitor  
trigger data. Below are the details of these measurements. 
\begin{table} 
\centering 
\begin{tabular}{l c}\hline\hline 
Cut & Acceptance \\ \hline 
Tracking in RS	&0.99996$\pm$0.00001 \\  
Tracking in UTC and target    &0.99568$\pm$0.00010 \\  
Beam selection criteria	        &0.50779$\pm$0.00074 \\  
Photon veto 	        &0.76784$\pm$0.00218 \\  
Track stop in RSSC	&0.98195$\pm$0.00015 \\  
Muon veto in RS        &0.99591$\pm$0.00057 \\ \hline 
$A_{K_{\mu2}}$          &0.3796$\pm$0.0013 \\ \hline\hline 
\end{tabular} 
\caption{Acceptances of the \KPPNN~selection cuts measured  
from the \KMUTWO~monitor trigger data.  
The acceptance of beam cuts does not include those using  
the energy measurement in the target.  
The errors are statistical.} 
\label{tab:acc_kmu2_all} 
\end{table} 
 
{\it Tracking~in~RS:} 
To measure the acceptance of RS tracking, additional requirements 
(``set-up cuts'') were applied to $K\mu2$ monitor data  
to ensure  
a  good track in the RS without using  RS measurements. The  
events were required to  have an IC hit with $t_{IC}-t_{C_{K}}>5$ ns,  
at least 1.2~MeV energy loss in B4, and  
successful tracking in both the UTC and target. All surviving   
\KMUTWO~events were examined for  consistency  with  RS tracking.  
The acceptance of the RS tracking cuts is given in the 
second row of Table~\ref{tab:acc_kmu2_all}.

{\it Tracking~in~UTC~and~Target:} 
Because there should be no photon activity for \KMUTWO~peak events,  
the $K\mu2$ monitor was an ideal sample to measure 
the acceptance of target pattern recognition criteria. The sample was  
taken from events surviving the RS tracking cuts discussed above. To eliminate  
possible beam background contamination, this sample was also required to  
meet  the timing requirements on the beam instruments. A  
5 ns timing consistency was also required between $t_{IC}$ and $t_{rs}$. 
A subset of the PV cuts was applied to suppress possible  
$K_{\mu2\gamma}$ contamination. The BV and BVL elements of the PV cuts were 
not applied to avoid self-vetoing by long range \KMUTWO~events.  
Events surviving the  set-up cuts were then checked  
with the UTC and target requirements except for those involved  
$\pi^+$ energy and range measurement in the target, giving a measurement of 
the acceptance of tracking in UTC and target(Table~\ref{tab:acc_kmu2_all}).

{\it Beam~selection~criteria:} 
The \KMUTWO~events were chosen from those satisfying    
the requirements on tracking in the UTC and target described above but 
without the timing requirements on the beam instruments.  
To suppress beam background  
contamination, the momentum deviation was required to be within 
two standard deviations of the \KMUTWO~peak with $|$cos$\theta|<0.5$. 
Also required was that there should be no discernible  
scattering of tracks in the RS. 
The remaining events then passed through  
the beam cuts except for the pathology cuts using $\pi^+$ energy and range measurement  
information described in Section~\ref{sec:beam_cuts},  
providing a measurement of the corresponding beam  
selection acceptance. Because $K_{\mu2}$ events were simple single tracks, 
the efficiency of the DC trigger was  
also measured and included in the acceptance  
of beam selection criteria (Table~\ref{tab:acc_kmu2_all}).

{\it Photon~veto:} 
The acceptance of the PV cuts included contributions from 
both the online and  offline PV. 
Ideally, \KMUTWO~events should not contain any photons, and could thus  
be used to measure the acceptance loss due to the PV   
cuts. In the first step of this procedure, the selection criteria   
were  applied to  
remove possible beam backgrounds. However, it was noted that some $\mu^+$'s  
could penetrate the whole RS and reach the PV counters, 
resulting in a time-coincident PV hit and therefore an over-counting in  
the acceptance loss due to the PV cuts. To avoid this problem, 
the selected \KMUTWO~sample was further required to have  
the stopping layers prior to the $19^{\rm th}$ RS layer. The PV cuts  
were then applied to the  surviving \KMUTWO~events, yielding 
a measurement of the acceptance loss due to the application of  
the PV cuts. 
Since the $K\mu2$(1) trigger did not have an online PV cut applied,  
this acceptance factor also included the contribution from the  
online PV (Table~\ref{tab:acc_kmu2_all}).

{\it Track~stop~in~RSSC:} 
This cut was classified into the fiducial cuts in  
Section~\ref{sec:kincuts} and aimed at vetoing possible  
associated photon activity detected by the RSSC, even though 
it was not included in the PV cuts.   
Using the $K\mu2$ monitor events and  
applying the above cut to those with hits in the second  
layer of RSSC gave a measurement of the corresponding  
acceptance value (Table~\ref{tab:acc_kmu2_all}).

{\it Muon~veto~in~RS:} 
The $\pi\nu\bar{\nu}(1)$ trigger condition $\overline{19_{ct}}$ 
was also called as a muon veto in RS. 
This trigger requirement could result in acceptance losses when an 
accidental hit happened in the $19^{\rm th}$ layer  
along with an  otherwise good signal 
candidate event. This loss was measured with the  
$K\mu2$ monitor events, which were selected by  
requiring the stopping layer to be RS layer 17 and the range to be longer  
than 40~cm, in addition to applying all 
the cuts used in the above studies except for the momentum cut.  
In this selected sample, the online trigger condition 
$\overline{19_{ct}}$ 
was checked, giving a measurement of  
the acceptance for the muon veto in RS (Table~\ref{tab:acc_kmu2_all}). 
 
\subsubsection{Acceptance Factors from \KPITWO~Events} 
\label{chap:acc_kpi2} 
 
The $K\pi2(1)$ monitor trigger data were used to measure  
the acceptances for the pathology beam cuts involving the $\pi^+$ energy and 
range measurement and the KIN cuts involving the range-energy consistency in IC 
and target. This was a complement to the measurement of acceptance factors from 
the $K_{\mu2}$ events. The \KPITWO~events  
were selected with all cuts applied except for those to be measured. 
To ensure good \KPITWO~events, the momentum, range and energy  
were required to be within two standard deviations of the \KPITWO~peak  
positions and  observation of a $\pi^0\to\gamma\gamma$ decay was required.    
The result was 
\begin{equation} 
A_{K_{\pi2}} =  0.8785\pm0.0029_{stat}.  
\end{equation} 
 
\subsubsection{Kinematic Acceptance from Beam $\pi^+$ Events} 
 
The $\pi_{scat}$ monitor trigger data provided a pure  
$\pi^+$ sample to measure the KIN cuts  
related to the particle type: 
the cut on $\pi^+$ stopping layer in the fiducial cuts,  
the cuts on tracking quality in UTC and RS,  
the cuts energy loss in RS and the cut on range-momentum  
consistency in UTC and RS.  
The events were required to pass the Pass 1 cuts and  
the TD cuts. The $K^+$ selection criteria in the beam instruments  
were inverted to select beam $\pi^+$'s.  
The $t_{IC}$ was required to be within $\pm5$ ns of  
$t_{rs}$. The signal phase space cuts were  
additionally applied to select the events. The acceptance was measured to be  
\begin{equation} 
A_{\pi_{scat}}=0.6161\pm0.0085_{stat}\pm0.0189_{sys}. 
\end{equation} 
Since these $\pi^+$'s came from the beam $\pi^+$'s scattering in  
the target and not from the $K^+$ decays at rest,  
classification of the $K^+$  
fibers and $\pi^+$ fibers could be complicated because  
of their nearly coincident times and differences in fiber energy deposits  
of scattered $\pi^+$'s and $K^+$ decays at rest. 
Both of these features would  
result in more uncertainties in the momentum, energy and range  
measurements in these $\pi^+$ events. Systematic uncertainties  
were therefore investigated  
by loosening or tightening the signal phase space cuts by $\pm$1 
standard deviation. The corresponding  
variation in acceptance was treated as the systematic  
uncertainty.

\subsubsection{\PIMUE~Decay Acceptance from Beam $\pi^+$ Events} 
 
The acceptance of \PIMUE~decay sequence cuts was 
measured by using the $\pi_{scat}$  
monitor trigger data. This acceptance measurement included the  
online and offline (TD) $\pi^+$ identification cuts. The online ones  
included $L1.1$ and $L1.2$ in trigger. The sample was selected  
using the same cuts as those used in  
measuring $A_{\pi_{scat}}$ except for the cuts to be measured here.   
It should be noted that 
this measurement included the acceptance loss due to  
the $\pi^+$ absorption and $\pi^+$ decay-in-flight.  
This loss was estimated to be 1.4\% using Monte Carlo in  
Section~\ref{sec:acc_mc} and should be corrected to remove 
the double counting problem in the acceptance. 
The final acceptance of \PIMUE~decay sequence cuts was given below 
\begin{equation} 
A_{\pi\to\mu\to e}=0.3523\pm0.0077_{stat}\pm0.0067_{sys}. 
\end{equation}  
Since the \PIMUE~decay sequence cuts might be correlated to  
particle identification KIN cuts when using  
information from the RS, the effect on acceptance was  
investigated with and without 
the RS energy loss cuts. The observed about 2\% variation on the  
relative acceptance was  
assigned as the systematic uncertainty. 
 
\subsubsection{Acceptance Factors from Monte Carlo Simulation} 
\label{sec:acc_mc} 
 
Monte Carlo simulations of the \KPPNN~ were used to evaluate   
the trigger acceptance, the phase space  
acceptances and the acceptance loss 
due to $\pi^+$ absorption, decay-in-flight, and nuclear interaction,  
which could not be measured directly by the monitor trigger data.  
The \KPPNN~Monte Carlo samples were generated with and without  
including the nuclear interaction.

{\it Trigger~Requirements:} 
All the trigger requirements as described in  
Section~\ref{sec:trigger} were simulated by Monte Carlo except for  
the $DC$, $L1.1$ and $L1.2$, which were already measured 
using the $K_{\mu2}$ and $\pi_{scat}$ monitor trigger data. 
From the Monte Carlo without including the nuclear interaction,  
the acceptance for the trigger requirements was  
\begin{equation} 
A_{trig}=0.1796\pm0.0010_{stat}\pm0.0084_{sys}, 
\label{eq:acc_trig} 
\end{equation} 
where the systematic uncertainty was estimated to be 4.7\%  
from the measurement on the branching ratio of $K_{\pi2}$ in 
Section~\ref{sec:brkp2}. 
It was noted that the trigger acceptance measured here was 
primarily due to  geometry.

{\it Phase~Space:} 
The phase space acceptance was used to determine  the acceptance of 
offline cuts on the  
momentum, range and energy.  
To measure the phase space acceptances ($A_{PS}$), events   
were first taken from those surviving from the trigger in the Monte Carlo 
simulation, and then  the phase space cuts were applied.  
The $\pi^+$ nuclear interaction was  not included in this 
simulation. The acceptance was measured to be 
\begin{equation} 
A_{PS} 	= 0.3630\pm0.0029_{stat}, 
\end{equation}  
which included the loss due to both  $\pi^+$ 
absorption and decay-in-flight.

{\it Correction~for~Nuclear~Interaction:} 
The nuclear interaction effect was investigated separately, in order  
to study the systematic uncertainty associated with it.  
\KPPNN~events were generated with and without nuclear interaction, 
respectively.  
The ratio between the trigger acceptances multiplied by  
the ratio between the  phase space acceptances gave 
the correction for nuclear interaction  
\begin{equation} 
A_{nucl.}=0.4953\pm0.0077_{stat}\pm0.0248_{sys}, 
\end{equation} 
where the systematic uncertainty took into account the  
observed 0.15~cm difference on the range resolution (Table~\ref{tab:res_pre}).  
This difference could affect the acceptance due to the Rdev cut 
and translated into a 5\% uncertainty in the  
acceptance. It should be pointed that this definition took into account  
loss associated with  nuclear interactions such  
 extra energy in the  
detector associated with  nuclear interactions that caused 
the PV counters to fire. This loss was not included in the  
$K_{\mu2}$-based acceptance $A_{K_{\mu2}}$ given in  
Table~\ref{tab:acc_kmu2_all}  because muon-nuclear interactions are rare.

\subsubsection{Correction to $T\cdot2$ Trigger Inefficiency} 
 
The $T\cdot2$ trigger required coincident hits both in  
the inner-most two RS layers  
and in the IC. It was noted that the $T\cdot2$ simulation  
result did not include  
the acceptance loss due to the geometrical and counter  
inefficiencies of the T-Counters.  
The geometrical inefficiency was 
due to tracks passing through azimuthal gaps 
between adjacent T-Counters.  
The counter inefficiency occurred if the scintillation  
light induced by the charged  
track was not detected by the PMT's.  
This inefficiency was measured by using  
the \KMUTWO~and \KPITWO~events in the  
$KB$ monitor data. UTC track extrapolation was required to give  
the expected $T\cdot2$ counter.   
The online $T\cdot2$ trigger condition of the reconstructed event data  were 
checked to measure the $T\cdot2$ efficiency for  
\KMUTWO~and \KPITWO~events, separately. Since the energy  
losses in the $T\cdot2$ counter from  
the \KMUTWO~and \KPITWO~events  
were different, simulations were done for  these 
decay modes to obtain the average energy loss for \KMUTWO~and \KPITWO. Using  
an  energy extrapolation gave the correction to the $T\cdot2$  
inefficiency for the signal 
\begin{equation} 
A_{T\cdot2} = 0.9358\pm0.0011(stat)\pm0.0140(sys), 
\end{equation} 
where the systematic uncertainty accounted for the fact that 
there was a 1.5\% variation when changing the $z$ requirement on 
the UTC track extrapolation to the $T\cdot2$ counter.

\subsubsection{Normalization to the  $K_{\mu2}$ Branching Ratio} 
\label{sec:fs} 
 
Since a beam $K^+$ could decay after the $\check{\rm C}$erenkov
counter with a daughter satisfying the B4 and target requirement 
in $KB$, or a beam $K^+$ could deposit energy in the B4 and target
but exit the target without stopping, the total number 
of $K^+$'s that satisfy the $KB$ trigger requirement should be corrected  
for the $K^+$ stopping fraction ($f_s$). 
This fraction was obtained by normalizing the  
total $K^+$ exposure to the $K_{\mu2}$ branching ratio.  
The $K_{\mu2}$ events were selected from the $K\mu2$ monitor  
trigger data with the same selection criteria as those used for  
signal, excluding all the cuts related to the  
$\pi^+$ particle type, the BV and the BVL.  
The momentum, range and energy cuts for signal were  
replaced by a minimum 40~cm range requirement on the $K_{\mu2}$ events.  
The $K_{\mu2}$ acceptance measurement was also performed in the same way  
as that for the signal. This stopping fraction was computed by 
\begin{eqnarray} 
f_s&=&\frac{N_{K_{\mu2}}}{N_K^{eff}(K_{\mu2})\cdot 
    Acc(K_{\mu2})\cdot {\cal B}(K^+\to\mu^+\bar{\nu}_\mu)},\nonumber \\ 
   &=&0.7740\pm0.0011_{stat}, 
\end{eqnarray} 
where the $N_{K_{\mu2}}=355,119$ and was the number of surviving   
$K_{\mu2}$ events. 
$N_K^{eff}(K_{\mu2})=4.1475\times10^6$ and was the total  
exposure of $K^+$'s ($N_K$) with a prescaling factor  
for  the $K\mu2$ monitor 
trigger during the data acquisition period.  
$Acc(K_{\mu2})=17.4\%$ and was the acceptance.

\subsubsection{Confirmation of the  \KPITWO~Branching Ratio} 
\label{sec:brkp2} 
 
\begin{figure} 
\centering 
\epsfxsize 0.8\linewidth 
\epsffile{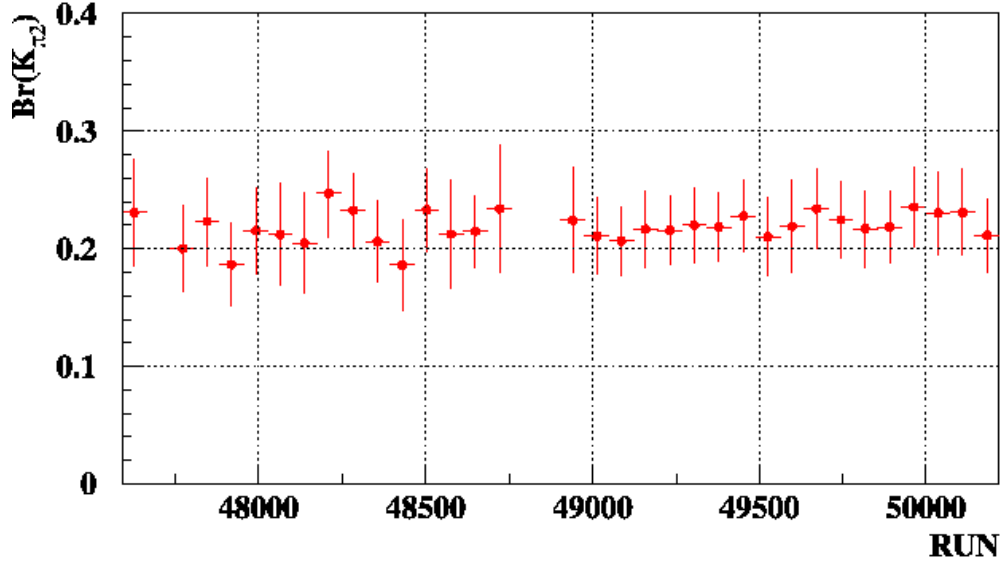} 
\caption{Measurement of the $K^+\rightarrow\pi^+\pi^0$ branching ratio  
as a function of run number in E949.} 
\label{brkpi2} 
\end{figure}

Measurement of the \KPITWO~branching ratio 
confirmed the validity of the  evaluation of the  
acceptance of the \KPPNN~selection cuts. 
Stopped \KPITWO~events were selected from the $K\pi2(1)$ monitor  
trigger data by imposing  cuts similar to the  
\KPPNN~selection criteria except for those used for the PV cuts  
and for defining the kinematic signal region.  
Good \KPITWO~events should also meet the requirements on the  
energy, momentum and range, which were defined to be within  
three standard deviations  of the \KPITWO~peaks. 
Fig.~\ref{brkpi2} shows the stability of the measured  
\KPITWO~branching ratio as  
a function of run number.  
The \KPITWO~branching ratio was measured to be 
\begin{eqnarray} 
{\cal B}(K^+\to\pi^+\pi^0)&=&\frac{N_{K_{\pi2}}}{N_K^{eff}(K_{\pi2})\cdot 
    Acc(K_{\pi2})\cdot f_s},\nonumber \\ 
   &=&0.219\pm0.005_{stat}, 
\end{eqnarray} 
where $N_{K_{\pi2}}=16,405$ and was the number of surviving $K_{\pi2}$ events. 
$N_K^{eff}(K_{\pi2})=1.3233\times10^6$ and was the total  
exposure of $K^+$'s ($N_K$) 
with a pre-scaling factor for the $K\pi2(1)$ monitor 
trigger during the data acquisition period.  
$Acc(K_{\pi2})=7.3\%$ and was the acceptance. 
This branching ratio was in agreement with the  
world average~\cite{Yao:2006px} value of 0.209$\pm$0.001  
within 4.7\%, indicating the validation of  
the acceptance measurement.  
This difference  was treated as the systematic  
uncertainty and assigned to the acceptance $A_{trig}$ in  
Equation~(\ref{eq:acc_trig}).   

\subsubsection{Summary of Acceptance and Sensitivity} 
 
The acceptances of the \KPPNN~decay were split into  
several parts as  given above. Table~\ref{tab:acc_and_sen} 
summarizes all the contributions to the total acceptance in  
the standard region $A^{standard}_{total}$. 
\begin{table} 
\centering 
\begin{tabular}{l c}\hline\hline 
Contribution           & Acceptance\\ \hline 
$A_{K_{\mu2}}$ 	       &$0.3796\pm0.0013_{stat}$ \\  
$A_{K_{\pi2}}$         &$0.8785\pm0.0029_{stat}$ \\  
$A_{\pi_{scat}}$       &$0.6161\pm0.0085_{stat}\pm0.0189_{sys}$ \\  
$A_{\pi\to\mu\to e}$   &$0.3523\pm0.0077_{stat}\pm0.0067_{sys}$ \\  
$A_{trig}$  	       &$0.1796\pm0.0010_{stat}\pm0.0084_{sys}$ \\  
$A_{PS}$               &$0.3630\pm0.0029_{stat}$ \\  
$A_{nucl.}$            &$0.4953\pm0.0077_{stat}\pm0.0248_{sys}$ \\  
$A_{T\cdot2}$          &$0.9358\pm0.0011_{stat}\pm0.0140_{sys}$ \\  
$f_{s}$                &$0.7740\pm0.0011_{stat}$ \\ \hline 
$A^{standard}_{total}(\times 10^{-3})$ &$1.69\pm0.05_{stat}\pm0.13_{sys}$ 
\\ \hline\hline 
\end{tabular} 
\caption{A breakdown of the acceptance for the \KPPNN~selection criteria.  
} 
\label{tab:acc_and_sen} 
\end{table} 
To get the acceptance in the extended signal region, the estimated  
acceptance gains given in Section~\ref{sec:extended} were applied to the  
acceptances for $A_{\pi\to\mu\to e}$, $A_{K_{\pi2}}$ and  
$A_{PS}$, yielding the final acceptance of  
\begin{equation} 
Acc.=(2.22\pm0.07_{stat}\pm0.15_{sys})\times 10^{-3}. 
\end{equation} 
This value is 10\% higher than that in E787. It is noted that  
the acceptance of the standard E787 signal region for 
the E949 data was 84\% of the acceptance for the E787 data 
due to losses incurred by the higher than expected instantaneous
rates (Section~\ref{subsect:overview}).
Based on the total exposure of $K^+$'s ($N_K$), 
the single event sensitivity ($SES$) of the  E949 2002 run was given by  
\begin{equation} 
SES=(2.55\pm0.08_{stat}\pm0.18_{sys})\times 10^{-10}. 
\end{equation} 
 
\subsection{Examining the Signal Region} 
\label{sec:candidate} 
 
After the background analysis and the acceptance measurement  
were completed and satisfactory,  
all the selection criteria were then be applied to the  
data. At the stage of examining the signal region, no cut could be changed.  
 
\begin{figure} 
\centering 
\epsffile{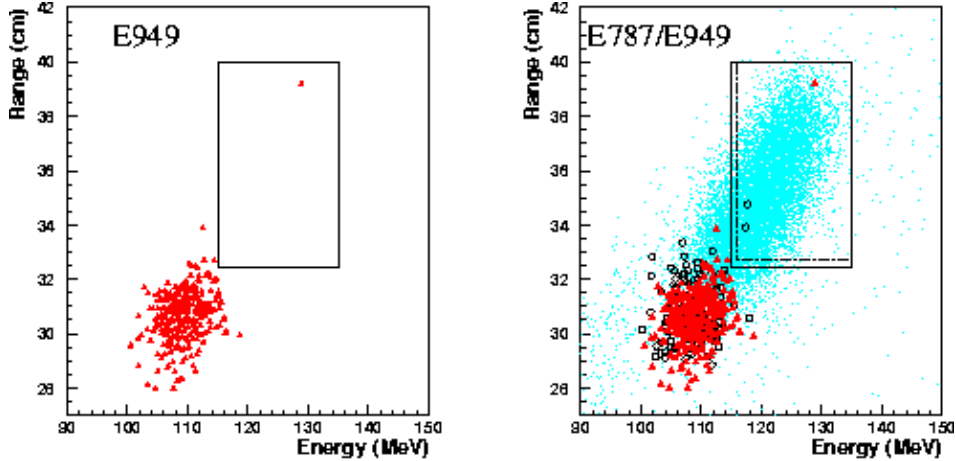} 
\caption{Range vs kinetic energy of the events satisfying  
all of the cuts, except for the phase space cuts 
on both the range and energy. The plots are shown separately for  
E949 only (left) and E787 plus E949 results.    
The rectangle represents the signal region defined  
in E787 (dashed lines) and E949 (solid lines). 
Events around $E=108$~MeV were  
due to \KPITWO, which were not removed by the photon veto cuts. 
The light points in the right hand plot represent the 
expected distribution of $K^+\to\pi^+\nu\bar{\nu}$ events from simulation.} 
\label{fig:candidate_R_vs_E} 
\end{figure}

\begin{figure} 
\centering 
\epsfxsize 0.8\linewidth 
\epsffile{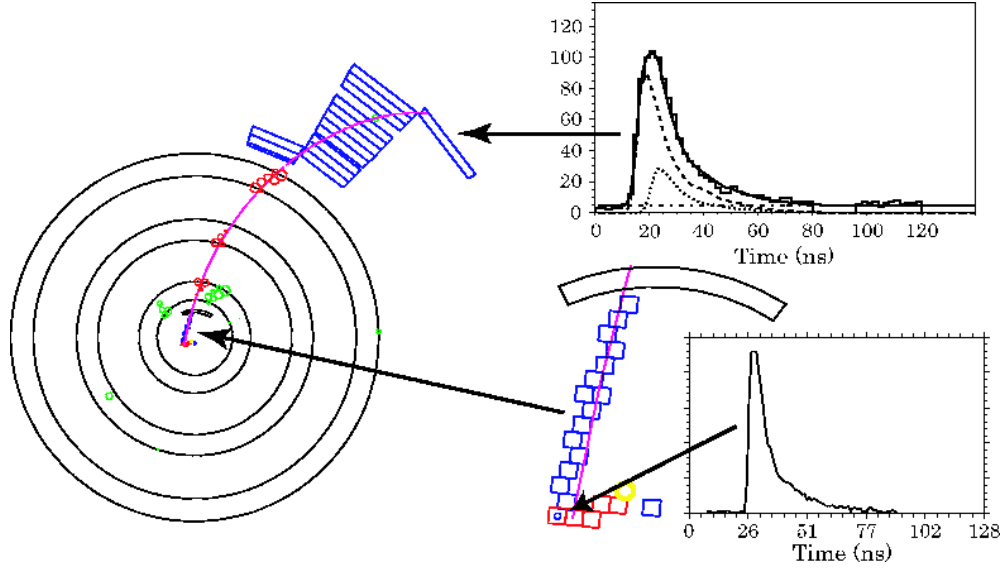} 
\caption{Reconstruction of the candidate event (end view).  
The clusters of squares indicate both the $K^+$ track and the  
$\pi^+$ track in the target. The hit IC sector is shown  
next to the $\pi^+$ cluster 
in the target. The curve is the result of the UTC track fit.  
The circles along the track are 
the hits in the UTC. The radius of each circle gives the drift distance.  
The RS and RSSC hit layers are shown outside the UTC.   
Also displayed are the TD data in the $\pi^+$ stopping 
counter, the reconstruction in the target and the CCD data in the  
$K^+$ stopping fiber. In the fit to the TD pulse shape, the $\pi^+$ pulse  
(dashed) and the $\mu^+$ pulse (dotted) are shown separately.  
No obvious $\pi^+$ pulse was observed in the CCD pulse shape 
for the $K^+$ stopping fiber.} 
\label{fig:event_display_candidate} 
\end{figure}

One candidate event was observed inside the signal region. 
A close check also found that this candidate was located in the  
$\pi^+\to\mu^+\to e^+$ extended region as described in 
Section~\ref{sec:extended}.      
Fig.~\ref{fig:candidate_R_vs_E} shows the range  
and kinetic energy of the events that passed  
all of the selection criteria, except for the phase  
space cuts on both the range and energy.  
This candidate 
together with the events observed in E787 are also shown 
in Fig.~\ref{fig:candidate_R_vs_E}. 
As indicated in the Figure, the signal box definition in E949 
was extended in comparison to that in E787.  
Fig.~\ref{fig:event_display_candidate} is an event display 
for this candidate. This event had a momentum of  
$227.3$~MeV/$c$, a kinetic energy of $128.9$~MeV and  
a range of $39.2$~cm. 
Kinematically, this candidate event  
agreed with all the requirements as a signal,  
though it existed near the limit expected for signal. 
The measured quantities of the observed candidate  
used in the selection criteria were compared to the  
expected distributions for signal to evaluate the signal probability 
distributions for the candidate.   
The probabilities for the single beam $K^+$  
requirements, the decay $\pi^+$ kinematic requirements and  
the $\pi^+\to\mu^+\to e^+$ decay sequence cuts showed 
a fairly flat distribution, which was consistent with the expected 
signal distribution. There was no observed photon activity  for 
this candidate.  

\section{Results} 
\label{chap:results}

In this section we describe the method used to obtain 
the \KPPNN~branching ratio and the impact of the E949 and E787  
\KPPNN~candidates on the unitarity triangle. We also describe the implication 
of the results  on the search for the hypothetical decay \KPIX\ where 
$X^0$ is a stable, massless, non-interacting particle~\cite{familon}. 
 
\subsection{Background Functions} 
 
\begin{figure} 
\centering 
\epsfxsize 1.0\linewidth 
\epsffile{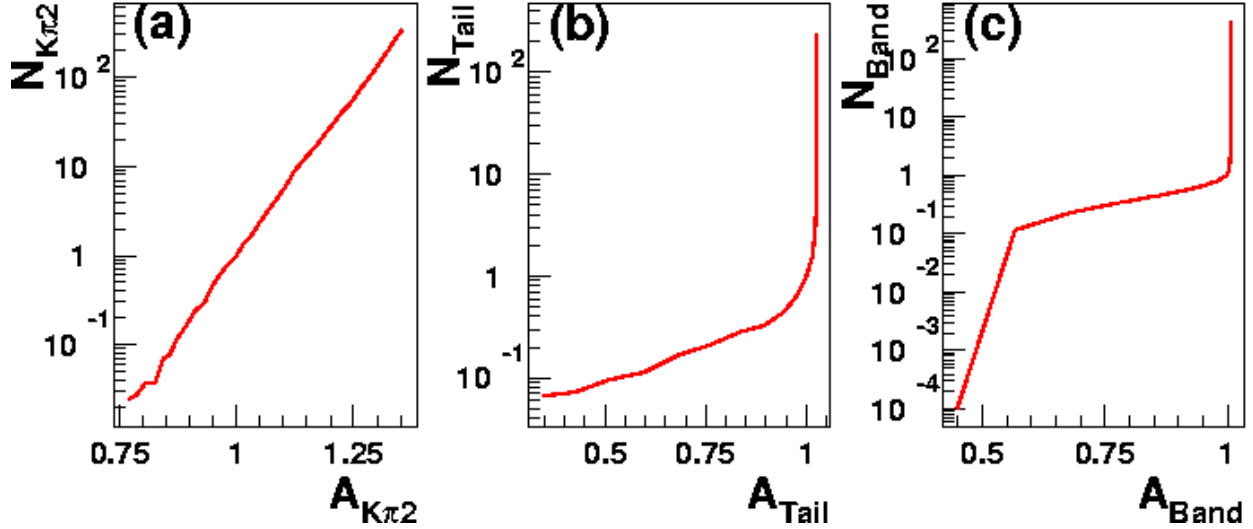} 
\caption{ 
The expected relative kinematic background rate in the normalization branch   
vs. the relative acceptance for the $K_{\pi2}$ background (a), the  
$K_{\mu2}$ range tail background (b) and  
the $\mu^+$ band background (c). 
Both background rates   
and acceptance curves were normalized to one at the  
cut positions.} 
\label{fig:kin_n_a} 
\end{figure}

\begin{figure} 
\centering 
\epsfxsize 1.0\linewidth 
\epsffile{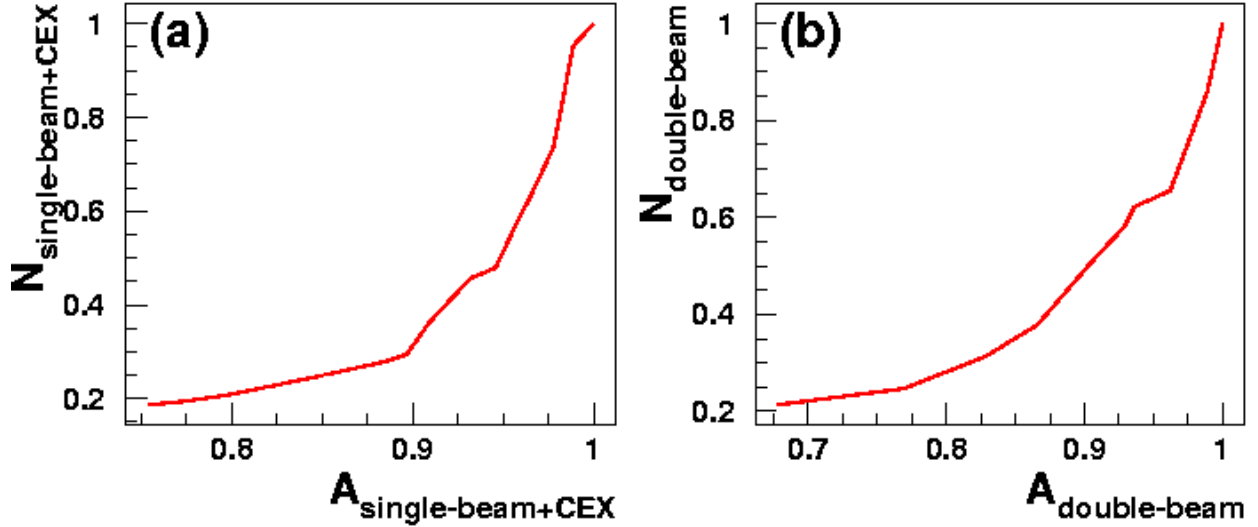} 
\caption{ 
The expected relative beam background rate  
vs. the relative acceptance for the single-beam and CEX backgrounds (a) 
and the double-beam background (b). 
Both background rates and acceptance curves were normalized to one at the  
cut positions.} 
\label{fig:beam_n_a} 
\end{figure}

We defined a number of cells in the extended signal region of differing 
signal/background, and calculated the expected 
signal/background ($dA/dN$) using 
individual $A_{\pi\nu\nu}^{type}$ and $N_{\pi\nu\nu}^{type}$ functions for 
each rejection or background type. In total, we had seven types: TD rejection, 
PV rejection, $K_{\pi2}$ background, $K_{\mu2}$ range tail background, $\mu^+$ 
band background, single-beam background and double-beam 
background. The variation of each type for different cut position or
cell could be expressed as the change of $N_{\pi\nu\nu}^{type}$ as a 
function of the corresponding acceptance $A_{\pi\nu\nu}^{type}$, yielding 
seven functions in total used in this analysis.  Two of them were  
the TD rejection vs.  acceptance function (Fig.~\ref{fig:nn_func}) and 
the PV rejection vs. acceptance  function (Fig.~\ref{fig:pv_tune_rva}). Three 
of them were the relative background rate in the normalization branch vs. 
the relative acceptance functions for the kinematic background 
backgrounds (Fig.~\ref{fig:kin_n_a}). The rest were the relative background 
rate vs. the relative acceptance functions for the beam  
backgrounds (Fig.~\ref{fig:beam_n_a}). Both relative background rates and the 
relative acceptance curves were normalized to one at the cut positions.

\subsection{Likelihood Method}  
 
The \KPPNN~branching ratio was determined using likelihood analysis 
incorporating the predicted background functions in the signal region.  
The likelihood ratio $X$ was defined as  
\begin{equation} 
X \equiv \prod^n_{i=1} 
\frac{e^{-(s_i+b_i)} (s_i+b_i)^{d_i}} 
     {{d_i}!} 
\Big{/} 
\frac{e^{-{b_i}}       b_i^{d_i}} 
     {{d_i}!} 
\ \ \ , 
\end{equation} 
where $s_i$ and $b_i$ were the estimated signal and background 
in the $i^{\rm th}$ cell,   
$d_i$ was the  number of signal candidates in the $i^{\rm th}$ cell  
and the product ran over all $n$ cells~\cite{junk}.  
In addition  the likelihood estimator  
$X_{obs}$ was defined as the value 
of $X$ given the 
observed candidates. 
Cells were defined based on the predicted  
background functions described in  
previous section. The predicted  
background functions showed that there was 
additional background rejection capability within the signal region that could 
be exploited by sub-dividing the signal region.  
The number of cells to be used 
for subsequent analysis, 3,781, was established prior to the  
examination of the signal region. 
 
The total background in the cell containing the signal candidate  
was estimated to be $5.75\times 10^{-5}$ dominated by a 
contamination of $4.92\times 10^{-5}$ events due to the  
$K_{\mu2}$ range tail background.   
The ratio of the acceptance in this cell to the total  
acceptance in the standard region 
($A^{standard}_{total}=0.001694$) was estimated to be 1.21$\times 10^{-4}$. 
The expected number of signal events in this cell was  
\begin{eqnarray} 
s_i & = & N_K \cdot {\cal B} \cdot A^{standard}_{total} \cdot A_i \nonumber \\ 
    & = & 1.77\times 10^{12} \times {\cal B} \times 0.001694 \times 1.21\times 10^{-4} \nonumber \\ 
    & = & 3.628\times 10^5 \times {\cal B}, 
\end{eqnarray} 
where \BR~was the \KPPNN~branching ratio.

\subsection{Branching Ratio of \KPPNN}

The central value of the branching ratio, defined 
as the value of \BR\ that maximized $X_{obs}$~\cite{central_value},  
was $0.96\times 10^{-10}$. 
Using only  the E949 data,  
 ${\cal B}(\KPPNN) = (0.96^{+4.09}_{-0.47})\times 10^{-10}$ where the 
quoted 68\% confidence level (CL) interval was determined from  
the behavior of $X$ as described in Ref.~\cite{junk} and included  
only the statistical uncertainty. 
The estimated probability that the E949 candidate was  
due to background alone was 0.074.

The results from  E787 and E949 were combined to calculate  
the branching ratio for \KPPNN. 
In the  E787 \KPPNN~analysis, two \KPPNN~candidate events  
were observed in the signal  
region~\cite{E787-1998}. 
The number of cells describing the E949 signal region were  
augmented by 488 cells that defined 
the signal region for the E787 analysis to produce a  
likelihood estimator $X_{obs}$ for 
the combined data.

The  confidence intervals for the combined E787 and E949 results  
took into account 
the estimated systematic uncertainties in the signal acceptance  
and the background rates.  
The systematic uncertainty of each background source was estimated 
to be about 15\% based upon the results of the correlation studies. 
From the study in Section~\ref{chap:acc_and_sen}, the systematic uncertainty  
on the acceptance was estimated to be about 8\%.  
The systematic uncertainty of each background 
component and the acceptance were assumed to be uncorrelated and 
to follow a normal distribution with the magnitudes given 
above regarded as one standard deviation. 
With these assumptions, the \KPPNN~branching ratio for the  
combined E787 and E949  
result was  
${\cal B}(\KPPNN) = (1.47^{+1.30}_{-0.89})\times 10^{-10}$  
where the uncertainty  
denoted the 68\% CL interval.  
The corresponding 90\% and 95\% CL intervals were 
$(0.27,3.84)\times10^{-10}$ and $(0.17,4.44)\times10^{-10}$, respectively. 
The estimated probability that all the \KPPNN~candidates  
observed in E787 and E949  
were due to background was 0.001.  
The inclusion of the estimated systematic uncertainties had a negligible 
effect on the CL intervals due to the relatively poor statistical precision 
inherent in a sample of three candidate events. 
 
\subsection{Search for  ${\cal B}($\KPIX)} 
 
The experimental signature of a \KPIX~decay was identical to that of \KPPNN\  
except that the kinematic signature afforded by the two-body decay  
 ($P_\pi=227.1$~MeV/$c$, $E_\pi=127.0$~MeV, $R_\pi=38.6$~cm)  
permitted the definition 
of a relatively high-acceptance, low-background signal region. 
The analysis method was identical to that used for the \KPPNN~analysis 
except that the signal region was defined to be 
within two standard deviations of the expected momentum, energy and range of 
the $\pi^+$ with the upper limits of the  
tightened to $P\le 229$~MeV/$c$, $E\le 135$~MeV,  
and $R\le 40$~cm  to suppress $K^+\to\mu^+ X$ background. 
The expected background level was small (0.05 events),  
because the region was far from the \KPITWO~peak. 
The acceptance studies for \KPIX\ decay paralleled  
those for the \KPPNN. The single event sensitivity for  
the E949 \KPIX~ decay analysis was estimated to be 
$(0.82\pm0.02_{stat}\pm0.06_{sys})\times 10^{-10}$. 
 
The candidate event observed in the signal region  
for \KPPNN~E949 analysis was also in the \KPIX~signal region. 
However, no candidates were observed in the \KPIX~signal region of  
E787~\cite{E787-1998}. 
The combined E787 and E949 sensitivity was 0.196$\times 10^{-10}$ and, 
using the one observed candidate event  
without subtraction of the estimated background,  
the upper limit on the branching ratio was  
${\cal B}(\KPIX) < 0.73 \times 10^{-10}$ at 90\% CL  
using the Feldman-Cousins method~\cite{Feldman}. 
This limit was larger than the  
previous 90\% CL limit of $0.59\times 10^{-10}$ 
of E787~\cite{E787-1998}  due to the   E949 candidate event.

\subsection{Impact on the Unitarity Triangle} 
 
\begin{figure} 
\centering 
\epsfxsize 0.9\linewidth 
\epsffile{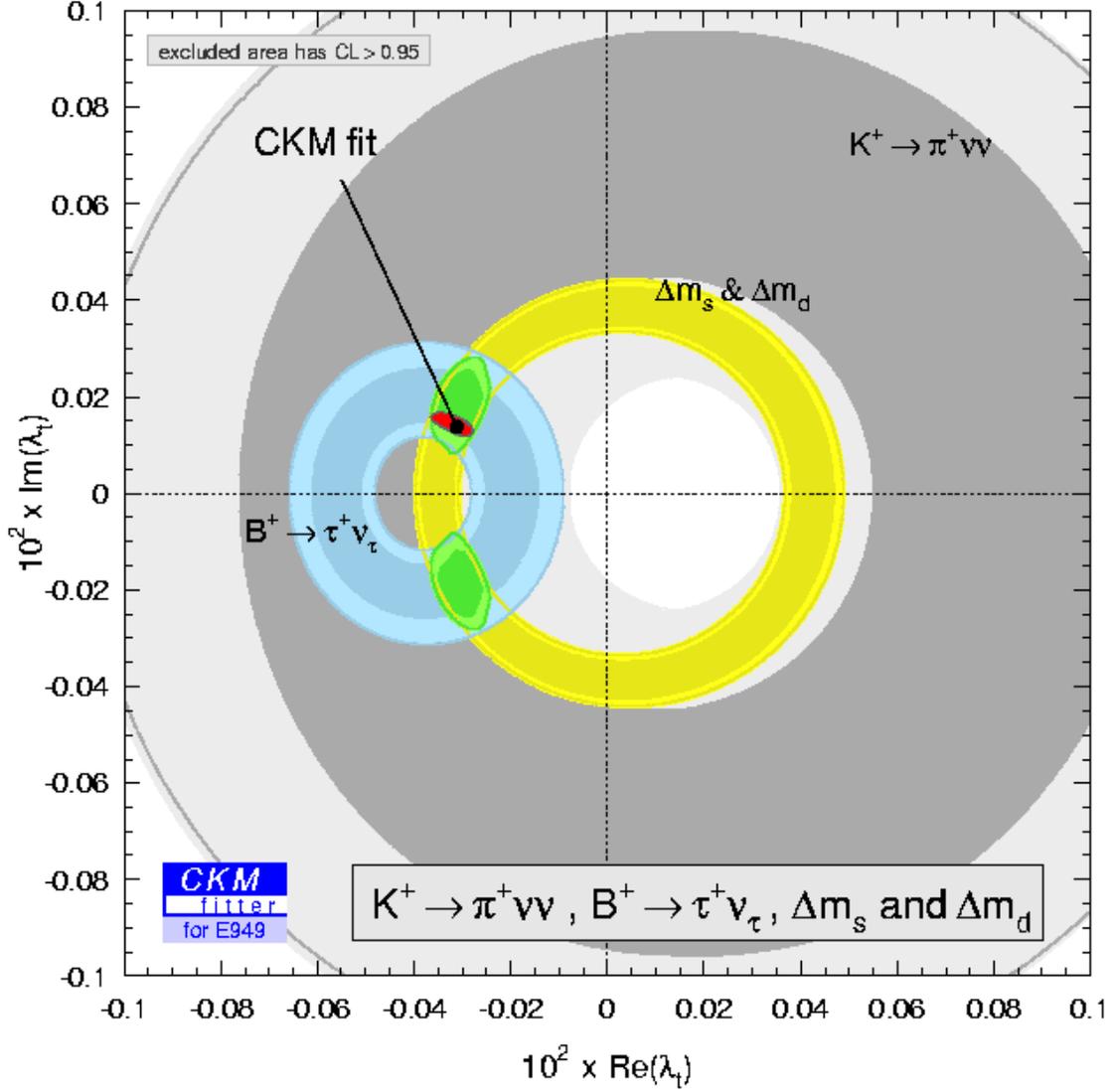} 
\caption{The allowed regions in the $\lambda_t$ plane allowed 
by the combined E787 and E949 determination of  
the \KPPNN~branching ratio (gray), 
$B^+\to\tau^+\nu$ (blue) and $B$-mixing measurements (yellow).  
The regions outside the lighter (darker) shading have CL$>$0.95 (0.68). 
The area excluded by \KPPNN\ at CL $>$0.95 is indicated by the gray line. 
The red-shaded region is allowed by the combination of these measurements and 
the small black region denotes the region allowed  by  
all CKM-related measurements  
as evaluated by the CKMfitter Group~\protect~\cite{CKMfitter}. 
} 
\label{fig:lambdatforE949} 
\end{figure}

 As described in Section~\ref{chap:intro}, the  
\KPPNN~branching ratio was directly 
related to the real and imaginary parts of  
$\lambda_t \equiv V^*_{ts}V_{td}$ (Equation~(\ref{eq:pnn_br})). 
In Fig.~\ref{fig:lambdatforE949} the regions of the  
complex $\lambda_t$ plane allowed by the \KPPNN~branching  
ratio determined from the combined E787 and E949 
results was compared to the regions allowed by other recent measurements with  
small theoretical uncertainties~\cite{CKMfitter}. The region  
favored by other CKM-sensitive measurements 
is at the edge of the 68\% CL region allowed by the \KPPNN\ measurement. 
 
  The other CKM-sensitive results~\cite{CKMfitter} used to produce the 
confidence level intervals in Fig.~\ref{fig:lambdatforE949}  
are dominated by measurements of $B$ meson decays.  
The possible discrepancy between the $\lambda_t$ regions allowed by 
the $B$-decay measurements and by ${\cal B}(\KPPNN)$ could be an  
indication of physics beyond the SM.  
As emphasized in Ref.~\cite{BBIL}, the  
clean theoretical interpretation of $K\to\pi\nu\bar{\nu}$ 
remains valid in most extensions of the SM in distinct  
contrast to the $B$-decay measurements currently used 
to determine the CKM parameters.  
Thus a precise measurement of ${\cal B}(\KPPNN)$  
would provide an  
unambiguous consistency test of the flavor sector of the SM. 
 
\section{Conclusion}

The rare decay \KPPNN~is a flavor-changing-neutral-current process and 
proceeds via 1-loop diagrams  
mediated mainly by the top quark.  
Measuring \BR(\KPPNN) is one of the cleanest ways to extract $|V_{td}|$. 
 
In this paper we have reported results from the 
BNL experiment E949, an upgraded version of the  
BNL-E787 experiment,  designed to improve the  
sensitivity for measurement of \KPPNN~decay.  
All the $K^+$ decays at rest were analyzed using  a  
blind analysis technique in which the  
signal region was masked  until 
the selection criteria were determined and the background  
levels were estimated. The development of the cuts and  
the estimation of the background levels were performed  using a  
bifurcation method and, a likelihood analysis method was  
developed for interpreting the quality of candidate events.  
Enlargement of the signal region compared to E787 analysis 
increased the acceptance by 30\% with a  
total background level in the signal region  
estimated to be $0.30\pm0.03$ events. 
 
An examination of the signal region yielded one event near the upper  
kinematic limit of the decay \KPPNN.  
Based on the candidate event, the branching ratio was determined to be 
${\cal B}(K^+\to\pi^+\nu\bar{\nu})=(0.96^{+4.09}_{-0.47})\times 10^{-10}$. 
E787 and E949 results were combined  
and the branching ratio was determined  to be  
$(1.47^{+1.30}_{-0.89})\times 10^{-10}$ at the 68\% CL level  
based on three events observed in the momentum region  
$211\le P\le 229$~MeV/$c$. 
The estimated probability that all the \KPPNN~candidates  
observed in E787 and E949  
were due to background was 0.001.  
The measured branching ratio
is in agreement with the SM prediction of \PNNSM within the uncertainty.
 
\begin{acknowledgments} 
We gratefully acknowledge the support and efforts of  
the BNL Collider-Accelerator Division 
for the high quality $K^+$ beam delivered. 
We wish to thank Jose Ocariz of the CKMfitter Group for producing  
Fig.~\ref{fig:lambdatforE949}. 
This research was supported in part by the U.S.  
Department of Energy, the Ministry of Education,  
Culture, Sports, Science and Technology of Japan through  
the Japan-U.S. Cooperative Research Program in High Energy  
Physics and under Grant-in-Aids for Scientific Research,  
the Natural Sciences and Engineering Research Council and  
the National Research Council of Canada, the Russian 
Federation State Scientific Center Institute for High Energy  
Physics, and the Ministry of Science and Education of the Russian Federation. 
S. Chen was also supported by Program for New Century Excellent 
Talents in University from the Chinese Ministry of Education. 
\end{acknowledgments}


\begin{thebibliography}{ }
\bibitem{central_value}  
 This definition of the central value of 
 the branching ratio has the desirable feature  
 that in the limit of $N_{obs}$ observed 
 candidates with high signal-to-background, the central value is  
 the same as that given by the product of the  
 sensitivity (Section~\ref{chap:acc_and_sen}) 
 and $N_{obs}$ which 
 is the fundamental definition of the branching fraction.
\end{thebibliography}
\end{document}